\documentclass[]{interact}

\usepackage{epstopdf}
\usepackage[caption=false]{subfig}

\usepackage[numbers,sort&compress]{natbib}
\bibpunct[, ]{[}{]}{,}{n}{,}{,}
\renewcommand\bibfont{\fontsize{10}{12}\selectfont}
\makeatletter
\def\NAT@def@citea{\def\@citea{\NAT@separator}}
\makeatother

\theoremstyle{plain}

\theoremstyle{definition}

\theoremstyle{remark}

\begin{document}


\title{Scaling invariance: a bridge between geometry, dynamics and criticality}

\author{
\name{$^1$Edson D. Leonel\thanks{CONTACT Edson D. Leonel Author. Email: edson-denis.leonel@unesp.br}, $^2$Diego F. M. Oliveira}
\affil{$^1$Departamento de F\'isica, Unesp - Universidade Estadual Paulista - 
Av.24A. 1515, 13506-900, Rio Claro, SP, Brazil\\
$^2$School of Electrical Engineering and Computer Science, University of North Dakota, Grand Forks, Avenue Stop 8357, 58202, ND, USA}
}

\maketitle

\begin{abstract}
Scale invariance is a central organizing principle in physics, underlying phenomena that range from critical behaviour in statistical mechanics to transport and chaos in nonlinear dynamical systems. Here we present a unified and physically motivated exploration of scaling concepts, emphasizing how invariance under rescaling transformations emerges across systems of increasing dynamical complexity. Rather than adopting a purely abstract approach, we combine simple geometrical constructions, analytical arguments, and prototypical dynamical models to build physical intuition. We begin with elementary, easily reproducible examples governed by a single control parameter, showing how power-law behaviour naturally arises when characteristic scales are absent. These examples illustrate how scaling ideas can be used to extract quantitative information from minimal data. We then extend the discussion to nonlinear dynamical systems exhibiting local bifurcations, where two scaling variables control the relaxation toward stationary states. In this context, scaling invariance manifests through critical exponents, crossover phenomena, and critical slowing down, allowing systems of different dimensionality to be grouped into universality classes. Finally, we address continuous phase transitions in chaotic dynamical systems, including transitions from integrability to non-integrability and from bounded to unbounded diffusion. By drawing on concepts traditionally associated with statistical mechanics -- such as order parameters, susceptibilities, symmetry breaking, elementary excitations, and topological defects -- we show how these transitions can be interpreted within a coherent scaling framework. Taken together, the examples discussed here demonstrate that scaling invariance provides a unifying language for understanding structure, transport, and criticality in nonlinear systems, bridging deterministic dynamics and nonequilibrium statistical physics in a transparent and physically intuitive manner.
\end{abstract}


\textbf{Outline of the paper.}
This paper is organized as a progressive exploration of scaling invariance and its
applications in nonlinear dynamical systems, with increasing levels of complexity.
Rather than presenting the subject in a purely abstract manner, we combine simple
illustrative examples, analytical arguments, and physically motivated models to
highlight the unifying role of scaling concepts across different contexts.

We begin by introducing scaling invariance in systems governed by a single control
parameter, using elementary and easily reproducible examples to build physical
intuition. These examples illustrate how power-law behaviour and homogeneous
functions emerge in the absence of characteristic scales.

The discussion is then extended to nonlinear dynamical systems exhibiting local
bifurcations, where two scaling variables control the relaxation towards stationary
states. In this context, we show how critical exponents, crossover phenomena, and
critical slowing down arise naturally and how systems of different dimensionality
can be grouped into universality classes.

Subsequently, we address continuous phase transitions in chaotic dynamical systems.
Focusing on transitions from integrability to non-integrability and from bounded to
unbounded diffusion, we demonstrate how concepts from statistical mechanics -- 
such as order parameters, susceptibilities, symmetry breaking, elementary excitations, and topological defects -- provide a coherent and physically transparent framework for interpreting the onset of chaos and diffusion.

The paper concludes with a synthesis of these results, emphasizing the universality
of scaling behaviour across mappings and billiard systems, and highlighting the
conceptual bridge between nonlinear dynamics and nonequilibrium statistical mechanics.

\tableofcontents

\section{Introduction}

Scale invariance is one of the most powerful and unifying concepts in modern physics. It provides a framework through which systems operating at vastly different length, time, and energy scales can be understood within a common language. Whenever a system exhibits no characteristic scale describing its dynamics, its macroscopic observables typically follow power-law behaviours. Such behaviour signals the presence of scale invariance and reveals that the underlying dynamics is governed by a set of scaling exponents rather than by microscopic details.

The importance of scale invariance spans a wide range of physical contexts. It plays a central role in the theory of critical phenomena in statistical mechanics \cite{ref1}, but it also appears in fluid turbulence \cite{ref2}, fracture processes \cite{ref3}, geophysical structures \cite{ref4}, biological and social networks \cite{ref5}, and even in financial markets \cite{ref6}. In all these cases, the absence of a privileged scale leads to self-similar behaviour under rescaling transformations \cite{ref7}. Systems that share the same set of scaling exponents, despite having different microscopic dynamics, are said to belong to the same universality class \cite{ref8}. This remarkable property allows seemingly unrelated systems to be described and understood within a unified conceptual framework.

Beyond its foundational role in understanding complexity \cite{ref9}, scale invariance has become a powerful diagnostic tool. The identification of power-law behaviour and scaling relations often provides early evidence of critical regimes, signalling the proximity of phase transitions \cite{ref10} or qualitative changes in the system's dynamics. In this sense, scaling ideas are not only descriptive but also predictive, offering insight into how systems respond to changes in control parameters and how different dynamical regimes are connected.

In this paper, we present different applications of the scaling formalism, organized in a progression of increasing complexity. We begin with simple systems described by a single control parameter, where scaling behaviour can be identified and analysed in a transparent manner. Two illustrative examples are discussed: (i) a folded paper boat, in which the maximum distance between the bow and the stern scales with the mass of the paper, and (ii) a crumpled paper ball \cite{ref11}, whose fractal dimension can be extracted from the scaling relation between its size and mass. Although elementary, these examples capture the essence of scaling invariance in one-parameter systems.

We then move to nonlinear dynamical systems that exhibit local bifurcations \cite{ref12}, focusing on discrete mappings. The discussion starts with one-dimensional mappings and is subsequently extended to two-dimensional cases. In such systems, bifurcations are typically controlled by two scaling variables, leading to the emergence of critical exponents that characterise the convergence towards stationary states \cite{ref10}. When the control parameter is tuned exactly to the bifurcation point, the relaxation dynamics is governed by a homogeneous and generalized function, giving rise to a set of three critical exponents. Away from the bifurcation, either before or after it, the convergence becomes exponential. The crossover between these regimes defines the phenomenon known as \textit{critical slowing down} \cite{ref13}, a clear dynamical manifestation of scale invariance. Systems that share the same critical exponents at bifurcation can be grouped into universality classes, highlighting the unifying power of the scaling approach.

Scaling invariance also plays a crucial role in the characterization of phase transitions \cite{ref8}. Within the framework of statistical mechanics \cite{ref1}, phase transitions are commonly classified as first order (discontinuous) or second order (continuous), according to the behaviour of derivatives of the free energy, following the Ehrenfest classification \cite{ref14}. A defining feature of continuous phase transitions is the emergence of scale invariance near the critical point: observables follow power-law behaviour, correlation lengths diverge, and no characteristic scale dominates the system. As a consequence, rescaling transformations leave the functional form of the governing equations unchanged in the vicinity of criticality.

We illustrate these ideas by discussing two classes of phase transitions in chaotic dynamical systems. The first corresponds to the transition from integrability to non-integrability in two-dimensional, area-preserving mappings \cite{ref15} and in billiard systems \cite{ref16}. The second concerns the transition from bounded to unbounded diffusion, observed in dissipative mappings such as the Chirikov-Taylor map \cite{ref17} and in time-dependent billiards \cite{ref18}, where fractional energy loss occurs at each collision. In this context, chaos is characterised by the exponential separation of nearby trajectories, quantified by positive Lyapunov exponents \cite{ref13}, while integrable systems possess as many constants of motion as degrees of freedom. Billiards \cite{ref19} provide a particularly clear physical realisation of these concepts, describing the motion of particles confined within closed boundaries and undergoing elastic or inelastic collisions.

The paper is organised as follows. In Section~\ref{sec2}, we discuss the basic properties of the scaling formalism for systems governed by a single control parameter. Section~\ref{sec3} presents applications to local bifurcations in discrete mappings, highlighting the robustness of critical exponents across different dimensionalities. Section~\ref{sec4} is devoted to the characterization of the transition from integrability to non-integrability, where we address a set of key questions required to identify continuous phase transitions from a scaling perspective. The discussion of the phase transition from bounded to unbounded diffusion is made in continuation of the section for the Chirikov-Taylor map and for a time-dependend billiard with particles suffering innelastic collisions with the boundary. Our final remarks and conclusions are presented in Section~\ref{sec5}.

\section{Scaling invariance for one parameter}
\label{sec2}

In this section we illustrate how scaling invariance emerges in systems governed by a single control parameter. Rather than starting with abstract models, we focus on two simple and easily reproducible examples that can be realised with minimal resources: (i) the folding of a paper boat, and (ii) the determination of the fractal dimension of a crumpled sheet of paper. Despite their simplicity, these examples capture the essential logic of scaling arguments and provide a clear physical intuition for the concepts introduced in later sections.

\subsection{Scaling for the paper boat folding}

We begin with a simple experiment based on the construction of a paper boat from a rectangular sheet of paper. Although familiar as a childhood exercise, the folding procedure leads to a non-trivial geometrical object whose characteristic size depends on the amount of material used.

The quantity of interest is the length $l$ of the boat, defined as the maximum distance between the forward extremity of the bow and the aft extremity of the stern. To investigate how this length depends on the amount of material, we proceed as follows. Starting from two identical sheets of paper, one sheet is kept as a reference, while the other, of mass $m$, is used to construct a paper boat according to the folding procedure described in Appendix \ref{app1}. The resulting length of the boat is denoted by $l_0$.

The remaining reference sheet is then cut into two equal parts, each with mass $m/2$. One half is used to construct a second boat of length $l_1$, while the other half is further divided into two pieces of mass $m/4$. This procedure is iterated successively, producing boats from sheets of mass $m/8$, $m/16$, and so on, until a total of seven boats is obtained following this protocol. Table~\ref{tab1} summarises the dimensionless values of the mass, $m/m_0$, and the corresponding length, $l/l_0$, where $m_0$ and $l_0$ refer to the original sheet and its associated boat. The original paper used in the procedure was USA Letter size of $215.9mm\times 279.4mm$ with a density of $75g/m^2$.

\begin{table}[t]
\centering
\caption{Table with the data obtained for the paper folding experiment. First column shows the boat. Second and third column show respectivelly the dimensionless mass $m/m_0$ and $l/l_0$ where $m_0$ and $l_0$ are the mass and the lenght of the boat for the original size paper.}
\begin{tabular}{|c|c|c|}
\hline
\textbf{boat} & \textbf{mass $m/m_0$} & \textbf{length $l/l_0$} \\
\hline
1  & $1$ & $1.0000(2)$ \\
2  & ${{1}/{2}}$ & $0.6495(2)$ \\
3  & ${{1}/{4}}$ & $0.50000(1)$ \\
4  & ${{1}/{8}}$ & $0.3224(1)$ \\
5  & ${{1}/{16}}$ & $0.2430(1)$ \\
6  & ${{1}/{32}}$ & $0.1542(1)$ \\
7  & ${{1}/{64}}$ & $0.1215(1)$ \\
\hline
\end{tabular}
\label{tab1}
\end{table}

A clear trend emerges from Table~\ref{tab1}: as the mass of the paper is reduced by successive factors of two, the length of the boat decreases accordingly, but not in a linear fashion. This observation suggests that the relationship between $l$ and $m$ is governed by a scaling law. We therefore assume that the length can be described by a homogeneous function of the form
\begin{equation}
l(m)=\ell\, l(\ell^a m),
\label{eq1}
\end{equation}
where $\ell$ is an arbitrary scaling factor and $a$ is a characteristic exponent.

Equation~(\ref{eq1}) encapsulates the essence of scaling invariance: a change in the mass of the paper can be compensated by an appropriate rescaling of the system's length, leaving the functional form of $l(m)$ unchanged. If $a=-1$, the relation between $l$ and $m$ would be strictly linear. However, the experimental data indicate a different behaviour. By choosing the scaling factor such that $\ell^a m = 1$, we obtain $\ell = m^{-1/a}$. Substituting this into Eq.~(\ref{eq1}) yields
\begin{eqnarray}
l(m) &=& m^{-1/a} l(1), \nonumber\\
l(m) &\propto& m^{-1/a},
\end{eqnarray}
where $l(1)$ is a constant.

\begin{figure}[t]
\centerline{(a)\includegraphics[width=0.5\linewidth]{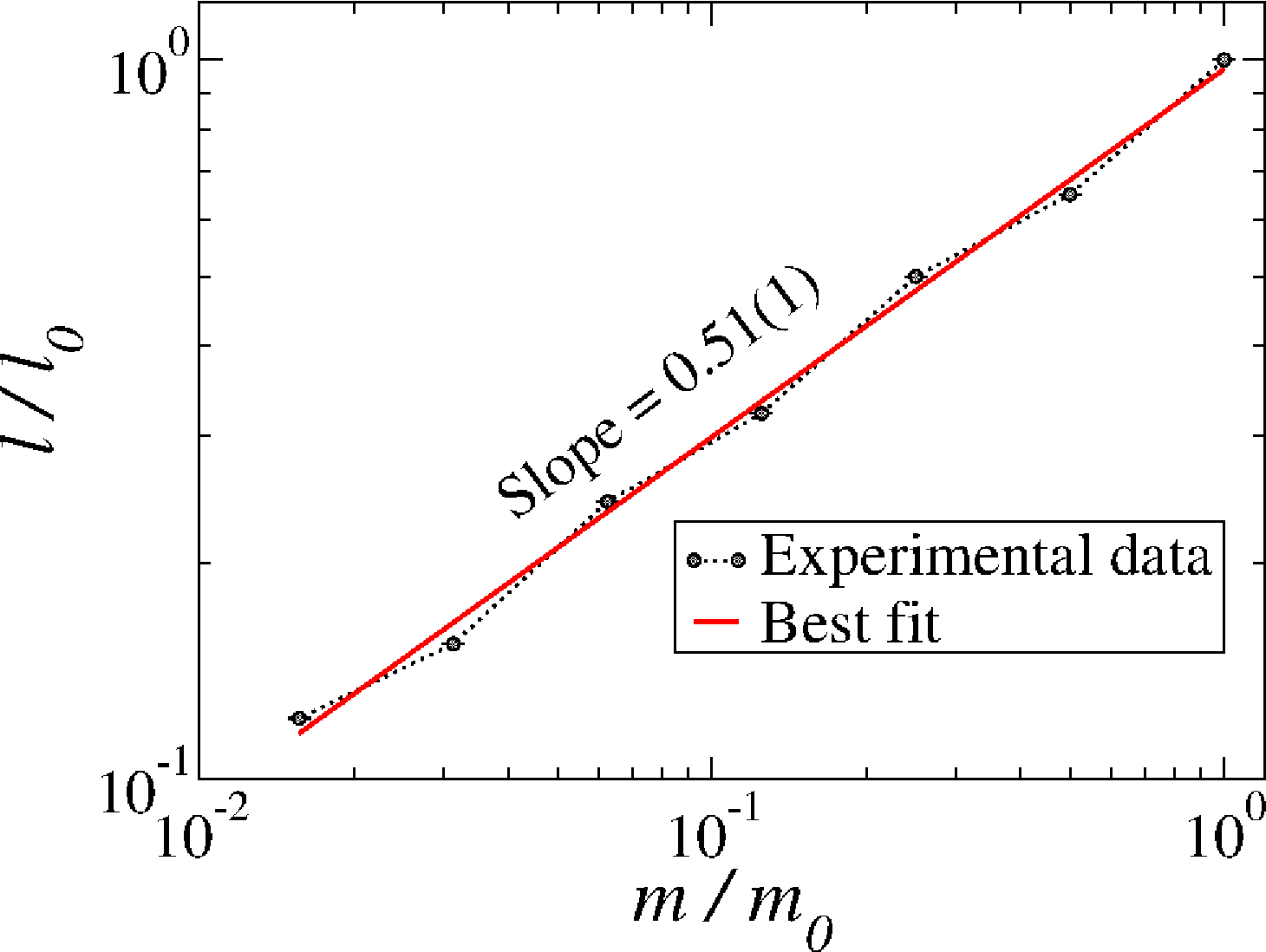}
(b)\includegraphics[width=0.45\linewidth]{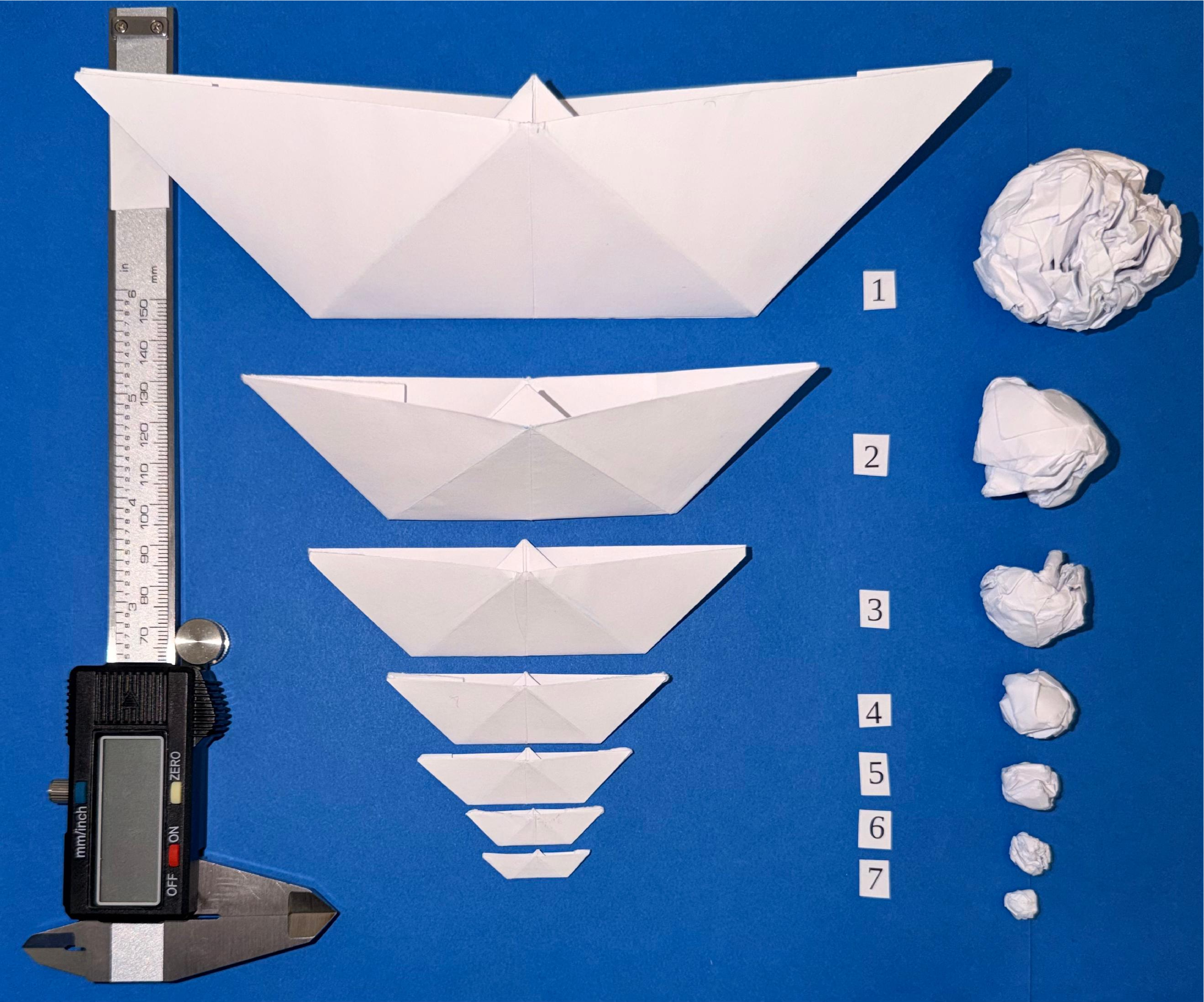}}
\caption{(a) Plot of $l/l_0$ versus $m/m_0$ for the paper boat folding experiment. A power-law fit yields a slope $0.51(1)\approx 1/2$, corresponding to a scaling exponent $a=-2$. (b) Illustation of the paper boat folding and the crumpled paper. Error bars in (a) represent the uncertainty associated with each measurement.}
\label{Fig1}
\end{figure}

The exponent $a$ can be extracted from a log-log plot of $l$ as a function of $m$. Figure~\ref{Fig1}(a) shows such a plot for the data in Table~\ref{tab1}. A power-law fit yields a slope of $0.51(1) \approx 1/2$, leading to the conclusion that $a=-2$. This result implies that the characteristic length of the paper boat scales as the square root of its mass, a clear manifestation of scaling invariance in a simple one-parameter system.

As a closing remark for this section, the paper boat experiment provides a pedagogical yet physically meaningful illustration of scaling invariance in a system governed by a single control parameter. Despite its simplicity, the experiment reveals a central idea that permeates many areas of physics: when a system lacks a characteristic scale, its macroscopic observables follow power laws, and their behaviour is fully encoded in scaling exponents rather than in microscopic details. The folding procedure acts only as a geometric constraint, while the observed non-linear dependence between length and mass emerges naturally from scale invariance. This example demonstrates how homogeneous functions and scaling arguments can be used to extract quantitative information from minimal data, providing clear physical intuition for more complex systems.

\subsection{Scaling for crumpled paper: determination of a fractal dimension}

To illustrate the effectiveness of the scaling formalism, we consider now a second one-parameter application to determine the fractal dimension of crumpled paper balls \cite{ref11}. We begin by recalling how scaling arguments naturally reproduce the Euclidean dimension of regular objects. For a one-dimensional object, such as a wire of linear mass density $\tilde{\lambda}$ and length $L$, the total mass is given by $m=\tilde{\lambda}L$. Since the mass depends linearly on $L$, it follows that
\begin{equation}
m \propto L^{D_o},
\end{equation}
where $D_o$ denotes the dimension of the object. In this case, $D_o=1$.

A similar argument applies to two-dimensional objects. Consider a surface mass distribution with superficial density $\sigma$. For simplicity, let the object be a square of side $L$, so that its mass is $m=\sigma L^{2}$. The same scaling relation $m \propto L^{D_o}$ holds, now with $D_o=2$.

The extension to three dimensions is straightforward. For a homogeneous object with volumetric mass density $\rho$ and cubic shape of linear size $L$, the mass reads $m=\rho L^{3}$, yielding $D_o=3$. In all these cases, the scaling exponent coincides with the Euclidean dimension of the embedding space.

However, not all physical objects exhibit such a simple correspondence between scaling and Euclidean dimension. Fractality emerges when an object develops a highly convoluted geometry, allowing, for instance, a very large surface area to be accommodated within a finite volume, as famously observed in biological structures such as the lungs \cite{Weibel}. In such cases, surface area must be maximized to ensure efficient gas exchange, while the total volume remains bounded. Despite this geometric complexity, the mass of the object continues to obey a well-defined scaling relation with its characteristic size.

We now turn to the specific case of crumpled paper balls \cite{ref11}. Each ball originates from a flat rectangular sheet of paper, which is a two-dimensional object with $D_o=2$. Upon crumpling, however, the sheet undergoes a highly complex topological deformation [see Fig. \ref{Fig1}(b)]. The resulting object is no longer planar, yet it does not fully occupy the three-dimensional space in which it is embedded. Consequently, its effective dimension lies between the original value $D_o=2$ and the embedding dimension $D_e=3$. We denote this intermediate exponent by $D_f$, the fractal dimension.

This generally non-integer value characterizes the fractal nature of the crumpled paper ball and can be quantitatively determined through scaling arguments relating its mass to its characteristic size. The experiment begins with two identical flat rectangular sheets of paper, each of mass $m$. We also used USA Letter papers of size $215.9mm\times 279.4mm$ with a density of $75g/m^2$.

From the perspective of Euclidean geometry, the initial sheet has dimension $D_o=2$. One sheet is manually crumpled into a roughly spherical ball. The diameter of the resulting object is measured along different directions -- we made an ensemble of 20 measures randomly chosen -- allowing an average radius $r$ to be defined. As reported in \cite{ref11}, the diameter of a crumpled paper ball exhibits a slow relaxation in time due to the viscoelastic properties of paper, stabilizing only after a sufficiently long resting period.

The second sheet is then cut in half, and one half, of mass $m/2$, is used to produce a new ball following the same procedure. The remaining half is again divided, and the process is repeated until a total of seven balls with progressively smaller masses are obtained.

The central question is how the mass of the ball relates to its radius. In analogy with the previous discussion, we assume that the mass satisfies a scaling relation of the form
\begin{equation}
m(r)=\ell\, m(\ell^a r),
\label{eq2}
\end{equation}
where $\ell$ is a scaling factor and $a$ is a characteristic exponent. Repeating the same scaling argument leads to
\begin{eqnarray}
m(r) &\propto& r^{-1/a}, \nonumber\\
m(r) &\propto& r^{D_f},
\label{eq3}
\end{eqnarray}
where $D_f=-1/a$ defines the fractal dimension of the crumpled paper ball. Measuring the dependence of $m$ on $r$ therefore allows a direct experimental determination of the fractal dimension, providing a concrete illustration of how scaling arguments can be used to characterize geometrical complexity in physical systems.

Table~\ref{tab2} presents the dimensionless values of $m/m_0$ and $r/r_0$ obtained from the crumpling experiments, where $m_0$ and $r_0$ correspond to the original sheet.

\begin{table}[ht]
\centering
\caption{Table showing the data for the ball experiment. First column indicate the ball, second column the dimensionless mass $m/m_0$ and third column for the dimensionless radius $r/r_0$.}
\begin{tabular}{|c|c|c|}
\hline
\textbf{ball} & \textbf{mass $m/m_0$} & \textbf{radius $r/r_0$} \\
\hline
1  & $1$ & $1.0000(3)$ \\
2  & ${{1}/{2}}$ & $0.7475(3)$ \\
3  & ${{1}/{4}}$ & $0.5495(2)$ \\
4  & ${{1}/{8}}$ & $0.4208(2)$ \\
5  & ${{1}/{16}}$ & $0.3168(2)$ \\
6  & ${{1}/{32}}$ & $0.2475(2)$ \\
7  & ${{1}/{64}}$ & $0.1832(2)$ \\
\hline
\end{tabular}
\label{tab2}
\end{table}

\begin{figure}[t]
\centerline{\includegraphics[width=0.7\linewidth]{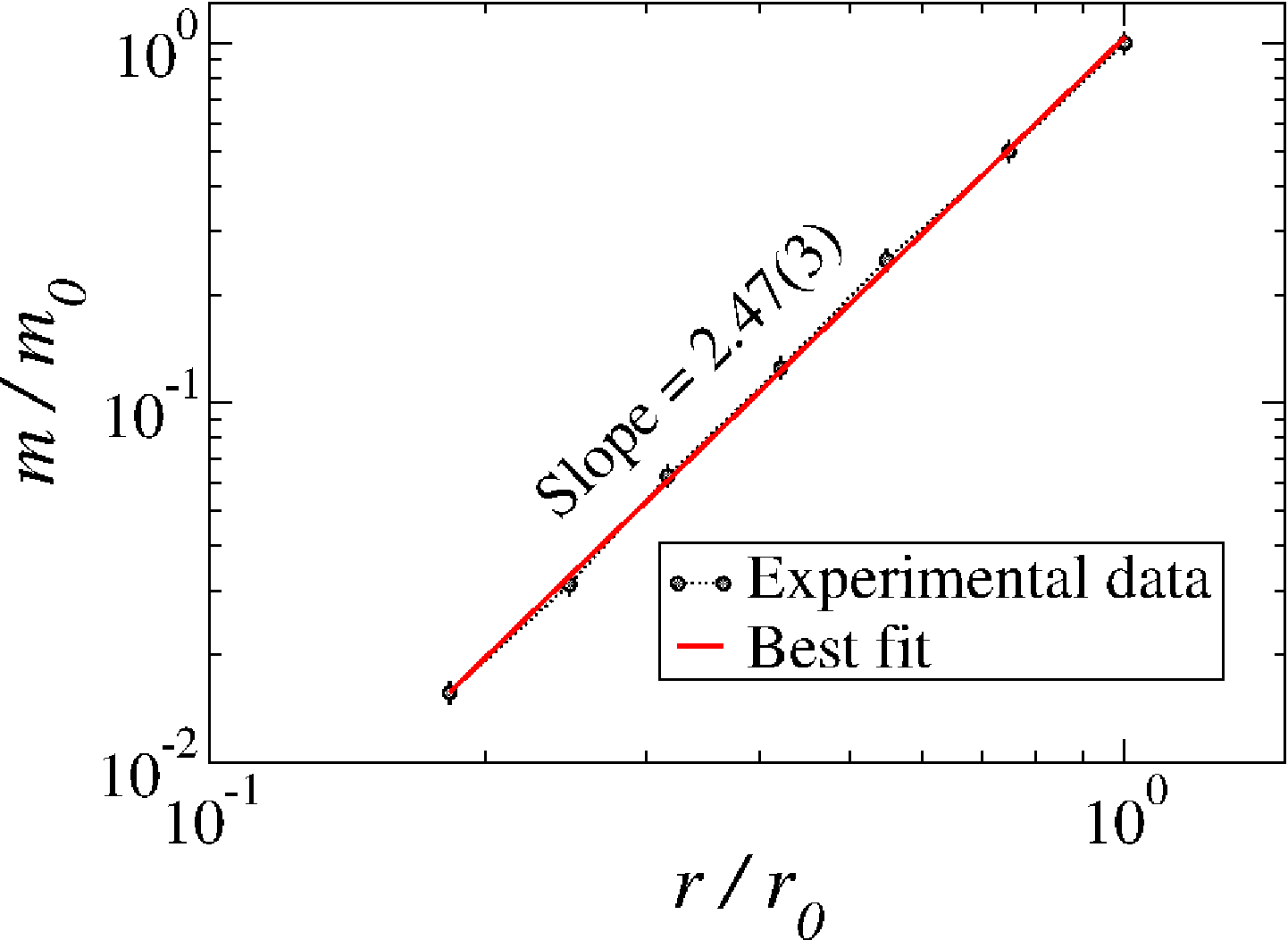}}
\caption{Plot of $m/m_0$ versus $r/r_0$ for the crumpled paper balls. A power-law fit yields a slope $D_f = 2.47(3)$.Error bars correspond to the standard deviation computed from 20 measurements.}
\label{Fig2}
\end{figure}

Figure~\ref{Fig2} shows a log--log plot of the dimensionless mass $m/m_0$ versus the dimensionless radius $r/r_0$ for the seven crumpled paper balls listed in Table~\ref{tab2}. Fig.~\ref{Fig1}(b) shows the balls used in the experiment. The data clearly exhibit a power-law dependence over the explored range of masses.

A power-law fit to the data yields a slope $D_f = 2.47(3)$, which can be identified with the fractal dimension of the crumpled paper ball. It is worth emphasizing that the value obtained satisfies $2 < D_f < 3$. This inequality has clear geometrical and topological interpretations. Although the original sheet is a two-dimensional object, it is embedded in a three-dimensional space. The crumpling process generates a highly complex structure that partially fills the surrounding volume without becoming a fully three-dimensional solid.

From a topological perspective, this intermediate fractal dimension indicates that the crumpled paper preserves the connectivity of a two-dimensional manifold while developing a hierarchy of folds, ridges, and cavities across multiple length scales. These features introduce nontrivial constraints that limit how the surface explores the embedding space, preventing it from uniformly filling the volume. The emergence of a fractal dimension between two and three thus provides a quantitative signature of how topology and geometry combine to generate complexity, illustrating the power of scaling arguments to capture structural properties beyond those accessible within classical Euclidean descriptions.

As a brief summary of this section, we have examined two simple, easily reproducible systems that exhibit scaling invariance as a function of a single control parameter. Despite their elementary nature, these examples illustrate how scaling arguments provide quantitative insight into geometrical and structural properties of physical systems. We now turn to situations in which scaling invariance emerges from the interplay of two control variables.

\section{Scaling invariance for two variables: bifurcation characterisation}
\label{sec3}

In this section we show that scaling invariance is not restricted to systems controlled by a single parameter, but also naturally arises when two variables simultaneously influence the dynamics. A particularly rich and well-studied context in which this occurs is that of bifurcations in nonlinear dynamical systems.

To facilitate the discussion, we consider discrete-time dynamical systems described by mappings \cite{ref20}. Given an initial condition $x_0$ at discrete time $n=0$, the application of the mapping generates a sequence $x_0 \rightarrow x_1 \rightarrow x_2 \rightarrow \ldots$, defining an orbit of the system. A bifurcation occurs when a control parameter is varied, leading to a qualitative change in the structure of the flow of solutions near a stationary state, obtained for $n\rightarrow\infty$.

Bifurcations are commonly classified into two broad categories. Local bifurcations are associated with changes in the stability of fixed points or periodic orbits and can be analysed through local properties of the dynamics \cite{ref12}. Global bifurcations, in contrast, involve large-scale reorganisations \cite{grebogi} of phase space and cannot be anticipated solely from the linear stability of stationary states. In the following, we focus on scaling properties associated with local bifurcations.

In close analogy with the discussion in Section~\ref{sec2}, we consider two classes of systems: (i) one-dimensional mappings of the form $x_{n+1}=f(x_n)$ with $f$ representing a nonlinear function of $x$, and (ii) two-dimensional mappings written as $(x_{n+1},y_{n+1}) = T(x_n,y_n)$, where $T$ denotes the mapping operator. We begin with the one-dimensional case.

\subsection{Scaling for local bifurcations in a one-dimensional mapping}

As a representative example, we consider a logistic-like map \cite{ref21} defined by
\begin{equation}
x_{n+1}=R x_n (1 - x_n^{\gamma}),
\label{c9_eq1}
\end{equation}
where $R \in \mathbb{R}$ with $R \ge 0$, $\gamma \ge 1$, and $x$ is the dynamical variable. The discrete time index is $n=0,1,2,\ldots$. The standard logistic map \cite{ref20} is recovered for $\gamma = 1$, while $\gamma = 2$ corresponds to a cubic map \cite{ref22}, and higher values of $\gamma$ generate increasingly nonlinear dynamics.

\begin{figure}[t]
\centerline{\includegraphics[width=1.0\linewidth]{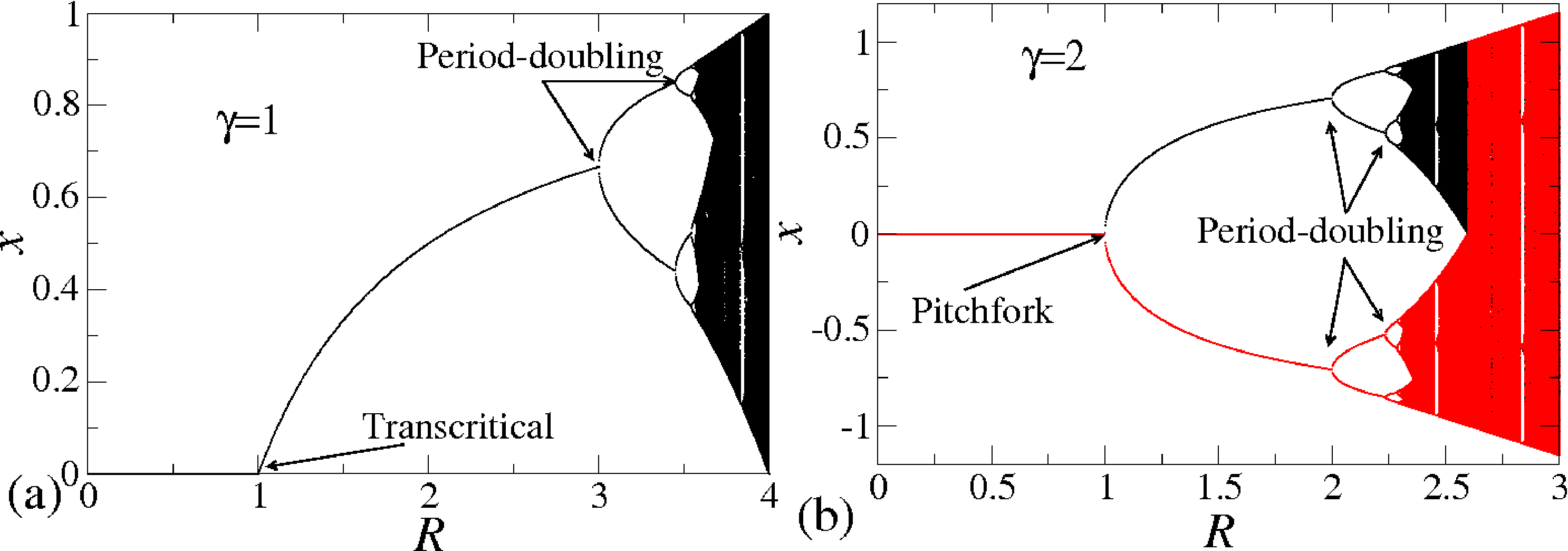}}
\caption{Orbit diagram of the mapping~(\ref{c9_eq1}) for two different values of $\gamma$: (a) $\gamma=1$ and (b) $\gamma=2$.}
\label{diagrama}
\end{figure}

Figure~\ref{diagrama} shows the orbit diagram of the logistic-like map for two representative values of $\gamma$. As the control parameter $R$ is varied, the system undergoes a sequence of qualitative changes in its long-term behaviour, characteristic of local bifurcations.

The stationary states (fixed points) of the map are obtained by imposing $x_{n+1} = x_n = x^*$. This condition leads to the fixed point $x_1^* = 0$, and, depending on the value of $\gamma$, to other fixed points. If $\gamma$ is not an even integer (that is, if it is odd, rational, or irrational), a single additional fixed point exists,
\begin{equation}
x_2^* = \left(1 - \frac{1}{R} \right)^{1/\gamma}.
\label{c9_eq2}
\end{equation}
If $\gamma$ is an even integer, the symmetry of the map gives rise to two additional fixed points,
\begin{equation}
x_{2,3}^* = \pm \left(1 - \frac{1}{R} \right)^{1/\gamma},
\label{c9_eq3}
\end{equation}
where the $(+)$ and $(-)$ signs correspond to $x_2^*$ and $x_3^*$, respectively.

The stability of the fixed points is determined by the condition $\left| \frac{df}{dx} \right|_{x^*} < 1$. The trivial fixed point $x_1^* = 0$ is asymptotically stable for $R \in [0,1)$, independently of the value of $\gamma$. The non-trivial fixed points $x_{2,3}^*$ are asymptotically stable for $R \in \left(1, \frac{2+\gamma}{\gamma}\right)$. The value $R=1$ therefore marks the occurrence of a bifurcation.

When $\gamma$ is even, the bifurcation at $R=1$ is a supercritical pitchfork bifurcation \cite{ref21}, characterised by the loss of stability of $x_1^*$ and the simultaneous emergence of two symmetric fixed points $x_{2,3}^*$, each with its own basin of attraction. When $\gamma$ is not even, the bifurcation at $R=1$ is transcritical \cite{ref21}: the fixed point $x_1^*$ loses stability while $x_2^*$ becomes asymptotically stable, leading to an exchange of stability between the fixed points. For larger values of $R$, the fixed points $x_{2,3}^*$ lose stability at $R=(2+\gamma)/\gamma$, giving rise to a period-doubling bifurcation.

In the following, we analyse the convergence towards the fixed points in these bifurcations, focusing on the emergence of scaling behaviour. We begin with the transcritical bifurcation.

\subsubsection{Transcritical bifurcation}
\label{c9_sec3}

\begin{figure}[t]
\centerline{\includegraphics[width=0.95\linewidth]{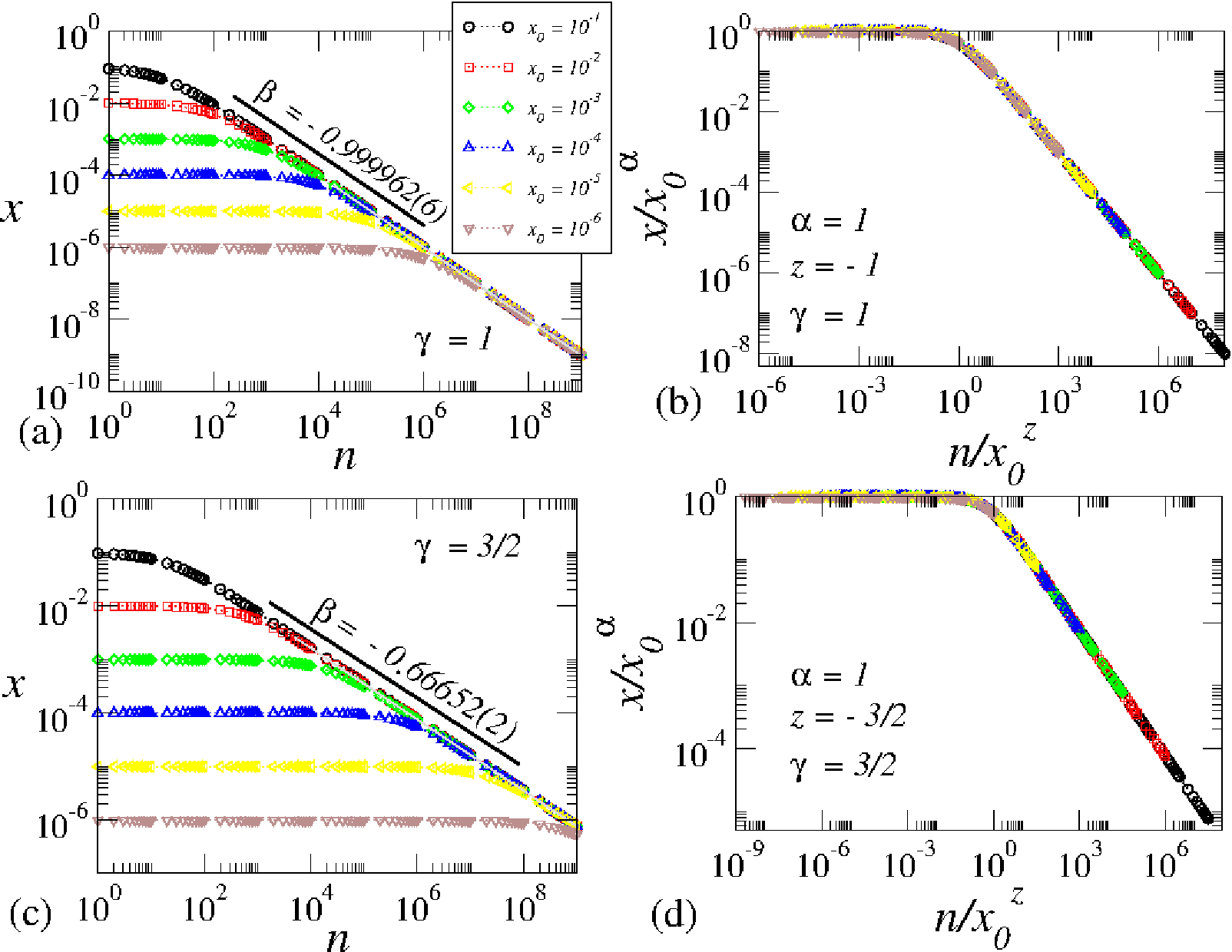}}
\caption{{(a,c) Plot of $x~vs.~n$ at the bifurcation point $R=1$, for $\gamma=1$ and $\gamma=3/2$, respectively, and for different initial conditions $x_0$. (b,d) Collapse of the curves in (a,c) onto universal plot under the transformations $x \rightarrow x / x_0^{\alpha}$ and $n \rightarrow n / x_0^z$.}}
\label{c9_Fig1}
\end{figure}

The transcritical bifurcation occurs at $R=1$ when $\gamma$ takes non-even values. At this point, the fixed point $x_1^*=0$ becomes unstable, while $x_2^* = \left(1 - \frac{1}{R}\right)^{1/\gamma}$ becomes asymptotically stable. The convergence towards the stable fixed point from different initial conditions is illustrated in Figure~\ref{c9_Fig1} for two representative values of the nonlinearity parameter: (a) $\gamma=1$ and (c) $\gamma=3/2$.

A close inspection of the curves reveals two distinct dynamical regimes. For short times, $n \ll n_x$, the orbit remains approximately constant, forming a plateau whose height depends on the initial condition. At a characteristic iteration number $n = n_x$, a crossover occurs, and the system enters a decay regime governed by a power law. The crossover time $n_x$ depends explicitly on the initial distance to the fixed point, which in this case coincides with the initial condition $x_0$, since $x^*=0$.

Based on these observations, we formulate the following scaling hypotheses.  
(1) For $n \ll n_x$, the orbit satisfies $x(n) \propto x_0^{\alpha}$ where $\alpha$ is a critical exponent. Numerical inspection of Figures~\ref{c9_Fig1}(a,c) yields $\alpha=1$.  (2) For $n \gg n_x$, the decay follows a power law, $x(n) \propto n^{\beta}$, where $\beta$ is the decay exponent. Importantly, the value of $\beta$ depends on $\gamma$, indicating that it reflects the degree of nonlinearity of the mapping.  
(3) The crossover iteration number scales with the initial condition as $n_x \propto x_0^{z}$, where $z$ is a third critical exponent.

These hypotheses suggest that the distance to the fixed point, $x(n,x_0)$, can be described by a homogeneous generalised function,
\begin{equation}
x(n,x_0) = \ell\, x(\ell^a n, \ell^b x_0),
\label{eq5}
\end{equation}
where $\ell$ is an arbitrary scaling factor and $a$ and $b$ are characteristic exponents.

Choosing $\ell^a n = 1$ leads to $\ell = n^{-1/a}$ and hence $x(n,x_0) = n^{-1/a} x\left(1, n^{-b/a} x_0\right)$. Assuming that $x(1,n^{-b/a}x_0)$ approaches a constant for $n \gg n_x$, direct comparison with the second scaling hypothesis yields $\beta = -1/a$.

Alternatively, choosing $\ell^b x_0 = 1$ gives $\ell = x_0^{-1/b}$ and $
x(n,x_0) = x_0^{-1/b} x\left(x_0^{-a/b} n, 1\right)$. Assuming that $x(x_0^{-a/b}n,1)$ is constant for $n \ll n_x$, a comparison with the first scaling hypothesis leads to $\alpha = -1/b$.

Combining the two expressions for $\ell$ yields $n_x^{-1/a} = x_0^{-1/b}$, or equivalently $n_x \propto x_0^{a/b}$. Comparing with the third scaling hypothesis, we obtain the scaling law $z = \frac{\alpha}{\beta}$. 

\begin{figure}[t]
\centerline{\includegraphics[width=0.5\linewidth]{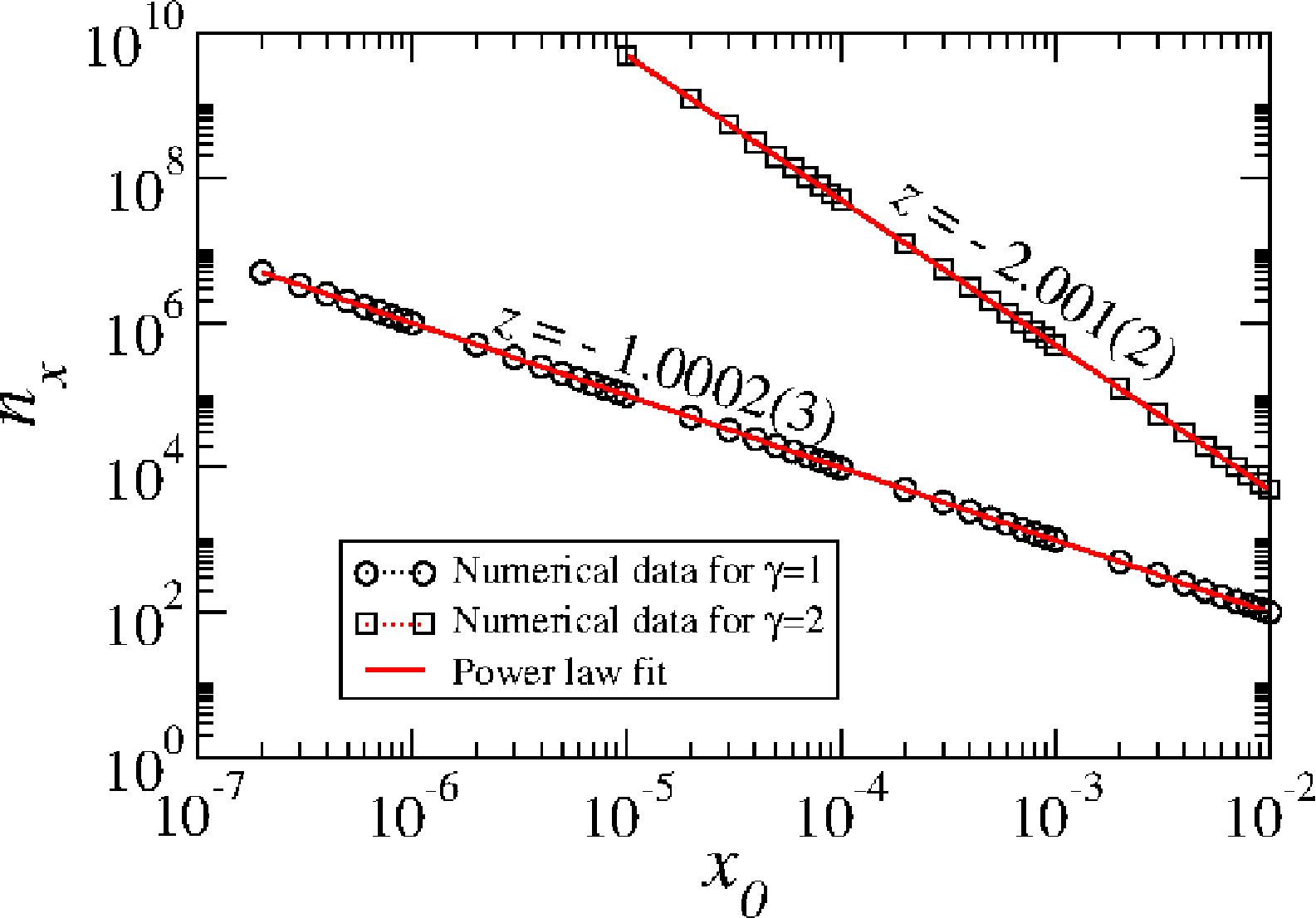}
\includegraphics[width=0.475\linewidth]{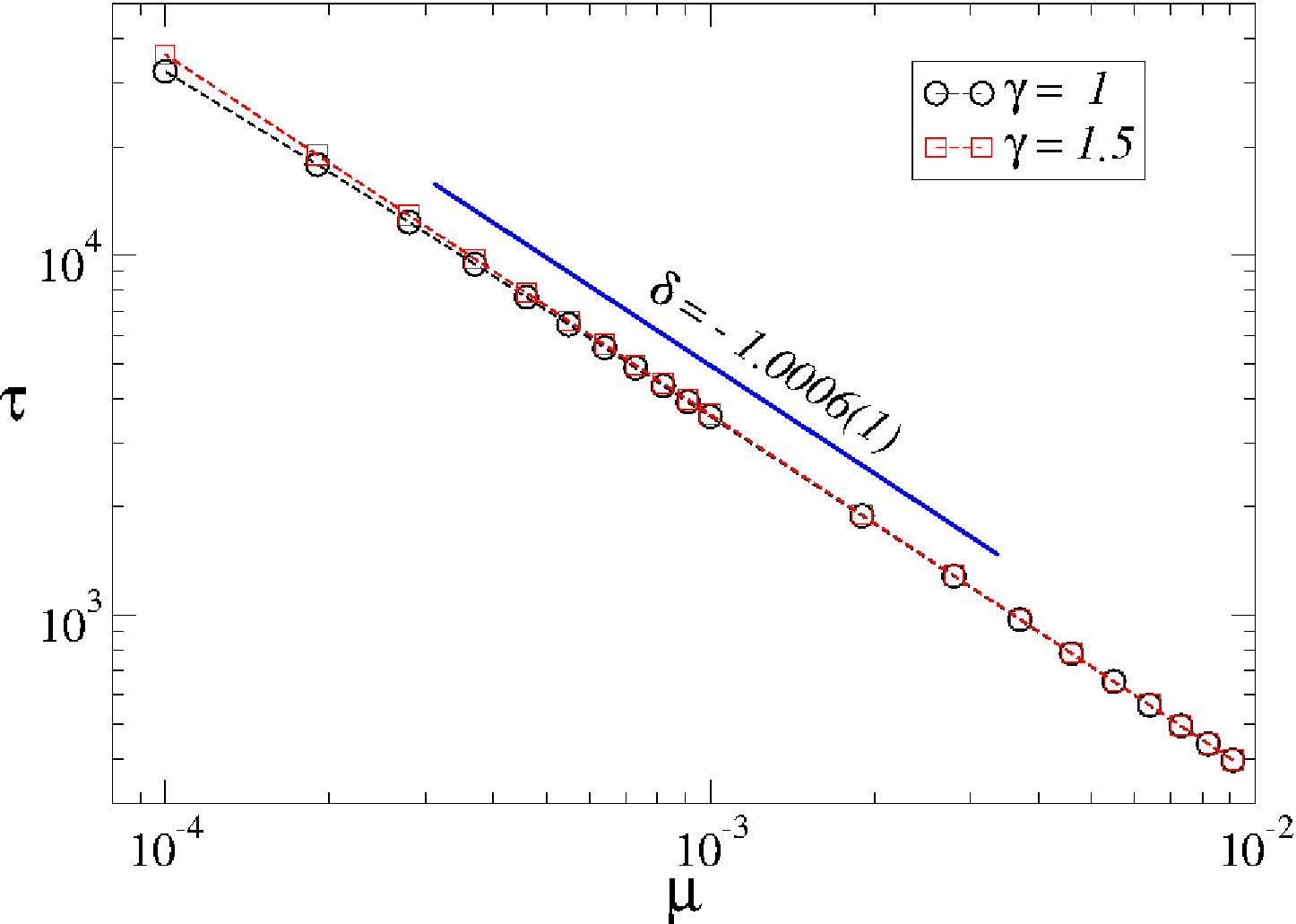}}
\caption{(a) Plot of the crossover iteration number $n_x$ with the initial condition $x_0$ for $\gamma=1$ and $\gamma=2$. Power-law fits yield $z=-1.0002(3)$ and $z=-2.001(2)$, respectively. (b) Plot of the relaxation time $\tau$ as a function of $\mu = R - R_c$ for $\gamma = 1$ and $\gamma = 3/2$. A power-law fit yields $\delta = -1$, independent of $\gamma$.}
\label{c9_Fig2}
\end{figure}

The exponent $\alpha$ characterises the short-time plateau and takes the value $\alpha=1$ for all cases studied. The decay exponent $\beta$, extracted from the long-time power-law regime, depends on the nonlinearity parameter $\gamma$. For $\gamma=1$, Figure~\ref{c9_Fig1}(a) yields $\beta=-0.999962(6)\approx -1$, while for $\gamma=3/2$, Figure~\ref{c9_Fig1}(c) gives $\beta=-0.66652(2)\approx -1/\gamma$. The crossover exponent $z$ can be obtained either from the scaling relation $z=\alpha/\beta$ or directly from fits of $n_x$ versus $x_0$, as shown in Figure~\ref{c9_Fig2}(a).

Finally, the validity of the scaling hypotheses is confirmed by applying the transformations $x \rightarrow x/x_0^{\alpha}$ and $n \rightarrow n/x_0^{z}$, under which all curves collapse onto a single universal curve, as illustrated in Figures~\ref{c9_Fig1}(b,d). This data collapse provides compelling evidence for scale invariance at the transcritical bifurcation.

The results discussed so far strictly apply at the bifurcation point, where the dynamics exhibits scale invariance with respect to the initial distance from the fixed point. In the vicinity of the bifurcation, however, the relaxation dynamics changes qualitatively. Instead of being governed by a generalised homogeneous function, the convergence towards the fixed point becomes exponential and can be described as
\begin{equation}
x(n) - x^* \cong (x_0 - x^*) e^{-n/\tau},
\label{c9_eq4}
\end{equation}
where $\tau$ is the relaxation time. Close to the bifurcation, $\tau$ diverges as a power law,
\begin{equation}
\tau \propto \mu^{\delta},
\end{equation}
with $\mu = R - R_c$, where $R_c$ denotes the critical value of the control parameter, and $\delta$ is a critical exponent.

To determine the exponent $\delta$, we consider an ensemble of initial conditions distributed close to a given value $x_0$ and follow their temporal evolution. When the distance between the dynamical variable and the fixed point becomes smaller than a prescribed tolerance, typically $tol < 10^{-8}$, the number of iterations required to reach this condition is recorded for each trajectory. At the end of the ensemble, an average relaxation time is computed for a given value of the control parameter. This procedure is repeated for different values of $\mu$.

The resulting behaviour of the relaxation time $\tau$ as a function of $\mu$ is shown in Figure~\ref{c9_Fig2}(b). A power-law fit of the numerical data yields $\delta \cong -1$, indicating a divergence of the relaxation time as the bifurcation point is approached.

An important result emerging from Figure~\ref{c9_Fig2}(b) is that the exponent $\delta$ does not depend on the nonlinearity parameter $\gamma$, nor does it distinguish between the different local bifurcations considered. A similar analysis performed for period-doubling bifurcations yields the same qualitative picture: the short-time exponent $\alpha = 1$, while the long-time decay exponent takes the value $\beta = -1/2$. The remaining exponents $z=-2$ and $\delta=-1$ assume fixed values that are independent of the order of the period-doubling bifurcation, whether it corresponds to the first, second, or any subsequent occurrence.

At this point, it is instructive to illustrate the practical implications of scaling and critical slowing down. For $\gamma = 1$, the orbit diagram of the logistic map, constructed after sufficiently long transients, is shown in Figure~\ref{diagrama}(a). Since such diagrams are typically obtained after discarding a very large number of iterations, one may ask how the orbit diagram changes when the transient time is finite.

\begin{figure}[t]
\centerline{(a)\includegraphics[width=0.475\linewidth]{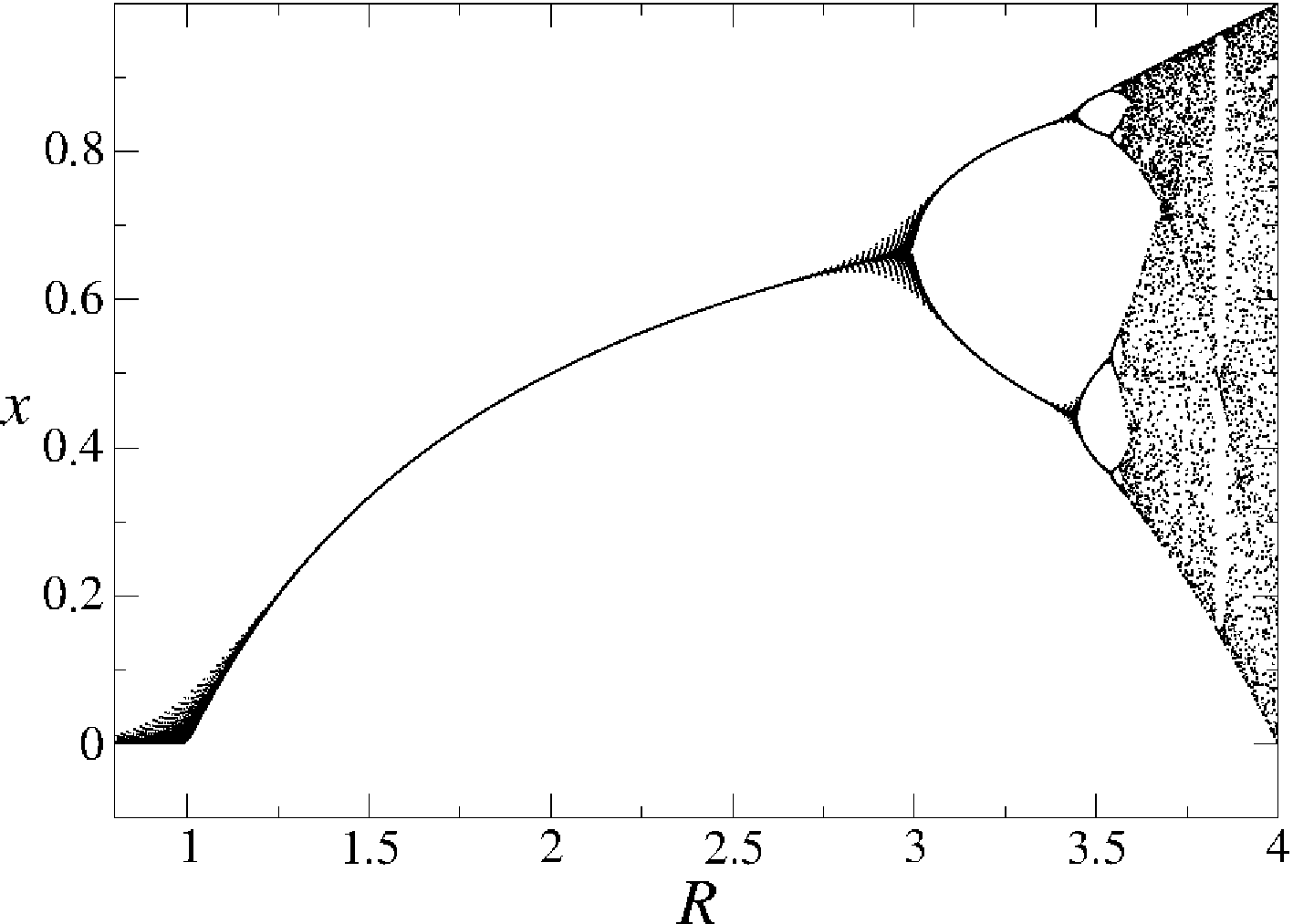}
 (b)\includegraphics[width=0.475\linewidth]{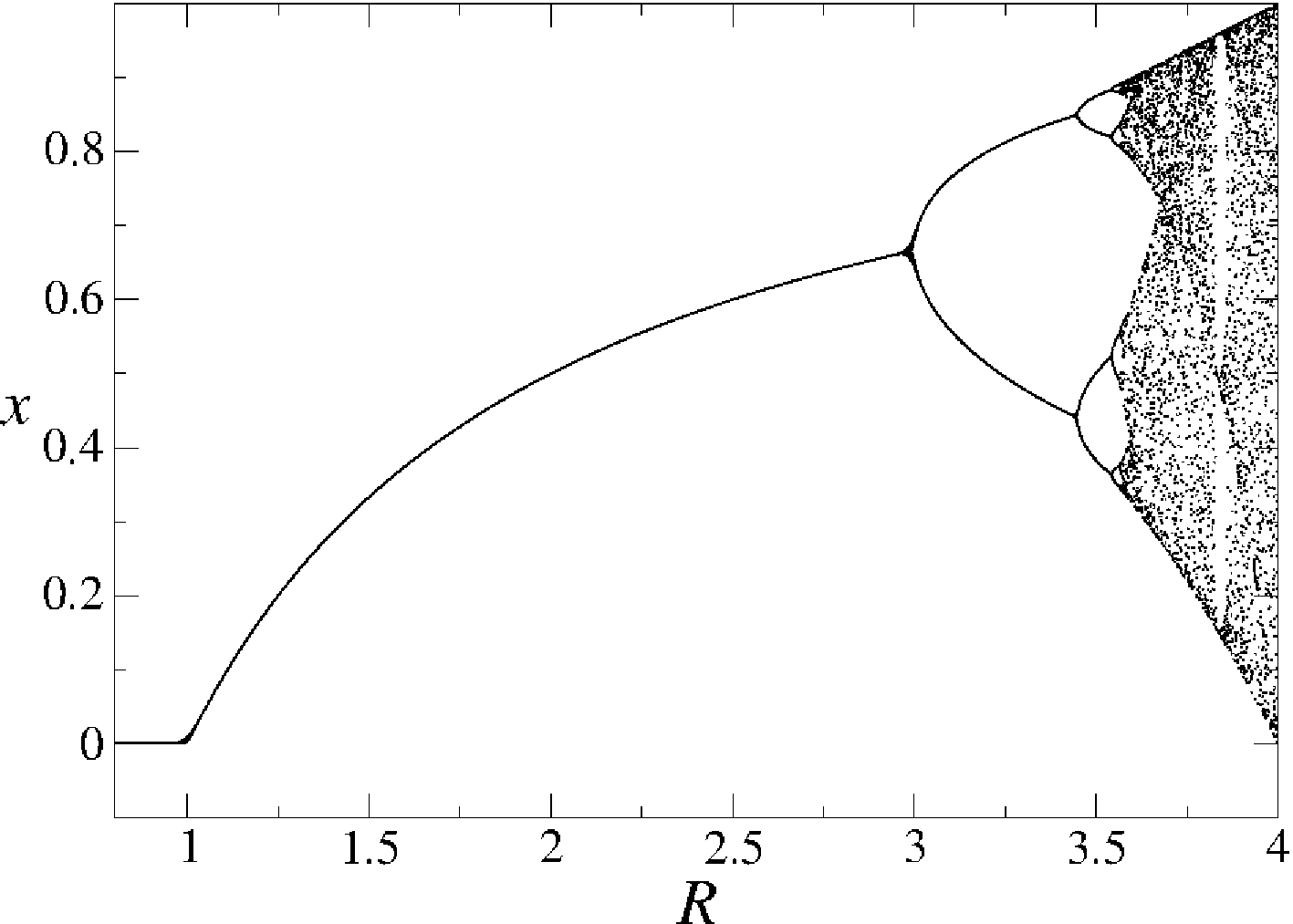}}
\centerline{(c)\includegraphics[width=0.475\linewidth]{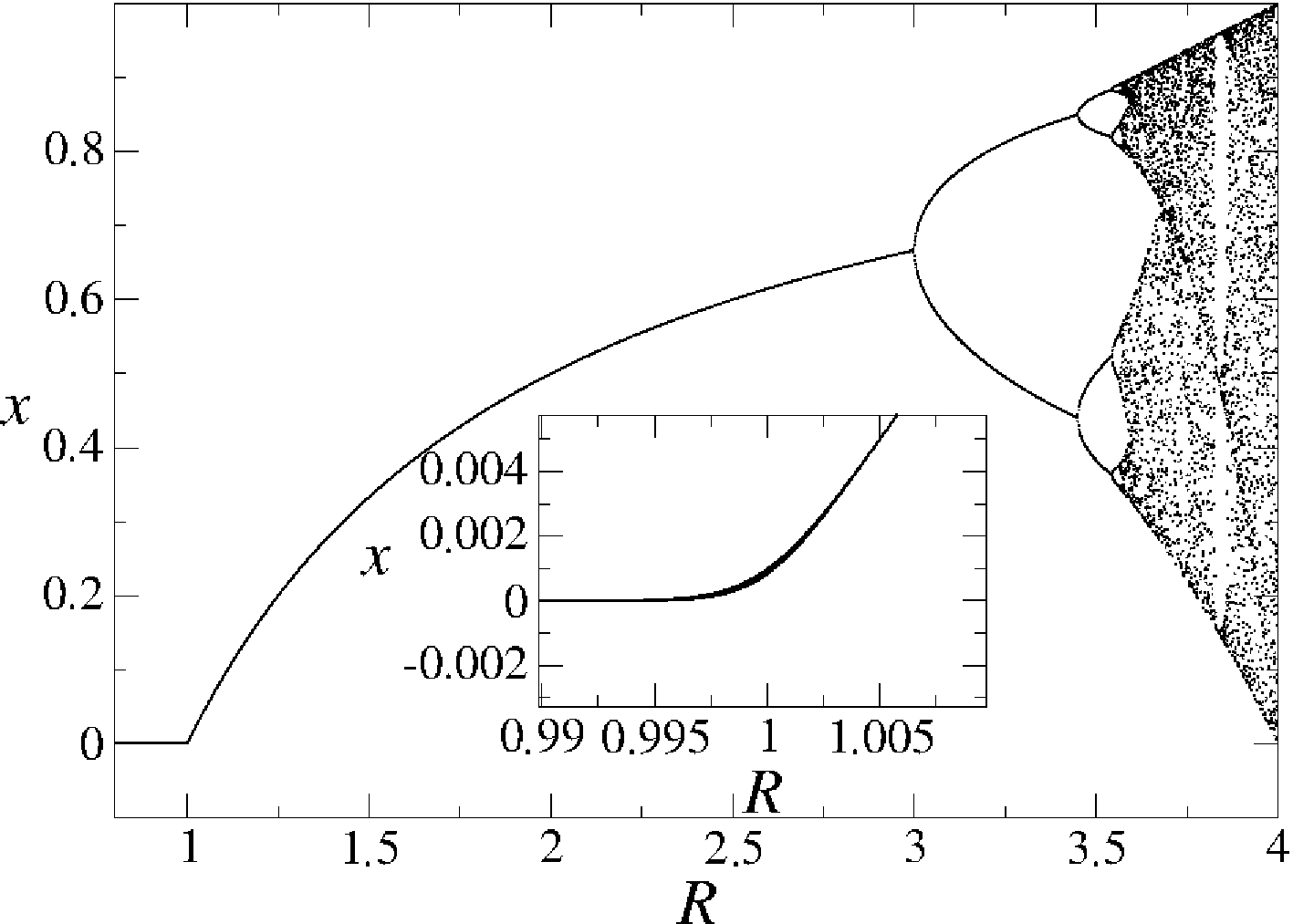}
(d)\includegraphics[width=0.475\linewidth]{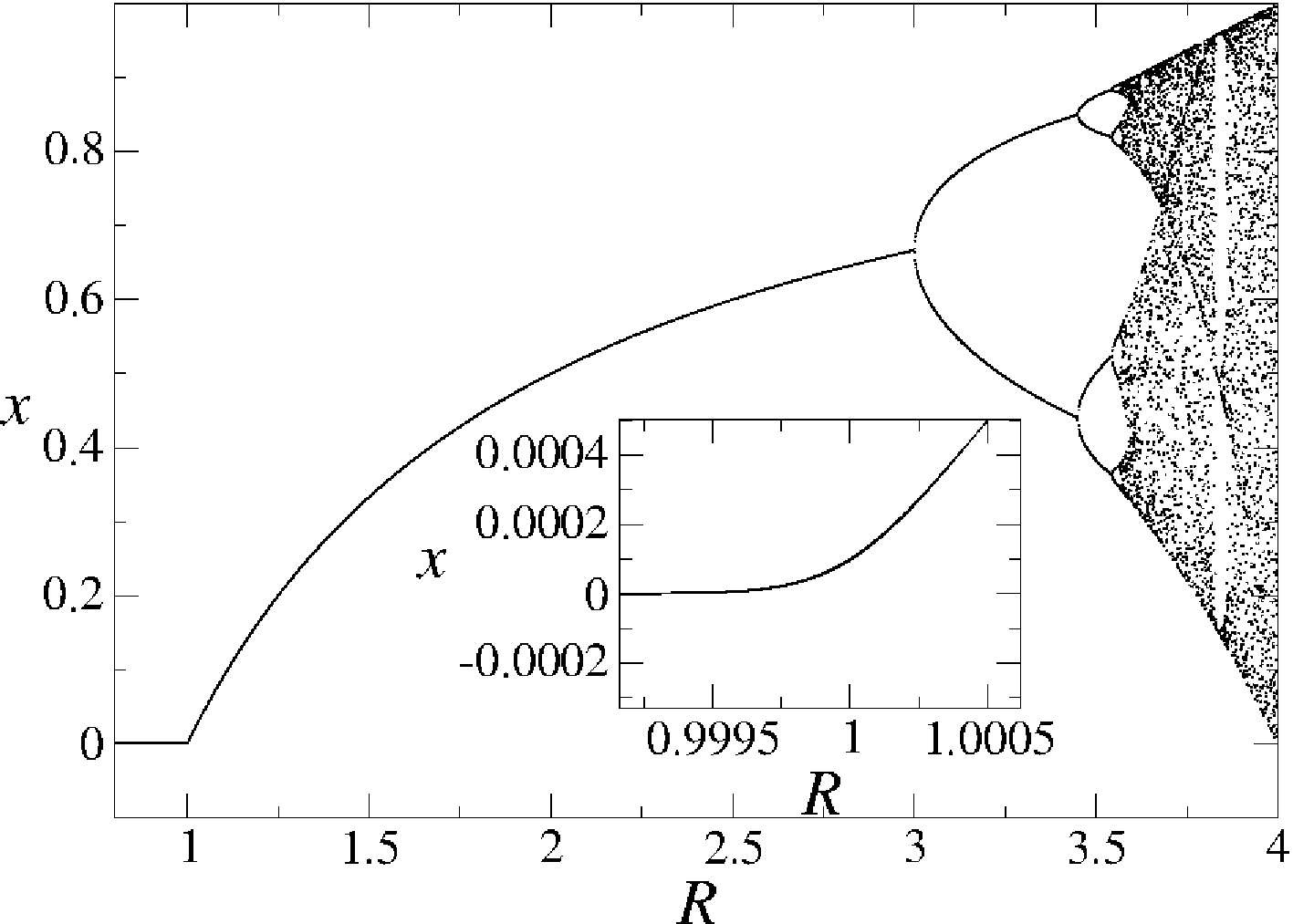}}
\caption{Orbit diagram of the logistic map constructed after finite transient times: (a) $n=10$, (b) $n=100$, (c) $n=1000$, and (d) $n=10000$ iterations. The insets in (c–d) show magnified views of the region near R=1.}
\label{c1_Fig5}
\end{figure}

Figure~\ref{c1_Fig5}(a) shows the orbit diagram constructed after only $10$ iterations. Near the bifurcation points, clusters of points remain visibly separated from the attractor (fixed point), reflecting the fact that the initial condition $x_0 = 0.5$ has not yet converged to the stationary state. When the number of iterations is increased to $100$, as shown in Figure~\ref{c1_Fig5}(b), these clusters become significantly smaller, although they are still present. For $1000$ iterations, the deviation from the attractor is barely visible in the global diagram, but becomes evident when the region near $R=1$ is magnified, as shown in the inset of Figure~\ref{c1_Fig5}(c). This effect is even seen for $10000$ iterations, as illustrated in Figure~\ref{c1_Fig5}(d).

These observations clearly demonstrate that convergence towards the attractor is increasingly slow as the system approaches the bifurcation. The spatial scale of the magnification required to visualise deviations from the attractor is inversely proportional to the number of iterations, providing a direct and intuitive illustration of critical slowing down. This example highlights a practical manifestation of scaling invariance in dynamical systems.

\subsection{Scaling for local bifurcations in a two-dimensional mapping}

\begin{figure}[t]
\centerline{\includegraphics[width=1\linewidth]{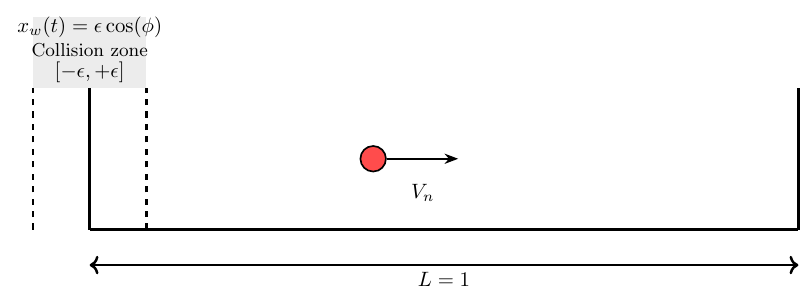}}
\caption{Illustration of the system representing the Fermi-Ulam model.}
\label{fum}
\end{figure}

We now extend the discussion to systems described by two dynamical variables. As a representative example, we consider a mechanical model consisting of a classical particle of mass $m$ confined between two infinitely heavy walls, often referred to as Fermi-Ulam model \cite{ref24}. One wall is fixed, while the other oscillates periodically around an equilibrium position. Figure \ref{fum} illustrates the system. 
Collisions with the moving wall allow the particle to exchange energy and momentum, while the fixed wall acts as a reinjection mechanism, returning the particle for subsequent collisions. 

The dynamics of the system is described by a discrete mapping that relates the particle's velocity and the phase of the moving wall after successive collisions. Depending on the phase of the wall at the instant of impact, the particle's velocity may increase or decrease. In order to simplify the analysis, we employ the static wall approximation \cite{karlis}, in which both walls are treated as fixed, but the energy exchange at the instant of collision is computed as if the wall were moving. This is an approximation used for small values of $\epsilon$.

The dynamics of the system is described by the mapping
\begin{equation}
T:\left\{\begin{array}{ll}
V_{n+1}=q^2 V_n - (1+q)\epsilon \sin(\phi_{n+1}) \\
\phi_{n+1}=\left[\phi_n + \frac{1}{V_n} + \frac{1}{q V_n}\right] \;\; \text{mod}\; 2\pi
\end{array}
\right.,
\label{deq1}
\end{equation}
where $q \in [0,1]$ is the restitution coefficient. The dynamical variables $V$ and $\phi$ represent the particle's velocity and the phase of the moving wall at $x=0$, respectively, while the index $n$ labels the collision number.

The term $1/V_n$ corresponds to the time interval required for the particle to travel from the wall located at $x=0$ to the wall at $x=1$, whereas $1/(q V_n)$ represents the return time after the inelastic collision. The factor $q^2$ in the velocity update accounts for the fractional energy loss associated with the double collision process. The determinant of the Jacobian matrix of the mapping is $\det J = q$. Since $q < 1$, Liouville's theorem \cite{ref8} is violated and phase-space volume is no longer conserved, leading to the formation of attractors.

\begin{figure}[t]
\centerline{\includegraphics[width=0.7\linewidth]{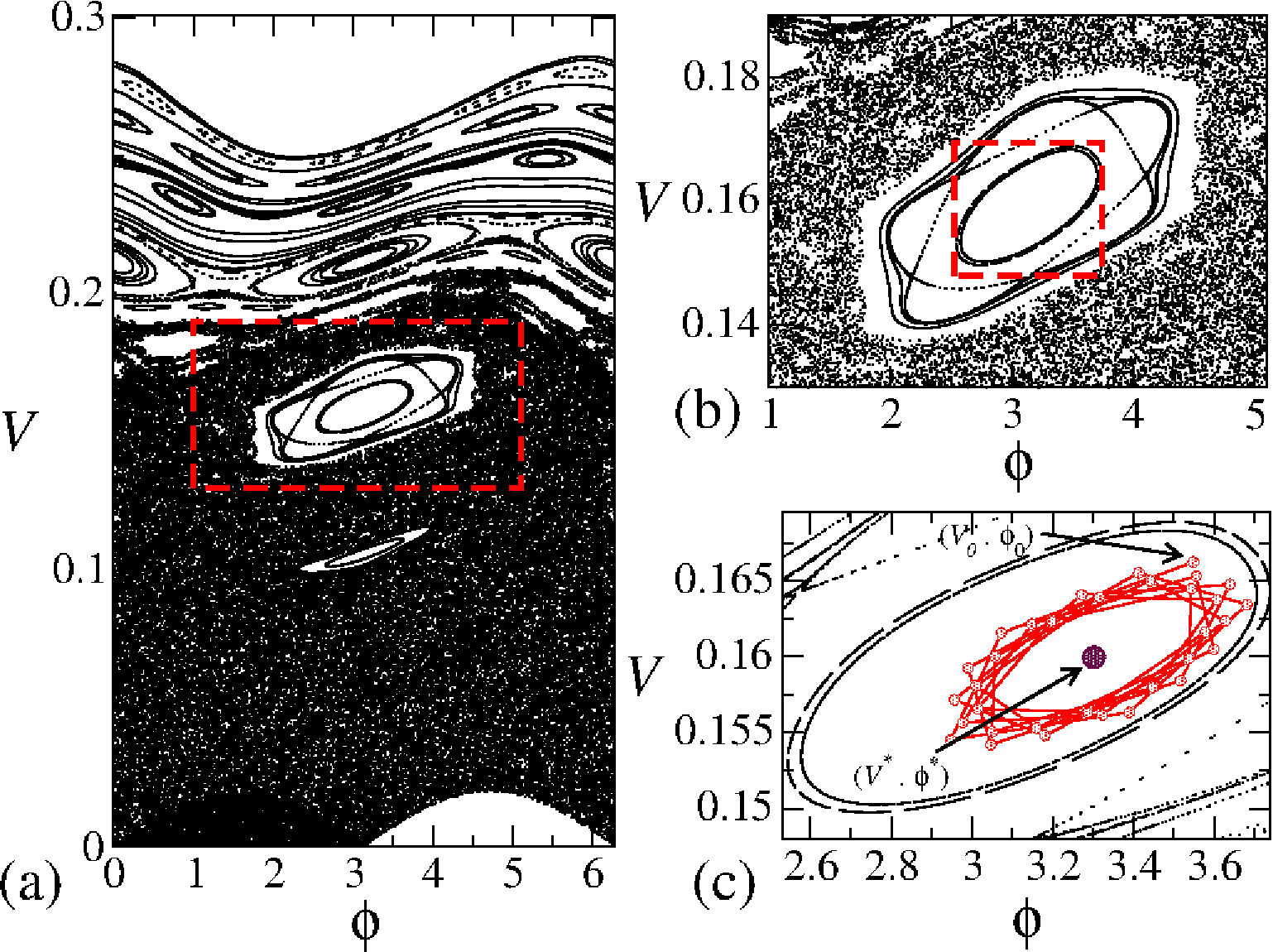}}
\caption{(a) Phase space for $\epsilon=10^{-2}$ and $q=1$. (b) Zoom-in highlighting a period-one region. (c) Further magnification showing the convergence towards the stationary state for $q=0.99$.}
\label{dFig4}
\end{figure}

Figure \ref{dFig4}(a) shows a plot of the phase space for the control parameter $\epsilon=10^{-2}$ and $q=1$ while (b) shows a zoom-in highlighting a period-one region. In both cases, the phase space displays a mixed structure, with chaotic regions coexisting with regular islands. (c) Further magnification showing the convergence towards the stationary state for $q=0.99$.

\begin{figure}[t]
\centerline{\includegraphics[width=0.9\linewidth]{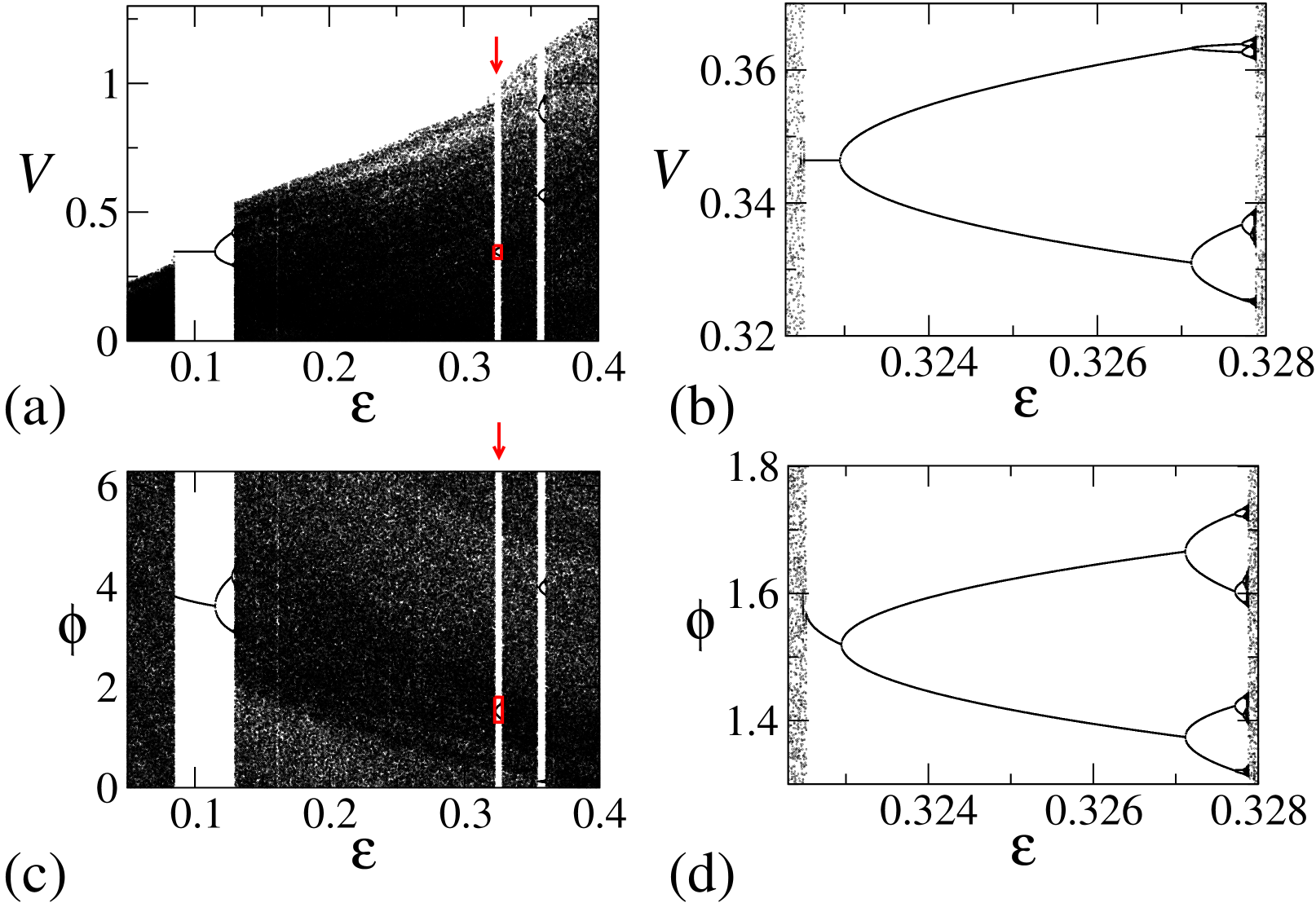}}
\caption{Orbit diagrams for the mapping~(\ref{deq1}) with $q=0.85$. Panels (a,c): $V$ versus $\epsilon$; panels (b,d): $\phi$ versus $\epsilon$. Panels (b) and (c) are enlarged views of the regions inside the red squares indicated by the arrows.}
\label{dFig2}
\end{figure}

When dissipation is introduced ($q<1$), the dynamical scenario changes dramatically. The mixed phase-space structure disappears and attractors dominate the long-term dynamics. Figure~\ref{dFig2} shows the orbit diagrams for the mapping~(\ref{deq1}) with $q=0.85$. Panels (a) and (b) display $V$ versus $\epsilon$, while panels (c) and (d) show $\phi$ versus $\epsilon$. The initial condition $(V_0,\phi_0)=(2.1,\pi+0.01)$ was iterated $10^6$ times to eliminate transients, and only the final $100$ collisions were recorded. Panels (b) and (d) correspond to magnifications of (a) and (c), respectively.

\begin{figure}[t]
\centerline{\includegraphics[width=0.7\linewidth]{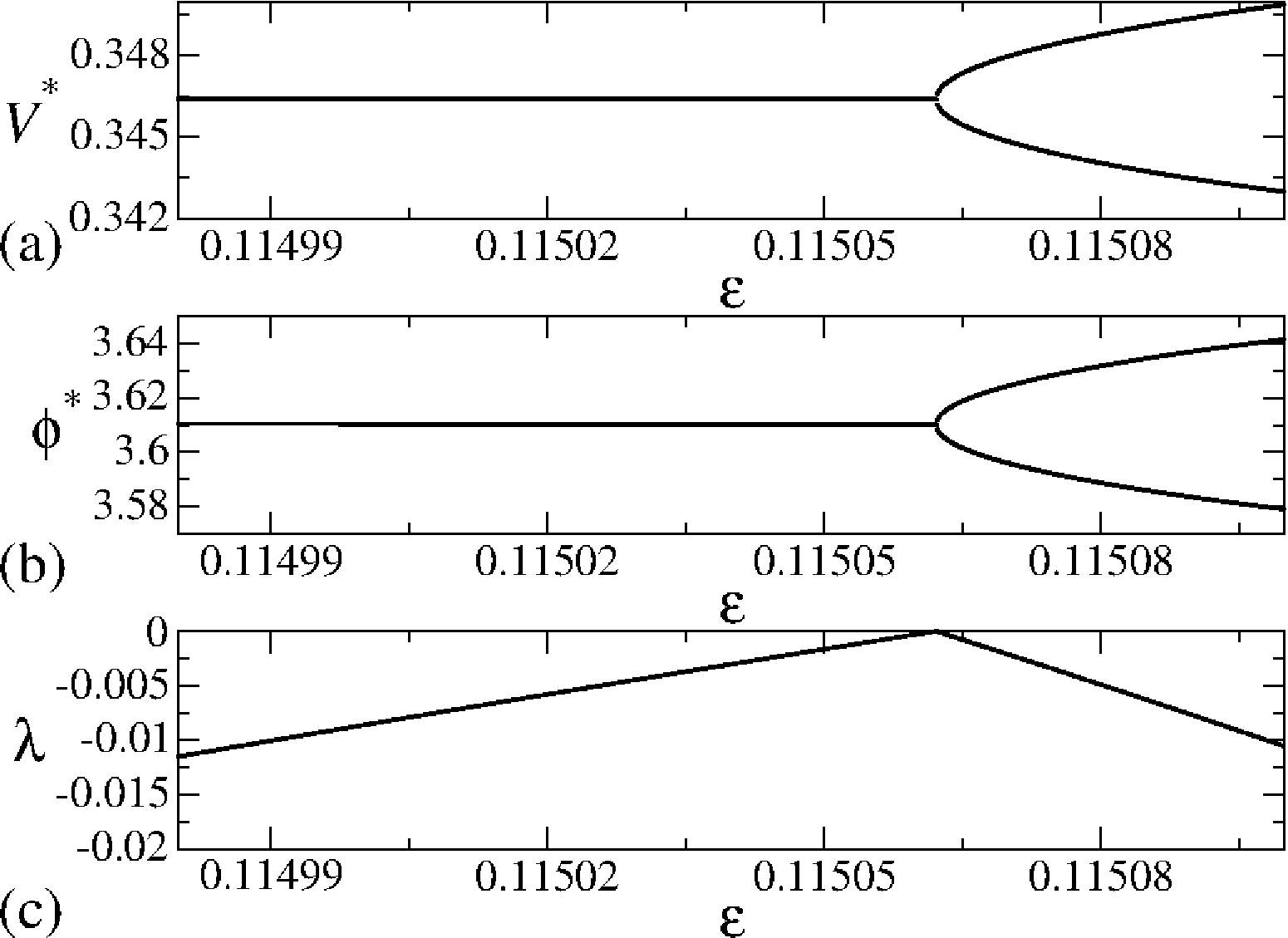}}
\caption{Orbit diagrams for $q=0.85$: (a) $V$ versus $\epsilon$, (b) $\phi$ versus $\epsilon$, and (c) largest Lyapunov exponent. A period-doubling bifurcation occurs at $\epsilon_c=0.11506218$.}
\label{dFig3}
\end{figure}

A further magnification of Figures~\ref{dFig2}(b,d) is shown in Figure~\ref{dFig3}. Panels (a) and (b) display $V$ and $\phi$ as functions of $\epsilon$, respectively, while panel (c) shows the largest Lyapunov exponent computed over the same parameter range. A period-doubling bifurcation \cite{danilo} is observed at $\epsilon_c = 0.11506218$. Since the leading eigenvalue of the Jacobian matrix equals unity at the bifurcation, the largest Lyapunov exponent vanishes at this point, as clearly indicated in Figure~\ref{dFig3}(c).

Since the dynamics evolves in the two-dimensional phase space $(V,\phi)$, it is necessary to define an appropriate metric to quantify the distance between a trajectory and the stationary state. To quantify this convergence, we define the distance between the particle and the fixed point as
\begin{equation}
d(n)=\sqrt{(V_n-V^*)^2 + (\phi_n-\phi^*)^2},
\label{deq_d}
\end{equation}
which measures the distance to the stationary state at the $n$th collision.

\begin{figure}[t]
\centerline{\includegraphics[width=0.7\linewidth]{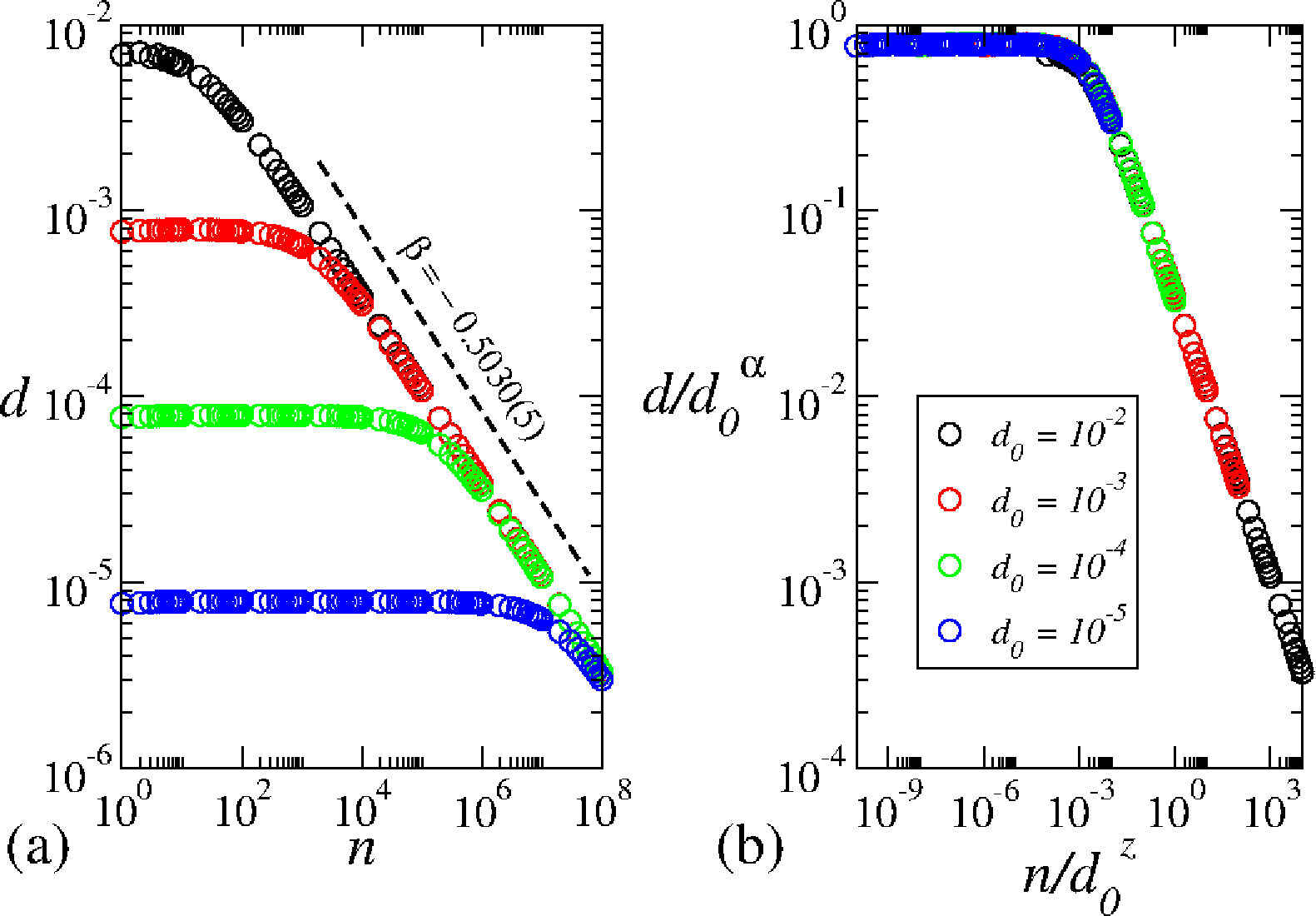}}
\caption{(a) Distance $d$ versus iteration number $n$ for different initial distances from the fixed point. (b) Collapse of all curves in (a) onto a single universal curve after appropriate scaling transformations.}
\label{dFig5}
\end{figure}

A careful stability analysis of the period-one orbit, combined with the behaviour of the Lyapunov exponent, confirms the existence of a period-doubling bifurcation at $\epsilon_c = 0.11506218$. The evolution of $d(n)$ for different initial conditions, computed at $\epsilon_c$ and $q=0.85$, is shown in Figure~\ref{dFig5}(a).

The curves shown in Figure~\ref{dFig5}(a) exhibit a clear and systematic behaviour. For short times, typically $n \ll n_x$, the distance to the fixed point remains confined to a stationary plateau, which can be described by
$d(n) \propto d_0^{\alpha}$,
where $\alpha$ is a critical exponent. From the figure, it is evident that the plateau value coincides with the initial distance $d_0$, implying $\alpha = 1$.

As time evolves, the dynamics eventually leaves the plateau after a characteristic crossover iteration number $n_x$ and enters a decay regime described by
$d(n) \propto n^{\beta}$,
valid for $n \gg n_x$, where $\beta$ is a second critical exponent. A power-law fit of the data in Figure~\ref{dFig5}(a) yields $\beta = 0.5030(5) \approx 1/2$. Finally, the crossover iteration number separating the plateau regime from the power-law decay scales as $n_x \propto d_0^{z}$, where $z$ is a third critical exponent.

Remarkably, the set of critical exponents obtained for this two-dimensional mapping coincides with those observed for the period-doubling bifurcation in the one-dimensional case. This result highlights the robustness of the scaling behaviour across different dynamical settings.

In the vicinity of the bifurcation, the dynamics is no longer governed by a generalised homogeneous function. Instead, the relaxation towards the fixed point becomes exponential, in close analogy with the one-dimensional case,
\begin{equation}
d(n) = d_0 e^{-n/\tau},
\label{d_expo}
\end{equation}
where $d_0$ denotes the initial distance from the fixed point and the relaxation time diverges as
$\tau \propto \mu^{\delta}$,
with $\mu = \epsilon_c - \epsilon$ for $\epsilon < \epsilon_c$. As discussed in \cite{danilo}, the exponent obtained is $\delta = -0.944(2) \approx -1$.

The presence of dissipation in the Fermi--Ulam model destroys the intricate mixed phase-space structure characteristic of the conservative case. Nevertheless, near criticality, both one- and two-dimensional mappings exhibit the same qualitative behaviour: the convergence towards the stationary state is governed by identical scaling laws, independent of the dimensionality of the system. The bifurcations therefore share the same set of critical exponents and belong to the same universality class.

\section{Phase transitions}
\label{sec4}

Another major success of the scaling formalism lies in the investigation of continuous phase transitions \cite{ref8}. A central concept in this context is the order parameter \cite{ref1}, an observable capable of distinguishing between different phases. In a second-order phase transition \cite{ref14}, the order parameter vanishes in one phase and assumes a finite value in the other. At the same time, the response of the order parameter to variations of an external control parameter diverges at the transition point.

Near criticality, the order parameter often exhibits scaling invariance, leading to the emergence of critical exponents and universality classes \cite{ref8}. Beyond the order parameter itself, a comprehensive characterization of a phase transition typically involves additional ingredients \cite{ref14}: the identification of a symmetry breaking, the existence of an elementary excitation responsible for transport or diffusion, and, in some cases, the appearance of topological defects in phase space or configuration space.

To illustrate these ideas, we consider two classes of phase transitions in dynamical systems: (i) the transition from integrability to non-integrability observed in families of Hamiltonian mappings \cite{ref23} and in billiards with static boundaries \cite{ref16}; and (ii) the transition from bounded to unbounded diffusion in dissipative standard mappings \cite{ref17} and in time-dependent billiard systems \cite{ref18}. We begin with the family of Hamiltonian mappings.

\subsection{Transition from integrability to non-integrability in a family of area-preserving mappings}

Hamiltonian systems often exhibit rich and intricate dynamics. A system with two degrees of freedom can be described in terms of two pairs of canonical variables $(I_i,\theta_i)$, with $i=1,2$, where $I$ denotes the action and $\theta$ the corresponding angle \cite{ref24}. The Hamiltonian governing the transition of interest can be written as
$H(I_1,\theta_1,I_2,\theta_2) = H_0(I_1,I_2) + \epsilon H_1(I_1,\theta_1,I_2,\theta_2)$,
where $H_0$ represents the integrable part and $H_1$ the non-integrable perturbation controlled by the parameter $\epsilon$.

For $\epsilon = 0$, the system is integrable, since both the energy and the actions are conserved. When $\epsilon \neq 0$, integrability is broken and only the energy remains conserved. Since the Hamiltonian is time independent and $H=E$ is a constant, one of the action variables can be eliminated, reducing the four-dimensional phase-space flow to a three-dimensional one. The intersection of this flow with a constant plane, defined as a Poincar\'e surface of section, leads to a two-dimensional, area-preserving mapping of the form \cite{ref24}
\begin{equation}
\left\{
\begin{array}{ll}
I_{n+1} = I_n + \epsilon h(\theta_n,I_{n+1}) \\
\theta_{n+1} = [\theta_n + K(I_{n+1}) + \epsilon P(\theta_n,I_{n+1})] \;\; {\rm mod}\; (2\pi)
\end{array}
\right.,
\label{eq0}
\end{equation}
where $K(I_{n+1})$, $P(\theta_n,I_{n+1})$, and $h(\theta_n,I_{n+1})$ are nonlinear functions. Area preservation holds provided the condition
${\partial P(\theta_n,I_{n+1})}/{\partial \theta_n} + {\partial h(\theta_n,I_{n+1})}/{\partial I_{n+1}} = 0$
is satisfied.

To illustrate the transition from integrability to non-integrability, we consider the following area-preserving mapping,
\begin{equation}
\left\{
\begin{array}{ll}
I_{n+1} = I_n + \epsilon \sin(\theta_n), \\
\theta_{n+1} = \left[\theta_n + \frac{1}{|I_{n+1}|^{\tilde{\gamma}}}\right] \;\; {\rm mod}\; (2\pi),
\end{array}
\right.
\label{teq1}
\end{equation}
where $h(\theta_n,I_{n+1})=\sin(\theta_n)$, $P(\theta_n,I_{n+1})=0$, and
$K = 1/|I_{n+1}|^{\tilde{\gamma}}$. The parameter $\tilde{\gamma}>0$ controls the rate at which the angular variable diverges in the limit of vanishing action $I$.

This particular choice is motivated by the onset of chaotic diffusion. When the action $I$ is small, successive values of the angular variable $\theta_n$ become effectively uncorrelated, leading to diffusive motion in phase space and allowing the action to grow. As $I$ increases, correlations between successive angles are gradually restored, and regular structures emerge in phase space in the form of periodic islands and invariant spanning curves. Since the determinant of the Jacobian matrix of the mapping is unity, Liouville's theorem holds and phase-space volume is preserved. As a consequence, chaotic trajectories cannot cross invariant spanning curves and remain confined within a finite region of phase space, giving rise to nontrivial scaling properties for the diffusion.

The control parameter $\epsilon$ plays a central role in the dynamics. For $\epsilon=0$, the system is integrable, as both the energy and the action $I$ are conserved. In this limit, the phase space is foliated by invariant curves of constant action, and the dynamics is completely regular and predictable.

When $\epsilon \neq 0$, integrability is broken and the phase space is no longer foliated. Instead, it acquires a mixed structure, in which a chaotic sea coexists with periodic islands and invariant spanning curves. Depending on the initial conditions and parameter values, trajectories may explore different regions of phase space. Due to area preservation and Liouville's theorem, periodic islands act as impenetrable barriers: trajectories inside them cannot escape, and chaotic trajectories outside cannot enter. These islands can therefore be regarded as analogues of topological defects \cite{ref23}, breaking the global ergodicity of the system.

\begin{figure}[t]
\centerline{\includegraphics[width=0.95\linewidth]{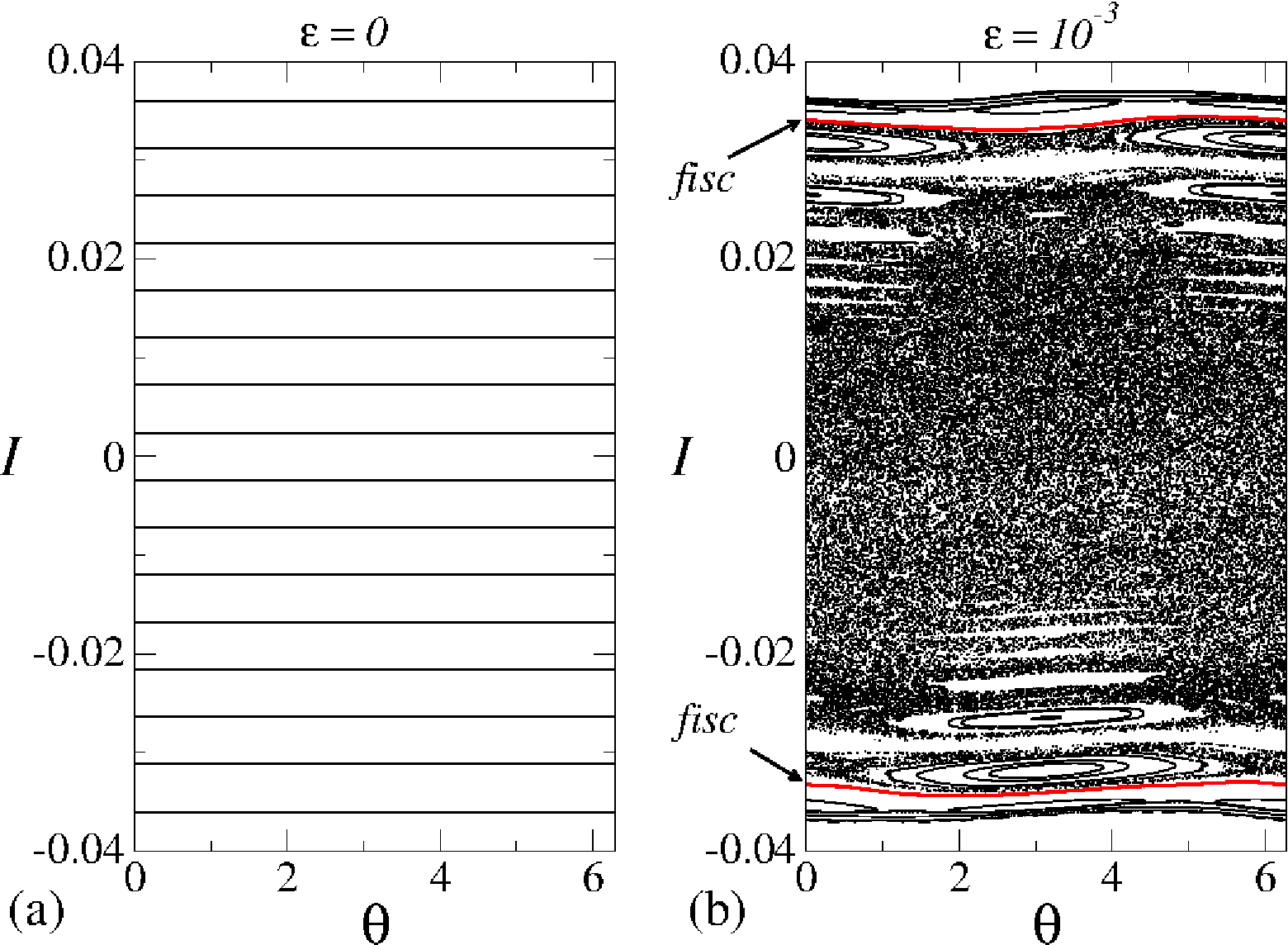}}
\caption{Phase space of the mapping~(\ref{teq1}) for (a) $\epsilon=0$ and (b) $\epsilon=10^{-3}$. The curves in (b) correspond to the first invariant spanning curves and scale as $\epsilon^{1/(1+\tilde{\gamma})}$.}
\label{tFig1}
\end{figure}

Invariant spanning curves play an equally crucial role. By preventing motion across different regions of phase space, they define the effective boundaries of the chaotic sea and determine its extent. Figure~\ref{tFig1} illustrates these features by showing the phase space of the mapping~(\ref{teq1}) for two values of the control parameter: (a) $\epsilon=0$ and (b) $\epsilon=10^{-3}$.

The transition between these two regimes can be interpreted in terms of symmetry breaking. For $\epsilon=0$, the dynamics preserves the action and the phase space exhibits full regularity, with no exponential separation of nearby trajectories. This symmetry corresponds to integrability. When $\epsilon \neq 0$, the nonlinear term $\sin(\theta)$ modifies the evolution of the system, destroying the regular foliation of phase space. The resulting mixed structure, composed of periodic islands, invariant spanning curves, and a chaotic sea, signals the breaking of this symmetry and the emergence of chaotic dynamics.

The chaotic sea that forms for $\epsilon \neq 0$ has a well-defined extension \cite{ref23} in the action variable. Initial conditions within the chaotic region can diffuse only within the interval
$I \in \left(-\left[\frac{\tilde{\gamma} \epsilon}{0.9716\ldots}\right]^{1/(1+\tilde{\gamma})},
\left[\frac{\tilde{\gamma} \epsilon}{0.9716\ldots}\right]^{1/(1+\tilde{\gamma})}\right)$,
whose boundaries are precisely set by invariant spanning curves. These curves act as barriers, confining chaotic motion and determining the size of the chaotic sea. The destruction of regularity therefore defines the broken symmetry associated with the transition from integrable motion to chaotic diffusion.

The mixed nature of the phase space has further consequences. Due to the coexistence of chaotic and regular regions, averages computed over time along a single trajectory generally differ from ensemble (microcanonical) averages taken over phase space. As a result, the fundamental assumption of ergodicity \cite{ref1} is violated. Moreover, different regions of phase space cannot freely intermingle: chaotic trajectories cannot invade periodic islands, and vice-versa. When a chaotic trajectory approaches a periodic structure or an invariant spanning curve, it may remain temporarily trapped in its vicinity, a phenomenon known as stickiness \cite{ref25}. This effect plays an important role in transport properties and further enriches the dynamical behaviour near the transition.

We now turn to the identification of a suitable order parameter for the transition from integrability to non-integrability. As discussed earlier, the dynamics is regular for $\epsilon=0$, while for $\epsilon \neq 0$ chaotic motion and chaotic diffusion may occur, depending on the initial conditions. Due to the presence of two invariant spanning curves, one on the positive and the other on the negative side of the action axis, chaotic diffusion is confined to a finite region of phase space. Therefore a finite diffusion.

Because of the symmetry of the phase space, the average value of the action is not an appropriate observable. Instead, a natural quantity is the root mean square of the action, whose long-time value characterises the saturation of chaotic diffusion. This quantity scales as
$I_{\rm sat} \propto \epsilon^{\alpha}$
and provides an excellent candidate for an order parameter. It vanishes continuously as $\epsilon \rightarrow 0$, marking an ordered (integrable) phase, and becomes finite for $\epsilon \neq 0$, signalling the onset of a chaotic phase.

A direct analogy can be drawn with continuous phase transitions in ferromagnetic systems \cite{Stanley}. The spontaneous magnetisation m plays the role of the order parameter. For temperatures below a critical value $T_c$, a finite magnetisation is observed, while for $T>T_c$ the ordered phase is destroyed and m$=0$. As the temperature approaches $T_c$ from below, the magnetisation vanishes continuously, and the response of the order parameter to an external field, quantified by the magnetic susceptibility diverges. These features are hallmarks of a second-order phase transition.

\begin{figure}[t]
\centerline{\includegraphics[width=1.0\linewidth]{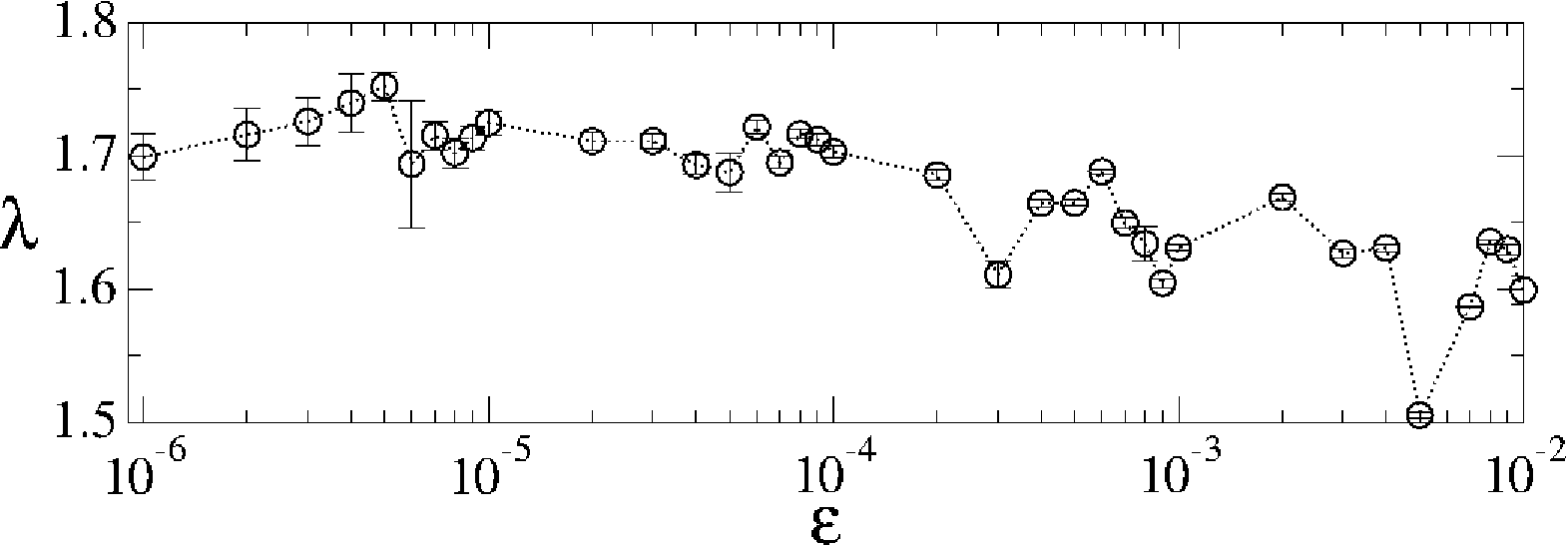}}
\caption{Positive Lyapunov exponent as a function of the control parameter $\epsilon$ over the interval $\epsilon \in [10^{-6},10^{-2}]$.}
\label{tFig2}
\end{figure}

Returning to the chaotic model, once the control parameter $\epsilon$ becomes nonzero, a chaotic sea of finite size is created. Figure~\ref{tFig2} shows a plot of the positive Lyapunov exponent computed over a wide range of the control parameter, $\epsilon \in [10^{-6},10^{-2}]$.

We observe that the Lyapunov exponent varies only weakly, remaining approximately within the interval $\lambda \in [1.5,1.75]$, despite the large variation of $\epsilon$. This behaviour supports the interpretation that the chaotic sea is born with a well-defined size and that the chaotic dynamics possesses a finite positive Lyapunov exponent. The near constancy of $\lambda$ is consistent with the fact that the chaotic sea itself is scale invariant with respect to the control parameter $\epsilon$.

\begin{figure}[t]
\centerline{\includegraphics[width=0.95\linewidth]{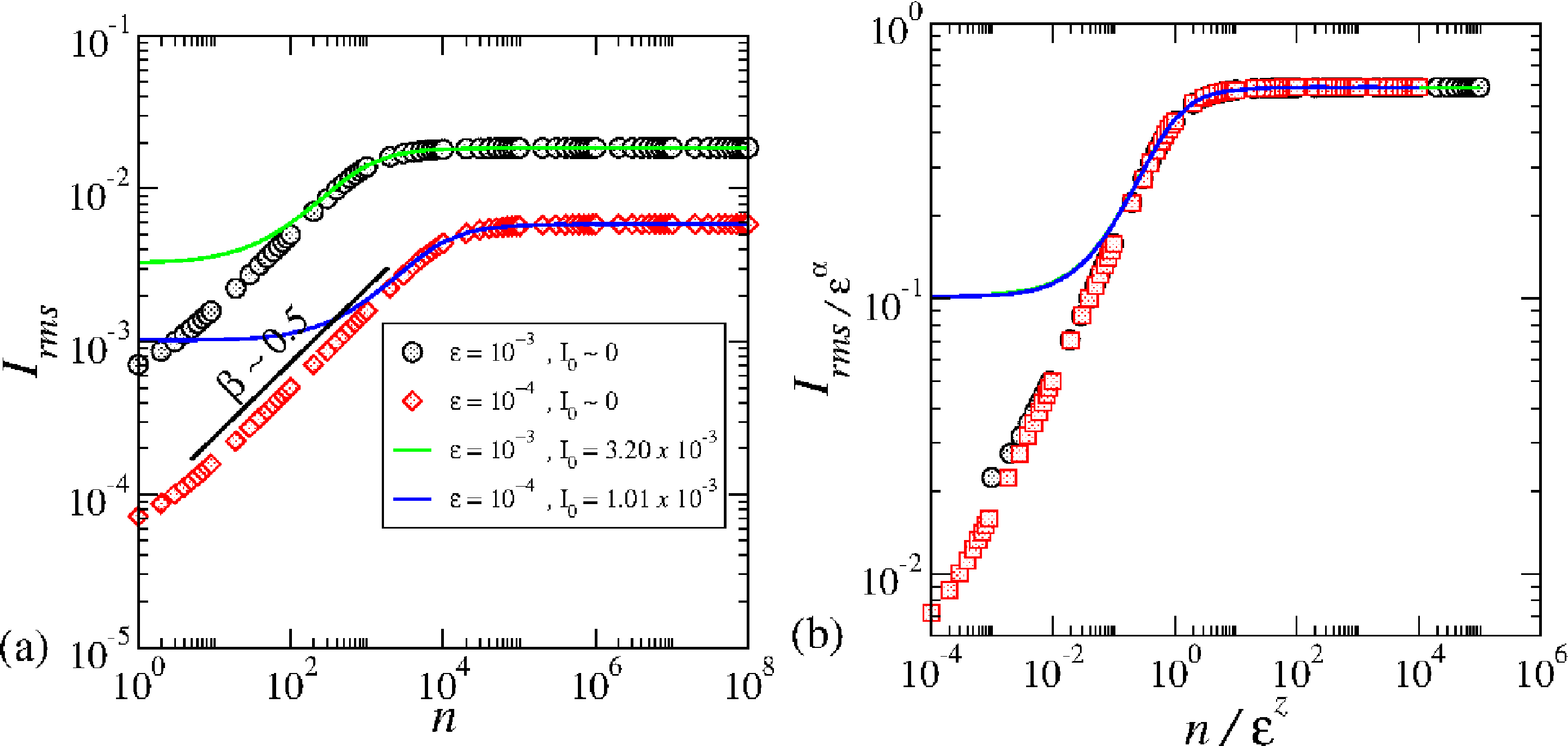}}
\caption{(a) Root mean square action $I_{\rm rms}$ as a function of time for different values of the control parameter $\epsilon$ and initial conditions. (b) Collapse of the curves in (a) onto a universal scaling curve.}
\label{tFig3}
\end{figure}

To characterise transport along the chaotic sea, we consider the root mean square of the action, defined as
\begin{equation}
I_{\rm rms} = \sqrt{\frac{1}{M} \sum_{i=1}^{M} \frac{1}{n} \sum_{j=1}^{n} I_{i,j}^2},
\label{teq2}
\end{equation}
where $M$ denotes the number of initial conditions in the ensemble and $n$ is the evolution time. The behaviour of $I_{\rm rms}$ is shown in Figure~\ref{tFig3}(a).

For initial actions typically $I_0 \approx 0$, the curves follow 
$I_{\rm rms} \propto (n \epsilon^2)^{\beta}$,
with $\beta \approx 1/2$, indicating that particle transport is equivalent to normal diffusion. The presence of the factor $\epsilon^2$ in this expression may at first appear ad hoc, but it has a clear dynamical origin. The nonlinear term $\sin(\theta_n)$ in the first equation of the mapping~(\ref{teq1}) acts as the elementary excitation of the dynamics. Assuming chaotic motion, statistical independence between $I$ and $\theta$, and small values of $I$, the dynamics reduces to an effective random walk with step size $\epsilon/\sqrt{2}$.

This interpretation can be made quantitative by squaring the first equation of the mapping~(\ref{teq1}), averaging over an ensemble of initial phases $\theta_0 \in [0,2\pi]$, and assuming statistical independence between $I$ and $\theta$. One then obtains
$\overline{I^2}_{n+1} = \overline{I^2}_n + \epsilon^2/2$,
which yields a diffusion coefficient $D = \epsilon^2/4$. Approximating the corresponding difference equation by a differential equation \cite{ref10} leads to
$\overline{I^2}(n) = \overline{I^2}_0 + n \epsilon^2/2$,
thereby confirming analytically the scaling with $\epsilon^2$.

At sufficiently long times, the presence of invariant spanning curves halts the growth of $I_{\rm rms}$, leading to saturation. The saturated value scales \cite{ref23} as
$I_{\rm rms,sat} \propto \epsilon^{\alpha}$,
with $\alpha = 1/(1+\tilde{\gamma})$. The crossover between the growth and saturation regimes occurs at a characteristic time
$n_x \propto \epsilon^{z}$,
where $z = -2\tilde{\gamma}/(\tilde{\gamma}+1)$. After appropriate scaling transformations, all curves collapse onto a single universal curve, as shown in Figure~\ref{tFig3}(b).

The excellent data collapse observed in Figure~\ref{tFig3}(b) provides strong evidence of scaling invariance in the chaotic dynamics near the transition from integrability to non-integrability for the mapping~(\ref{teq1}). We also note that scaling behaviour persists even when the initial action is not vanishingly small leading to a second time scale, as illustrated by the continuous curves obtained from the analytical solution of the diffusion equation under appropriate boundary conditions.

We note that in the limit $\epsilon \rightarrow 0$ the order parameter
$I_{\rm sat} \propto \epsilon^{\alpha}$ vanishes continuously. The susceptibility, defined as the response of the order parameter to variations of the control parameter, must diverge in this limit. The susceptibility is given by
\begin{equation}
\chi = \frac{\partial I_{\rm sat}}{\partial \epsilon}
= \frac{1}{1+\tilde{\gamma}}\,\frac{1}{\epsilon^{\tilde{\gamma}/(1+\tilde{\gamma})}}.
\end{equation}
Since $\tilde{\gamma}$ is non-negative, it follows that $\chi \rightarrow \infty$ as
$\epsilon \rightarrow 0$. This divergence provides a clear and unambiguous
signature of a second-order phase transition.

A final and essential element in the characterization of the transition is the
role played by topological defects. This terminology refers to the mechanism responsible for the breakdown of ergodicity. If the dynamics were fully chaotic, with no periodic
structures, and if time averages coincided with microcanonical averages, the
system would be ergodic. This is not the case. The phase space is mixed, with
the presence of periodic islands embedded in a chaotic sea. These islands act
as topological obstructions to transport and are therefore interpreted as
topological defects that destroy ergodicity. When chaotic trajectories pass
sufficiently close to such structures, stickiness occurs \cite{ref25}, modifying escape
rates and transport properties.

In summary, we have identified the fundamental ingredients required to classify
a second-order phase transition in a dynamical system. The scaling properties
of chaotic diffusion are linked to the finite size of the chaotic domain and are
characterized by a set of critical exponents that allow the curves of
$I_{\rm rms}$ to collapse onto a universal scaling function. The order
parameter was identified as $I_{\rm sat} \propto \epsilon^{\alpha}$, with
$\alpha = 1/(1+\tilde{\gamma})$, and vanishes continuously as $\epsilon \rightarrow 0$. The associated susceptibility diverges in the same limit, providing a second
independent signature of criticality. The elementary excitations responsible
for transport originate from the nonlinear term in the mapping, leading to an
effective random-walk dynamics at low action. Finally, periodic islands were
interpreted as topological defects in phase space, giving rise to stickiness
and also for anomalous transport. These results demonstrate that the transition from
integrability to non-integrability in mapping~(\ref{teq1}) is fully analogous
to a second-order phase transition, a halmark for the scaling invariance.

\subsection{Transition from integrability to non-integrability in a static billiard}

We now turn to a second example of a dynamical phase transition, observed in a
static billiard system. We consider an oval-like billiard \cite{berry} whose boundary is
defined by $R(\theta) = 1 + \epsilon \cos(p\theta)$. For $\epsilon = 0$, the
system is integrable and the phase space is foliated \cite{ref19}: the reflection angle is conserved, leading to periodic motion when the angle is commensurate with $\pi$
and to quasiperiodic motion otherwise. For small values of $\epsilon$, a thin
chaotic layer emerges, separating rotational from librational motion. As
$\epsilon$ increases, this chaotic layer grows in width and progressively
occupies a larger fraction of phase space, providing a clear picture of how
chaos develops as a function of the control parameter \cite{ref16}.

We show that the width of the chaotic layer scales with $\sim \epsilon^{\tilde{\alpha}}$, where the critical exponent $\tilde{\alpha}$
coincides with those obtained for the Fermi-Ulam model \cite{ref26}, periodically corrugated waveguides \cite{ref27}, and families of two-dimensional nonlinear area-preserving mappings with $\tilde{\gamma} = 1$, as discussed in the previous section.

To characterize this transition quantitatively, we analyze the root mean square
deviation of the reflection angle from its mean value,
$\omega_{\rm rms,sat}$, which plays the role of an order parameter. As
$\epsilon \rightarrow 0$, $\omega_{\rm rms,sat}$ vanishes continuously, while
its susceptibility $\chi = d\omega_{\rm rms,sat}/d\epsilon$ diverges. This
behaviour mirrors that of a continuous phase transition: a smooth suppression
of chaotic fluctuations accompanied by a divergent response function.
Stickiness, symmetry breaking emerge near criticality, and
diffusion within the chaotic layer. These results establish the
integrability--chaos transition in classical billiards as a genuine
second-order phase transition with universal features, unifying stochastic
layers, anomalous transport, and critical behaviour near the onset of chaos.

\begin{figure}[t]
\centerline{(a)\includegraphics[width=0.47\linewidth]{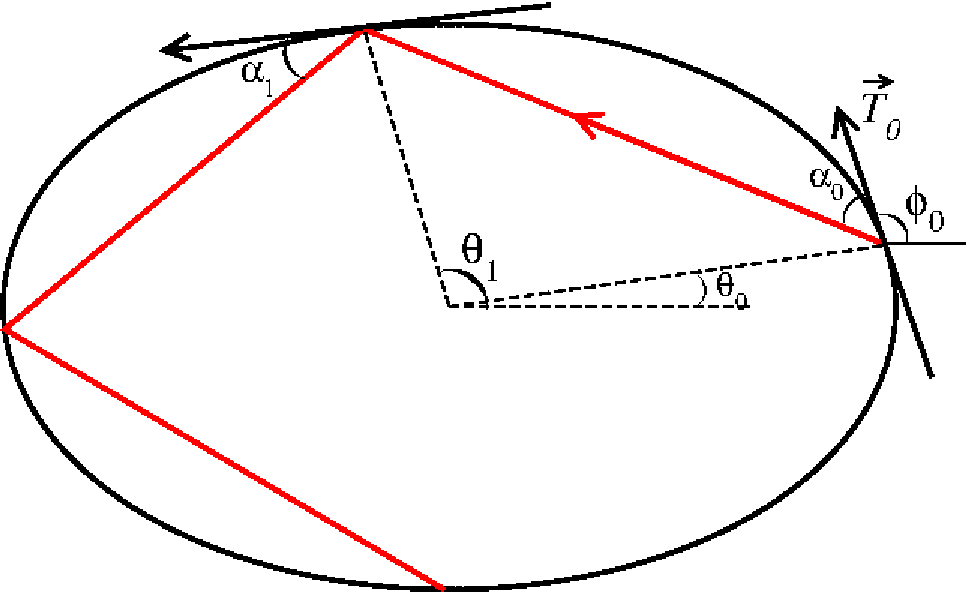}
(b)\includegraphics[width=0.47\linewidth]{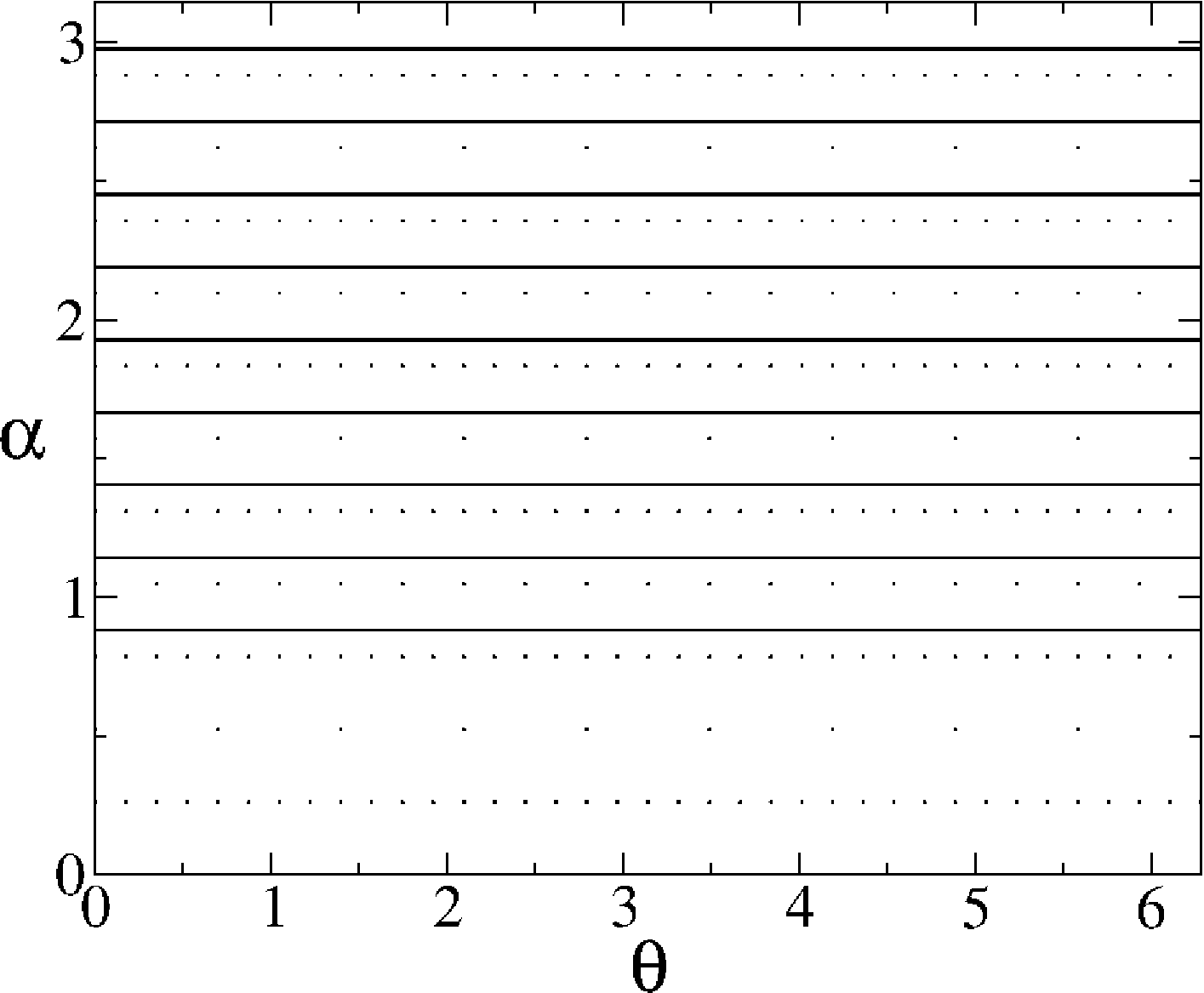}}\vspace{0.5cm}
\centerline{(c)\includegraphics[width=0.47\linewidth]{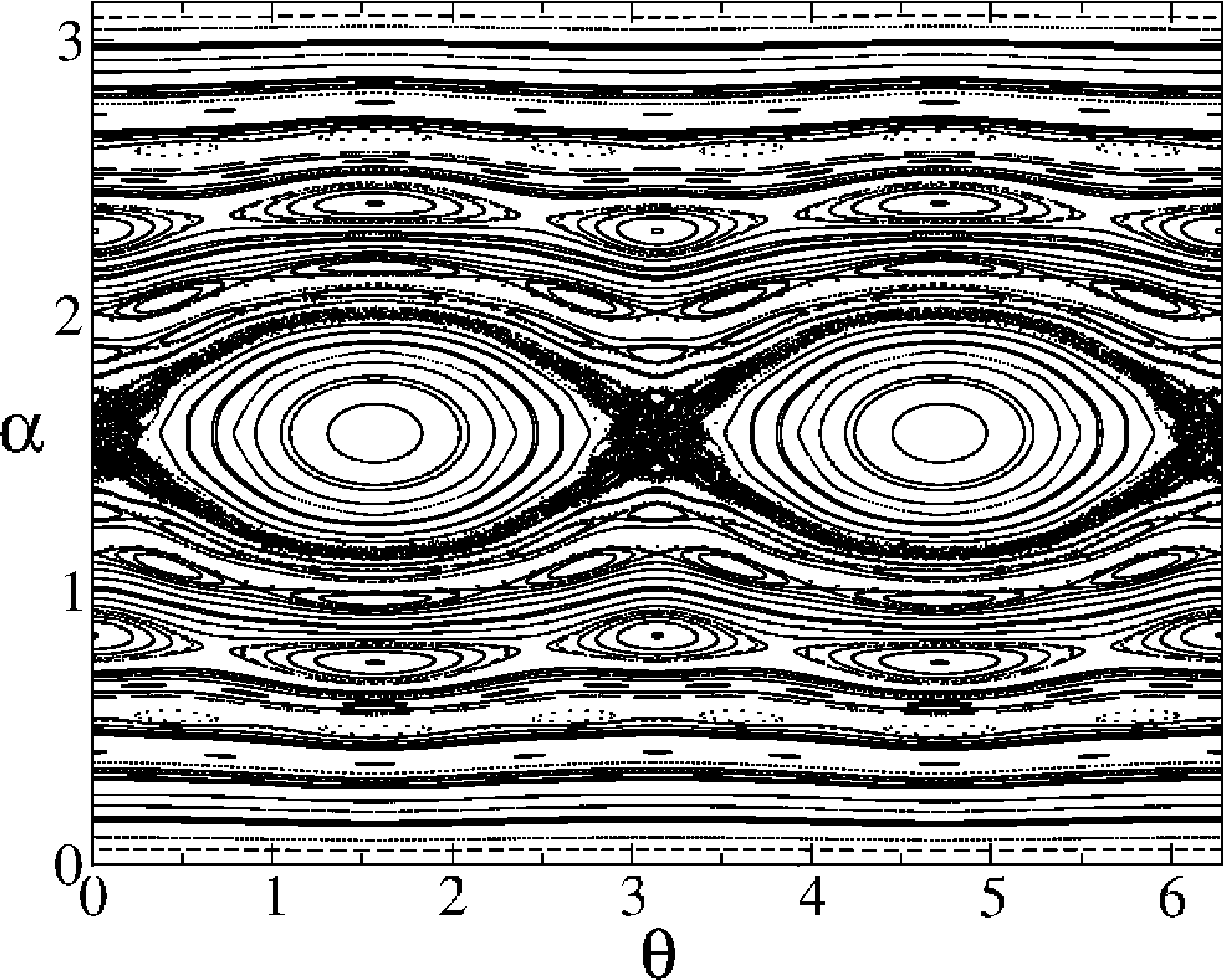}
(d)\includegraphics[width=0.47\linewidth]{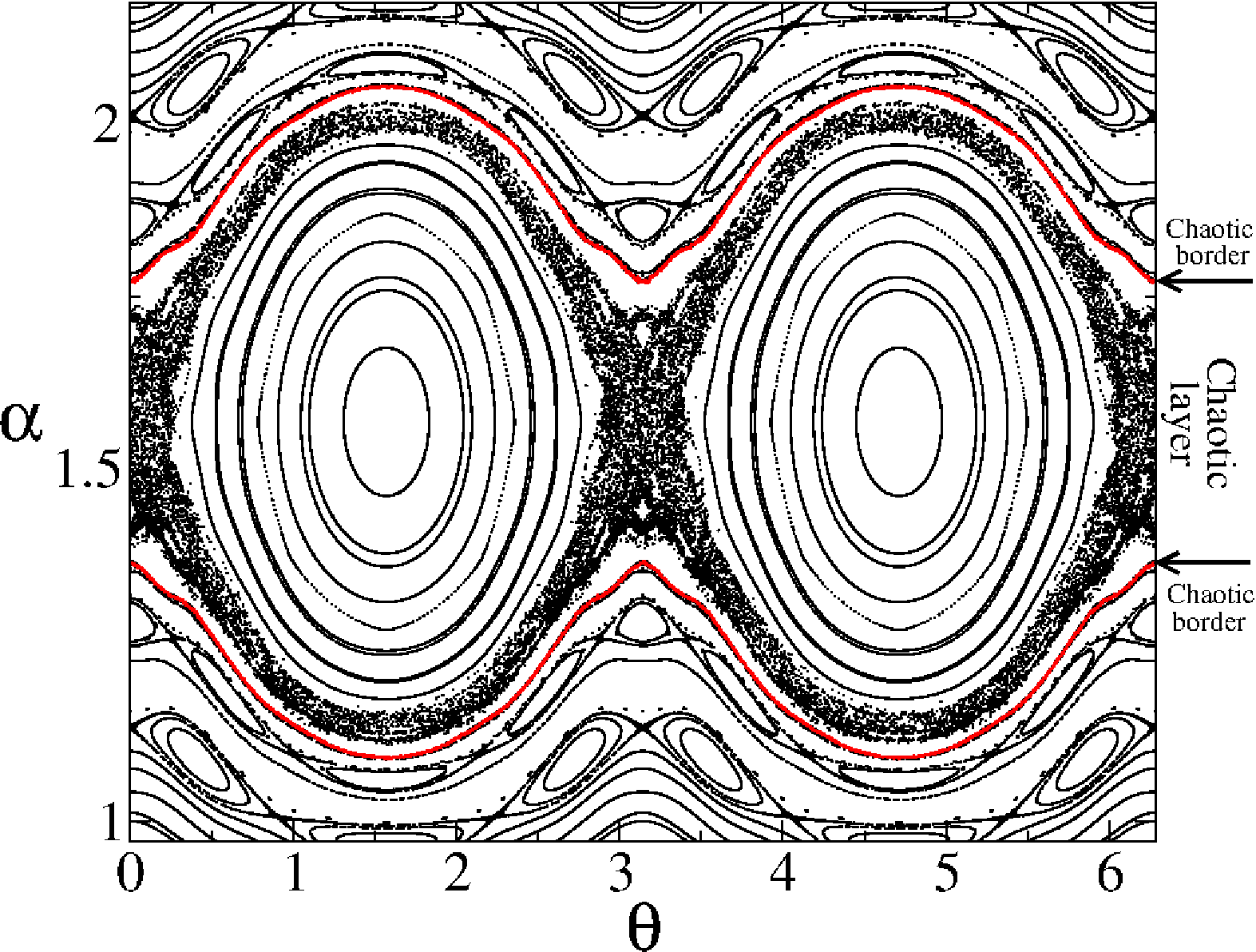}}
\caption{(a) Sketch of the angles describing the dynamics of the billiard. The boundary was constructed using $R=1+\epsilon\cos(p\theta)$ with $p=2$ and $\epsilon=0.1$. In the figure, $\theta_0$ denotes the polar angle at the initial collision, $\vec{T}_0$ is the tangent vector at $\theta_0$, and $\alpha_0$ is the angle of the trajectory measured with respect to $\vec{T}_0$. (b) Phase space for $\epsilon=0$ showing a symmetric and ordered dynamics. (c) Phase space for $\epsilon=0.05$, displaying a mixed structure with a chaotic layer near $\alpha=\pi/2$, periodic islands and invariant spanning curves. The periodic domain confined by the chaotic layer corresponds to librational motion, while the invariant spanning curves above and below it correspond to rotational motion. (d) Zoom of (c) showing the chaotic domain bounded by the upper and lower chaotic borders (continuous red curves).}
\label{bFig1}
\end{figure}

The model considered is the oval billiard whose boundary in polar
coordinates is given by $R(\theta) = 1 + \epsilon \cos(p\theta)$, where $\theta$
is the polar angle, $\epsilon$ controls the deformation from the circular shape,
and $p$ is an integer. The dynamics of a particle confined within the boundary
is described by a discrete mapping for the variables $(\theta_n, \alpha_n)$,
where $\theta_n$ denotes the angular position of the particle at the $n$th
collision and $\alpha_n$ is the angle between the particle trajectory and the
tangent vector to the boundary at $\theta_n$, as illustrated in
Fig.~\ref{bFig1}(a).

Since the boundary is expressed in polar coordinates, the position of the
particle at the $n$th collision in rectangular coordinates is given by
$X(\theta_n)=R(\theta_n)\cos(\theta_n)$ and
$Y(\theta_n)=R(\theta_n)\sin(\theta_n)$. For a given initial condition
$(\theta_n,\alpha_n)$, the angle between the tangent vector at the boundary
point $[X(\theta_n),Y(\theta_n)]$ and the horizontal axis is
$\phi_n=\arctan\!\left[Y'(\theta_n)/X'(\theta_n)\right]$. In the absence of
external forces between collisions, the particle moves along straight-line
segments with constant velocity. The trajectory connecting two successive
collisions satisfies
\begin{eqnarray}
Y(\theta_{n+1}) - Y(\theta_n) =
\tan(\alpha_n + \phi_n)\left[
X(\theta_{n+1}) - X(\theta_n) \right],
\label{B_eq2}
\end{eqnarray}
where $X(\theta_{n+1})$ and $Y(\theta_{n+1})$ denote the rectangular coordinates
of the next collision point, obtained numerically by solving
Eq.~(\ref{B_eq2}). The angle between the trajectory and the tangent vector at
$\theta_{n+1}$ is then given by
$\alpha_{n+1} = \phi_{n+1} - (\alpha_n + \phi_n)$.

The discrete mapping describing the billiard dynamics is given by
as
\begin{equation}
\left\{
\begin{array}{ll}
H(\theta_{n+1}) = R(\theta_{n+1})\sin(\theta_{n+1}) - Y(\theta_n) - \tan(\alpha_n + \phi_n)\left[
R(\theta_{n+1})\cos(\theta_{n+1}) - X(\theta_n) \right], \\
\alpha_{n+1} = \phi_{n+1} - (\alpha_n + \phi_n),
\end{array}
\right.
\label{B_eq4}
\end{equation}
where $\theta_{n+1}$ is obtained numerically from the condition
$H(\theta_{n+1})=0$, and
$\phi_{n+1}=\arctan[Y'(\theta_{n+1})/X'(\theta_{n+1})]$.

Figure~\ref{bFig1}(b) shows the phase space for $\epsilon=0$, corresponding to
the circular billiard with $R=1$. In addition to energy conservation, angular momentum is
also conserved, providing the two constants of motion required for
integrability \cite{ref19}. The phase space is fully symmetric, with the reflection angle
$\alpha$ remaining constant along the dynamics. In contrast,
Fig.~\ref{bFig1}(c) displays the phase space for $\epsilon=0.05$, where a mixed
structure emerges, consisting of chaotic regions, stability islands, and
quasiperiodic trajectories. The symmetry present in the integrable case is
clearly broken, confirming that symmetry breaking accompanies the transition
from integrability to chaos.

The chaotic dynamics develops within a well-defined layer in phase space,
whose width depends on the strength of the control parameter $\epsilon$ and is
bounded by two limiting curves, as shown in the magnified view of
Fig.~\ref{bFig1}(d). The monotonic growth of this chaotic band with increasing
$\epsilon$ provides a geometric measure of how chaos progressively invades the
phase space as regularity is destroyed.

\begin{figure}[t]
\centerline{(a)\includegraphics[width=0.47\linewidth]{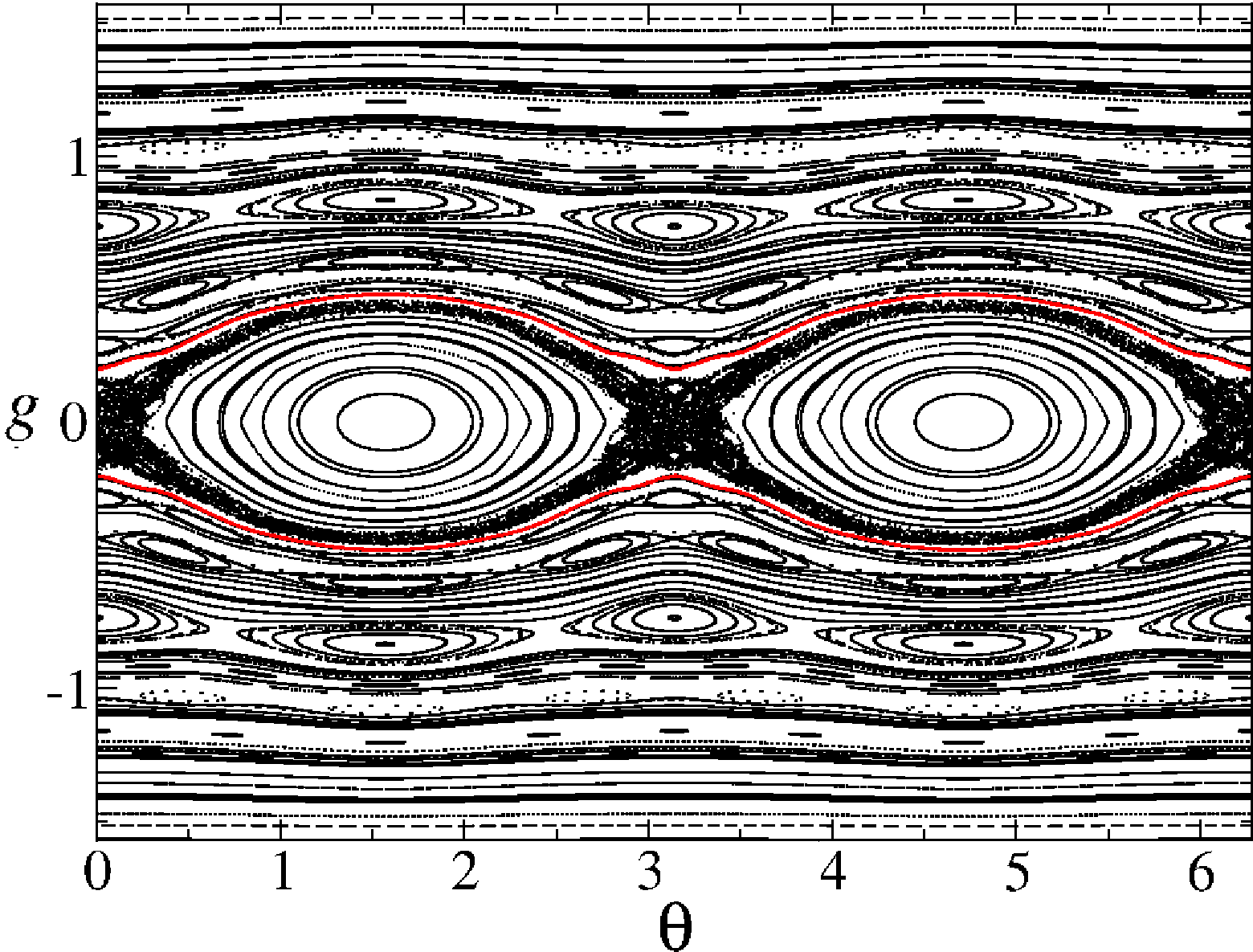}
(b)\includegraphics[width=0.47\linewidth]{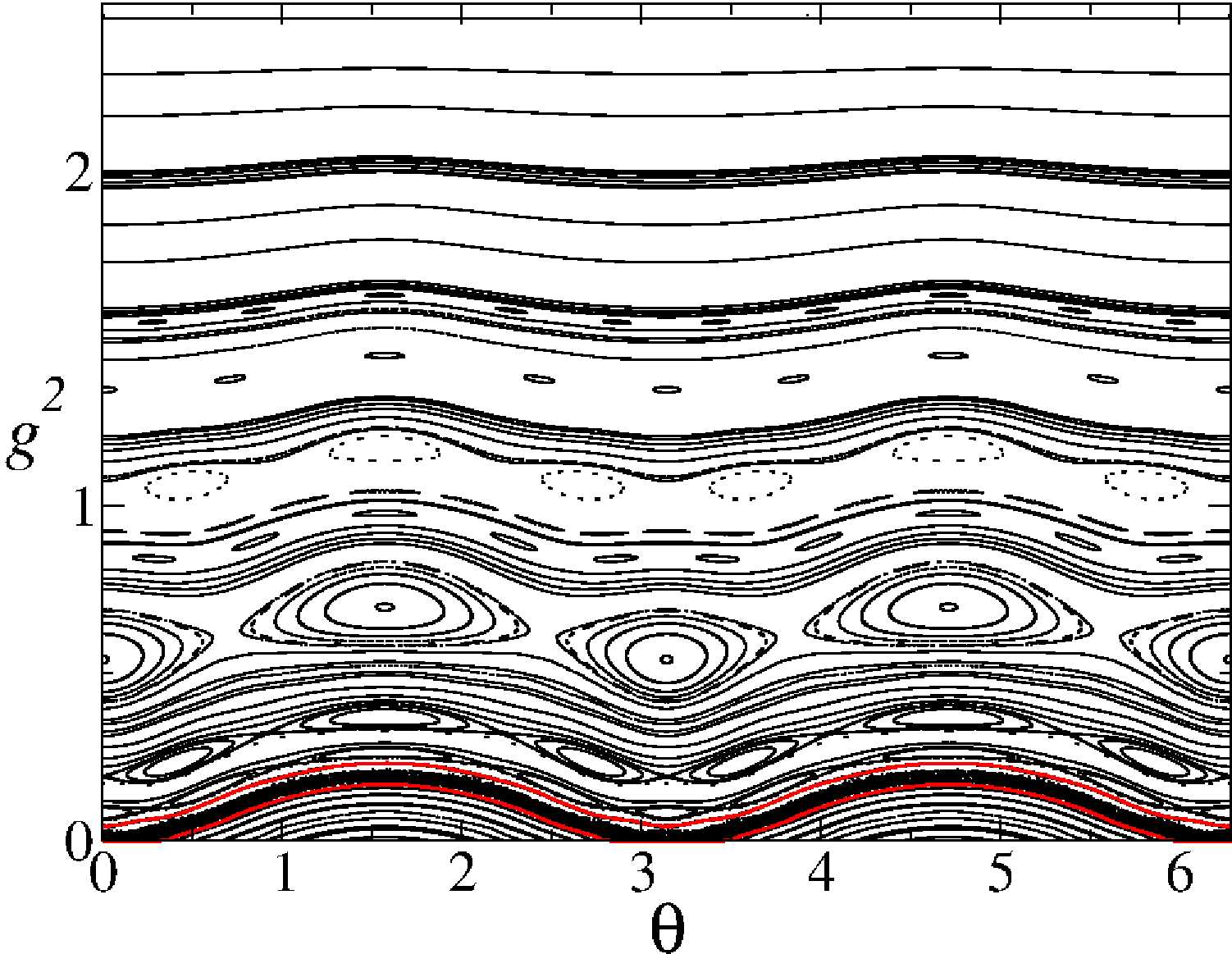}}\vspace{0.5cm}
\centerline{(c)\includegraphics[width=0.47\linewidth]{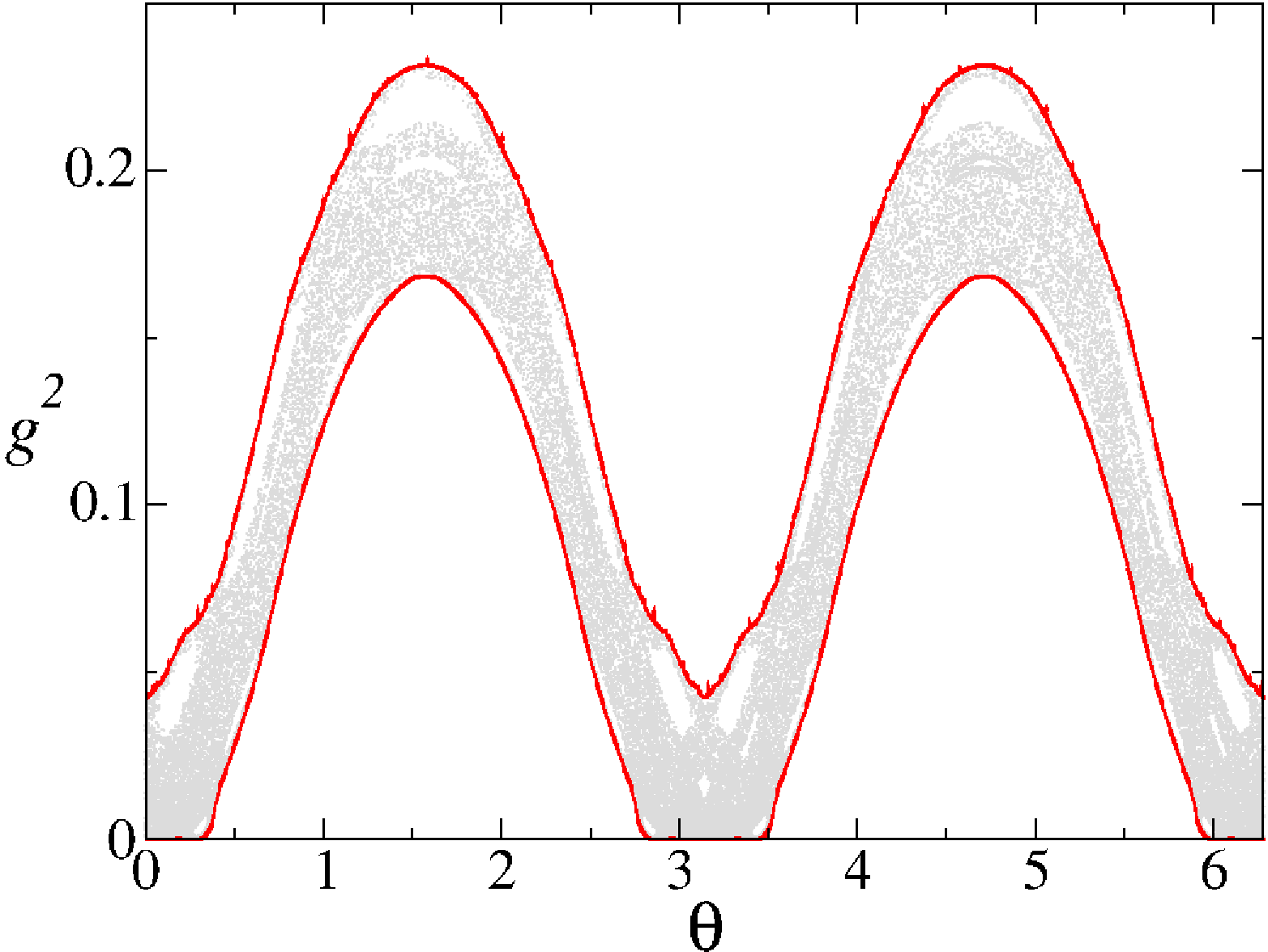}
(d)\includegraphics[width=0.5\linewidth]{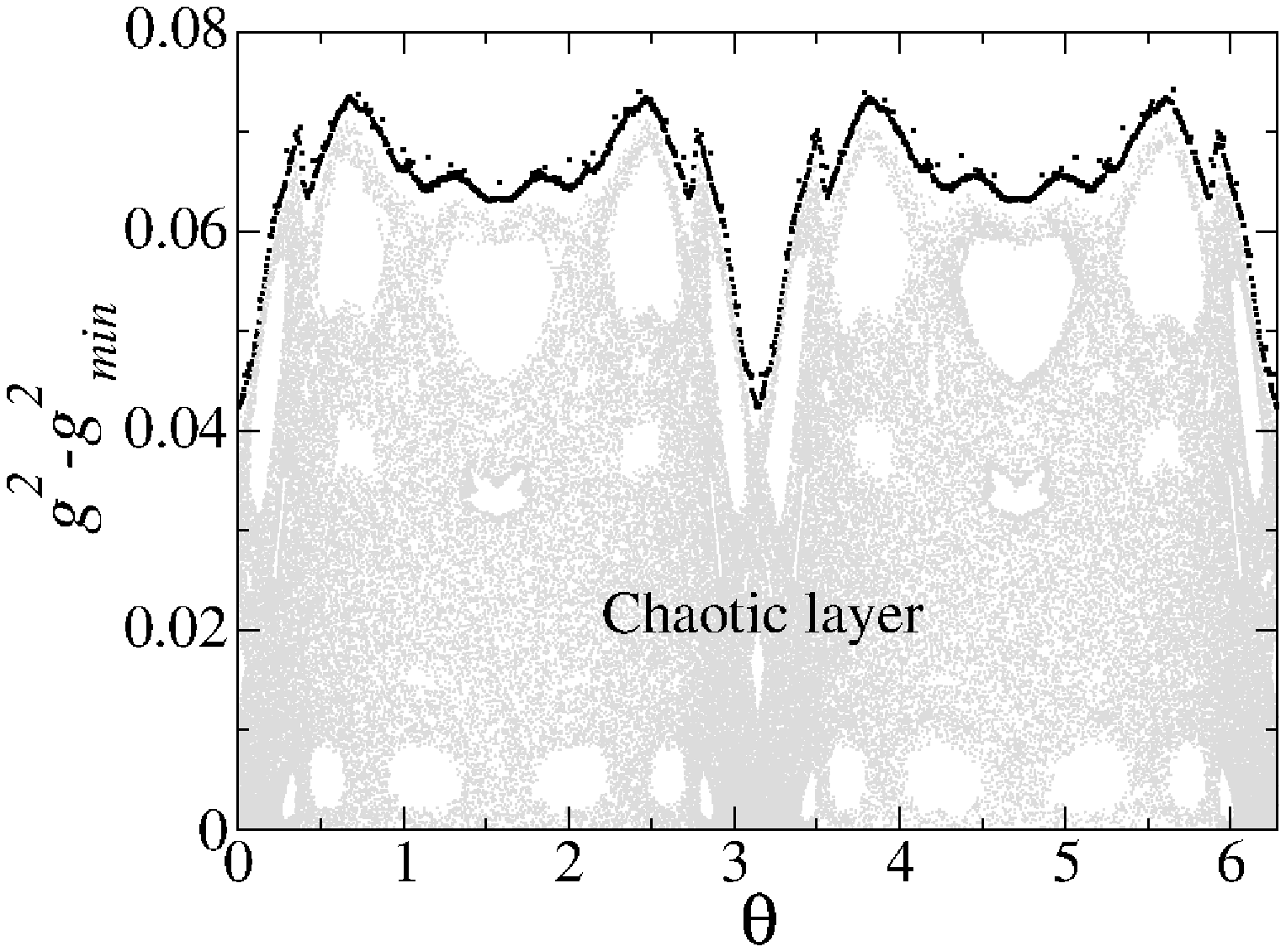}}
\caption{(a) Phase space for $\epsilon=0.05$ under the
transformation $g=\alpha-\pi/2$. The phase space is now symmetric about
$g=0$. (b) Plot of $g^2$ versus $\theta$ for the same control
parameter. (c) Chaotic stripe bounded by two limiting curves. (d) Chaotic layer
after the transformation $\omega=g^2-g_{\min}^2$, where the bullet
points indicate a numerical approximation of the chaotic border.}
\label{bFig2}
\end{figure}

To quantify the width of the chaotic layer, it is convenient to introduce a
more suitable set of variables. Inspection of Fig.~\ref{bFig1}(c) reveals that
the phase space is symmetric with respect to $\alpha=\pi/2$. It is therefore
natural to shift this symmetry axis to the origin by defining the variable
$g=\alpha-\pi/2$, as illustrated in Fig.~\ref{bFig2}(a).

With this transformation, the phase space becomes symmetric about $g=0$,
so that the positive region mirrors the negative one. While this symmetry is
conceptually useful, it implies $\langle g \rangle = 0$, making $g$
itself unsuitable for characterizing diffusion. We therefore consider its
square, $g^2$, shown in Fig.~\ref{bFig2}(b), which renders the entire phase
space positive. In these coordinates, the chaotic layer appears clearly as a
stripe bounded by two curves, as shown in Fig.~\ref{bFig2}(c). A further
transformation, $\omega=g^2-g_{\min}^2$, proves particularly useful
for analyzing the width of the chaotic diffusion, as illustrated in
Fig.~\ref{bFig2}(d).

As the control parameter $\epsilon$ increases, the chaotic layer widens,
forming a stripe along which chaotic motion develops. New periodic islands
emerge as a consequence of the boundary deformation, while invariant spanning
curves associated with grazing trajectories -- those corresponding to nearly
tangent collisions  --  become progressively less prominent. These curves
disappear entirely when the control parameter reaches the critical \cite{ref28} value
$\epsilon_c=1/(1+p^2)$. Beyond this point, all invariant spanning curves are
destroyed, allowing chaos to permeate most of the accessible phase space and to
invade the remaining periodic structures.

As the chaotic layer grows, increasingly large regions of phase space become
chaotic, providing an ideal setting for identifying a dynamical phase
transition. Our goal is to identify an observable associated with
chaotic diffusion that can act as an order parameter, distinguishing the
ordered (integrable) phase from the disordered (chaotic) regime, vanishing
continuously at the transition, and whose response to variations of the
control parameter diverges as the nonlinearity $\epsilon$ approaches zero.

We now investigate the behaviour of the fluctuations around the chaotic stripe
as a function of the control parameter $\epsilon$. To this end, we introduce
the observable
\begin{equation}
\Omega(n,\epsilon)=\frac{1}{M}\sum_{i=1}^{M}
\sqrt{\bar{\omega_i^2}(n,\epsilon)-\bar{\omega_i}^2(n,\epsilon)},
\end{equation}
where $\bar{\omega}(n,\epsilon)=\frac{1}{n}\sum_{k=1}^n\omega_k$. We consider an
ensemble of $M=10^3$ initial conditions, with the initial angles $\alpha$
chosen very close to the lower boundary of the chaotic stripe and
$\theta_0$ uniformly distributed in the interval $\theta_0\in[0,2\pi]$.
This choice allows the particles to experience the maximum possible diffusion
within the chaotic region.

\begin{figure}[t]
\centerline{(a)\includegraphics[width=0.45\linewidth]{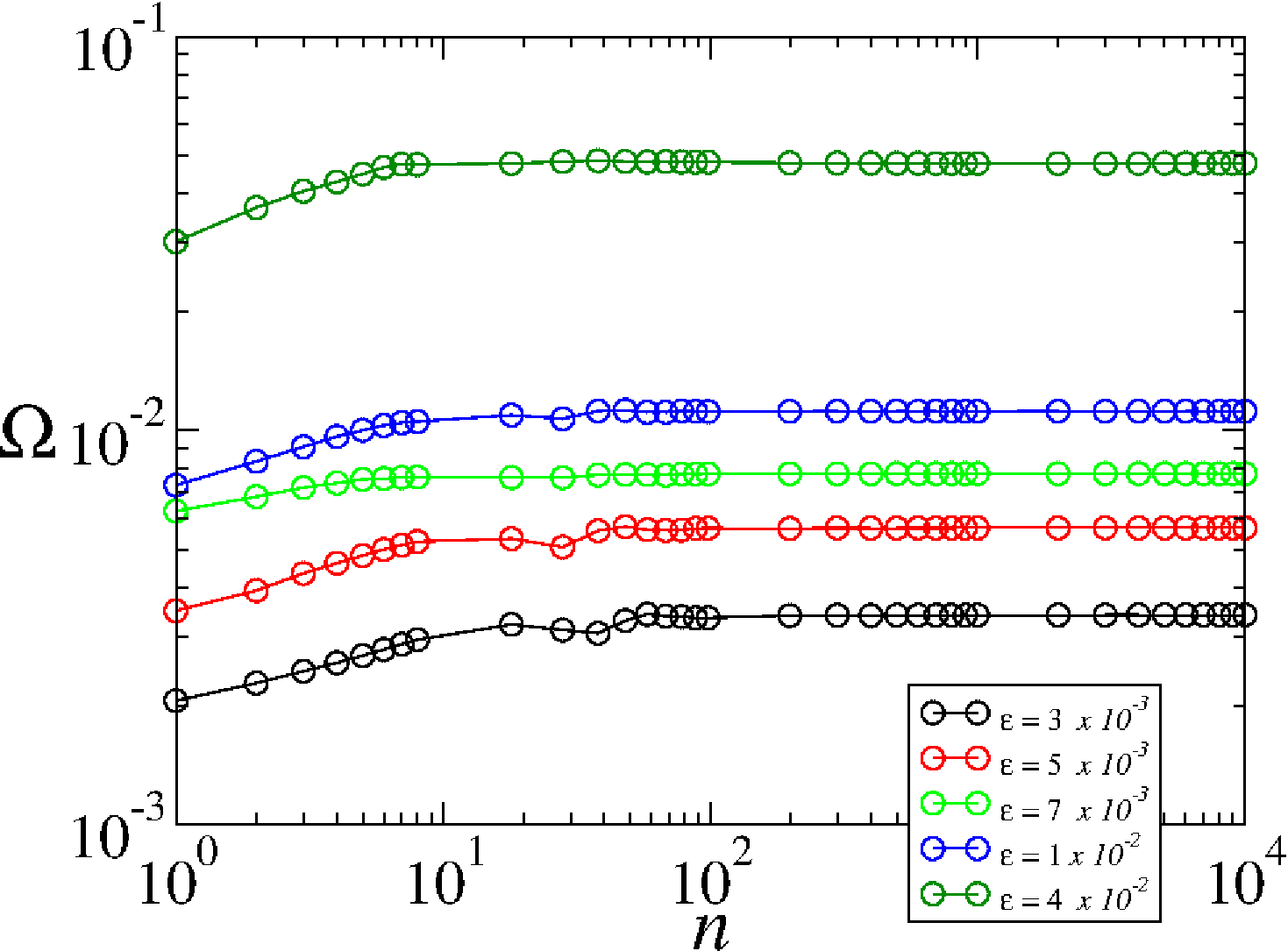}
(b)\includegraphics[width=0.45\linewidth]{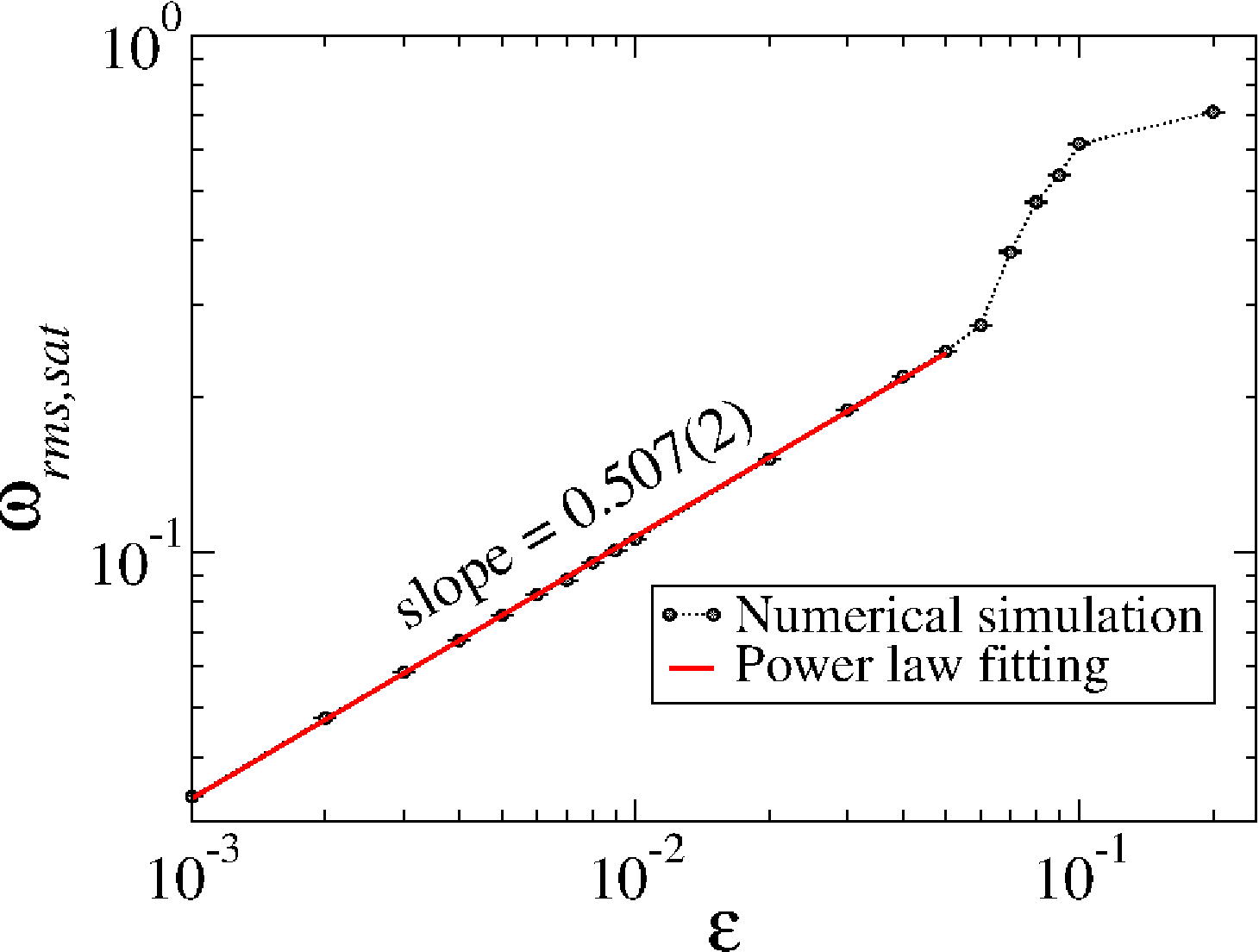}}
\caption{ (a) Plot of $\Omega~vs.~n$ for different control
parameters, as labelled. (b) Plot of $\omega_{rms,{\rm sat}}~vs.~\epsilon$.
A power-law fit yields $\tilde{\alpha}=0.507(2)$. Error bars indicate ensemble
averages.}
\label{bFig3}
\end{figure}

Figure~\ref{bFig3}(a) shows $\Omega$ as a function of the number of collisions
$n$ for several values of the control parameter $\epsilon$. The curves exhibit
an initial growth associated with diffusion, followed by a crossover and
eventual saturation, which marks the steady state of the diffusive process.

Since $\Omega$ is defined in terms of the squared angle, it is convenient to
analyze instead the quantity $\omega_{rms}=\sqrt{\Omega}$. For sufficiently
large $n$, all curves reach a saturation regime. The dependence of the
saturation value on the control parameter is shown in
Fig.~\ref{bFig3}(b). A clear power-law behaviour is observed for small $\epsilon$,
except in the region $\epsilon\gtrsim0.07$, where the chaotic domain expands
rapidly and dominates the phase space. This latter regime lies far from the
vicinity of integrability and is not relevant for the critical
analysis.

A power-law fit of the data closest to criticality yields
$\omega_{rms,{\rm sat}}\propto\epsilon^{\tilde{\alpha}}$ with
$\tilde{\alpha}=0.507(2)$. Because we are dealing with diffusive dynamics, it is
natural that the order parameter should quantify the extent of diffusion along
the chaotic stripe. The observable $\omega_{rms,{\rm sat}}$ satisfies all the
requirements for this role: it is well defined within the chaotic layer, it
follows a power law in $\epsilon$ near the transition, and it vanishes
continuously as $\epsilon\rightarrow0$.

The response of the order parameter to variations of the control parameter is given by
$\chi=d\omega_{rms,{\rm sat}}/d\epsilon\propto
\tilde{\alpha}/\epsilon^{\,1-\tilde{\alpha}}$. Since
$0<\tilde{\alpha}<1$, the susceptibility diverges as $\epsilon\rightarrow0$.
These results demonstrate that the transition from integrability to
non-integrability in the oval-like billiard displays all the hallmarks of a
second-order phase transition.

Regarding the role of elementary excitations, when $\epsilon=0$ the phase space
is entirely regular and no diffusion occurs. For small but finite $\epsilon$,
a chaotic layer emerges and spreads along a stripe in phase space. This
behaviour is only possible because the control parameter is nonzero. We interpret $\epsilon$ as setting the elementary step length that
enables diffusive dynamics to develop in the reflection angle $\alpha$.

Concerning topological defects, Fig.~\ref{bFig2}(d) reveals regions of phase
space that remain unvisited by chaotic trajectories. These regions correspond
to stability islands associated with periodic orbits. When chaotic trajectories
wander sufficiently close to such islands, they may become temporarily trapped
before escaping and later returning leading to stickiness \cite{ref25}.
The coexistence of stability islands and chaotic regions breaks ergodicity and
modifies transport properties in phase space. Nevertheless, because the chaotic
layer has a finite extent, bounded by the lower and upper limits of chaotic
motion, the deviation around the mean angle eventually saturates at long times.

As a brief discussion, the phase transition observed for mapping (\ref{teq1})
with $\tilde{\gamma}>0$, shows that the diffusive dynamics saturates due to the
existence of invariant spanning curves in phase space. The position of these
curves can be estimated as
\begin{equation}
I_{fisc}\cong
\left[{{\tilde{\gamma}\epsilon}\over{0.9716\ldots}}\right]^{{1}/{(1+\tilde{\gamma})}},
\label{fisc}
\end{equation}
which plays exactly the same role as $\omega_{fisc}$ in the oval billiard model
discussed above.

The particular case $\tilde{\gamma}=1$ recovers both the Fermi--Ulam model \cite{ref26} and the periodically corrugated waveguide \cite{ref27}. In general, the critical exponent
governing the saturation of the diffusive behaviour is
$\tilde{\alpha}=1/(1+\tilde{\gamma})$. For $\tilde{\gamma}=1$, one finds
$\tilde{\alpha}=1/2$, precisely the same value obtained for the transition
reported in the present work. This remarkable agreement demonstrates that,
despite the apparent differences between the systems -- whether a Fermi-Ulam
model, a periodically corrugated waveguide, or a family of nonlinear
area-preserving mappings such as Eq.~(\ref{teq1}) -- the corresponding phase
transitions belong to the same universality class.

\subsection{Phase transition from bounded to unbounded diffusion}

We now turn to the characterization of a transition from bounded to unbounded
diffusion, using the diffusion equation applied to the dissipative standard
mapping \cite{ref17}. The dynamics is described by
$I_{n+1}=(1-q)I_n+\epsilon\cos(\theta_n)$ and
$\theta_{n+1}=(\theta_n+I_{n+1})~{\rm mod}(2\pi)$, where $q\in[0,1]$ is the
dissipation parameter and $\epsilon$ controls the strength of the nonlinearity.

In the conservative limit $q=0$, this system exhibits two well-known
transitions \cite{ref24}: (i) a transition from integrability at $\epsilon=0$, where the phase space is foliated, to non-integrability for $\epsilon\neq0$, where a
mixed phase space composed of chaotic seas, periodic islands, and invariant
spanning curves emerges; and (ii) a transition at the critical value
$\epsilon_c=0.9716\ldots$, from locally chaotic dynamics for
$\epsilon<\epsilon_c$ to globally chaotic motion for $\epsilon>\epsilon_c$,
where invariant spanning curves are destroyed and chaotic diffusion becomes
unbounded, depending on the initial conditions.

For $q\neq0$, the determinant of the Jacobian matrix is
$\det J=(1-q)$, implying a violation of Liouville's theorem and the
appearance of attractors in phase space. For sufficiently large values of
$\epsilon$, typically $\epsilon>10$, sinks are no longer observed and the
dynamics is dominated by chaotic attractors in the limit of small dissipation.
In this regime, one observes a transition from bounded diffusion for
$q\neq0$ to unbounded diffusion in the conservative limit $q=0$.
This is precisely the transition addressed in the present work.

\begin{figure}[t]
\centerline{\includegraphics[width=0.9\linewidth]{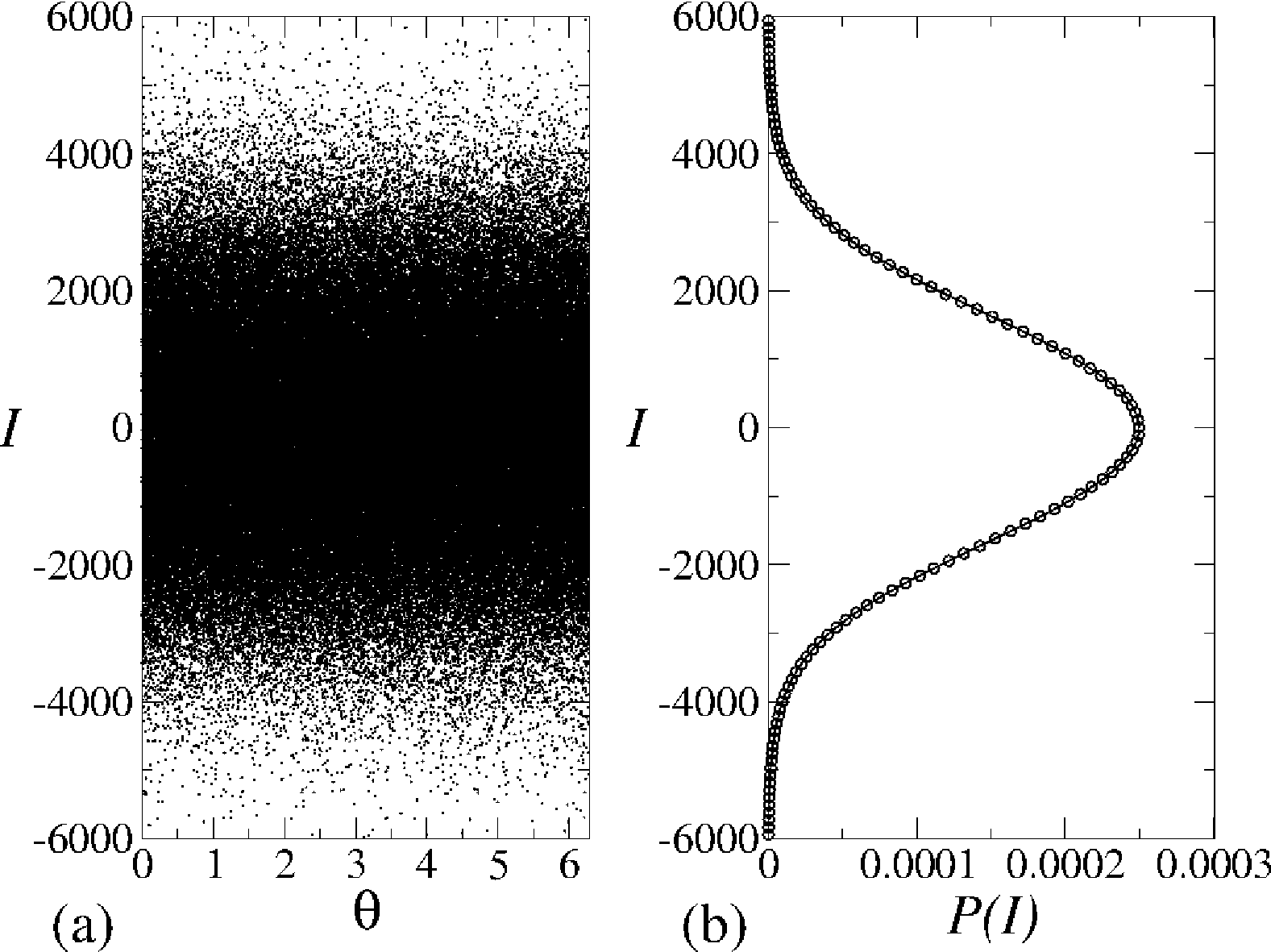}}
\caption{(a) Plot of the phase space for the standard dissipative mapping 
considering the parameters $\epsilon=100$ and $q=10^{-3}$. (b) Normalized 
probability distribution for the chaotic attractor shown in (a).}
\label{smFig1}
\end{figure}

Our main goal here is to give an analytical characterization of the scaling invariance
associated with the transition from bounded diffusion ($q\neq0$) to
unbounded diffusion ($q=0$) in the regime of strong nonlinearity
($\epsilon\gg1$). The parameter range relevant for this transition corresponds to small but finite dissipation, typically $q\in[10^{-5},10^{-2}]$ and $\epsilon>10$. In this regime, the system displays bounded
diffusion for $q\neq0$ and unbounded diffusion in the conservative limit.
A representative phase-space portrait is shown in Fig.~\ref{smFig1}(a) for
$\epsilon=10$ and $q=10^{-3}$, illustrating a chaotic attractor. The
corresponding normalized probability distribution along the attractor is shown
in Fig.~\ref{smFig1}(b). The density of points is concentrated around $I\cong0$
and is symmetric with respect to the vertical axis, decaying rapidly away from
the origin.

Given an initial condition close to $I\cong0$, the particle diffuses along the
chaotic attractor measured with a positive Lyapunov exponent $\lambda=3.9120(1)$. A natural observable to characterize this diffusion is the
root-mean-square action,
$I_{rms}(n)=\sqrt{{{1}\over{M}}\sum_{i=1}^MI_i^2}$, where $M$ denotes an ensemble
of initial conditions distributed along the chaotic attractor. To obtain this
quantity, we solve the diffusion equation that governs the probability of
observing a given action $I$ at iteration $n$, namely $P(I,n)$. The diffusion
equation \cite{ref29} reads
\begin{equation}
{{\partial P(I,n)}\over{\partial n}}=
D{{\partial^2P(I,n)}\over{\partial I^2}},
\label{smeq1}
\end{equation}
where the diffusion coefficient $D$ can be derived from the first equation of
the mapping as
$D={{\overline{I^2}_{n+1}-\overline{I^2_n}}\over{2}}$. Assuming statistical
independence between $I_n$ and $\theta_n$ within the chaotic domain, a
straightforward calculation yields
\begin{equation}
D(q,\epsilon,n)=
{{q(q-2)}\over{2}}\overline{I^2}_n+
{{\epsilon^2}\over{4}}.
\label{smeq2}
\end{equation}

The evolution of $\overline{I^2}_n$ can also be obtained directly from the first
equation of the mapping by noting that
$\overline{I^2}_{n+1}-\overline{I^2}_n \cong d\overline{I^2}/dn$, which leads to
${{d\overline{I^2}}\over{dn}}=q(q-2)\overline{I^2}+{{\epsilon^2}\over{2}}$. 
The solution of this equation is
\begin{equation}
\overline{I^2}(n)=
{{\epsilon^2}\over{2q(2-q)}}+
\left(I_0^2+{{\epsilon^2}\over{2q(q-2)}}\right)
e^{-q(2-q)n}.
\label{smeq3}
\end{equation}

To compare with the experimentally accessible observable, Eq.~(\ref{smeq3})
must be averaged along the orbit, yielding
\begin{equation}
<\overline{I^2}(n)>={{1}\over{n+1}}\sum_{i=0}^n\overline{I^2}(i)={{
q(q-2)}\over{2(n+1)}}
\left[I^2_0+{{\epsilon^2}
\over { 2q(q-2)}}\left({{1-e^{-(n+1)q(2-q)}}\over{1-e^{
-q(2-q) } } } \right)\right].
\label{smeq4}
\end{equation}

To obtain a unique solution of Eq.~(\ref{smeq1}), we impose the boundary
conditions $\lim_{I\rightarrow\pm\infty}P(I,n)=0$ together with the initial
condition $P(I,0)=\delta(I-I_0)$, which ensures that all particles start from
the same action $I_0$ but with $M$ different initial phases
$\theta\in[0,2\pi]$. Although the diffusion coefficient $D$ depends weakly on
$n$, its variation from one iteration to the next is small, allowing it to be
treated as approximately constant during the solution of the diffusion
equation from $n$ to $n+1$. Once the solution is obtained, the explicit expression for $D$ given
in Eq.~(\ref{smeq2}) is reinstated.

The diffusion equation is solved using Fourier transforms \cite{ref29}. Since the
probability density is normalized,
$\int_{-\infty}^{\infty}P(I,n)\,dI=1$, we define
\begin{equation}
R(k,n)=\mathcal{F}\{P(I,n)\}=
{{1}\over{\sqrt{2\pi}}}
\int_{-\infty}^{\infty}P(I,n)e^{ikI}dI.
\label{smeq5}
\end{equation}
Differentiating $R(k,n)$ with respect to $n$ and using the property
$\mathcal{F}\{\partial^2P/\partial I^2\}=-k^2R(k,n)$, we obtain
${dR}/{dn}(k,n)=-Dk^2R(k,n)$, whose solution is
\begin{equation}
R(k,n)=R(k,0)e^{-Dk^2n}.
\label{smeq6}
\end{equation}
For the chosen initial condition,
$R(k,0)=\mathcal{F}\{\delta(I-I_0)\}=
{{1}\over{\sqrt{2\pi}}}e^{ikI_0}$. Inverting the transform yields
\begin{eqnarray}
P(I,n)&=&{{1}\over{\sqrt{2\pi}}}
\int_{-\infty}^{\infty}R(k,n)e^{-ikI}dk,\nonumber\\
&=&{{1}\over{\sqrt{4\pi Dn}}}
e^{-{{(I-I_0)^2}\over{4Dn}}}.
\label{eq7}
\end{eqnarray}

Equation~(\ref{eq7}) satisfies the diffusion equation
(\ref{smeq1}), the boundary and initial conditions, and is normalized by
construction. The observable of interest is
$\overline{I^2}(n)=\int_{-\infty}^{\infty}I^2P(I,n)\,dI$, leading to
$\overline{I^2}(n)=2D(n)n+I_0^2$. Substituting $D(n)$ from Eq.~(\ref{smeq2}), we
obtain
\begin{equation}
I_{rms}(n)=\sqrt{I_0^2+
{{nq(q-2)}\over{n+1}}
\left[I_0^2+
{{\epsilon^2}\over{2q(q-2)}}\right]
\left[{{-(n+1)q(2-q)}\over{
1-e^{-q(2-q)}}}\right]}.
\label{smeq8}
\end{equation}

We now discuss relevant limits of Eq.~(\ref{smeq8}). For $n=0$, one trivially
recovers $I_{rms}(0)=I_0$, consistent with the initial condition. In the limit
$n\rightarrow\infty$, we find
\begin{equation}
I_{rms}=\sqrt{I_0^2+
q(q-2)\left[
I_0^2+
{{\epsilon^2}\over{2q(q-2)}}
{{1}\over{1-e^{-q(2-q)}}}
\right]}.
\label{smeq9}
\end{equation}
Expanding the denominator to first order,
$1-e^{-q(2-q)}\cong q(2-q)$, yields
\begin{equation}
I_{rms}={{1}\over{\sqrt{2(2-q)}}}
\epsilon\,q^{-1/2}.
\label{smeq10}
\end{equation}

This result is particularly significant. From scaling theory, the stationary
state is expected to behave as
$I_{rms}\propto\epsilon^{\alpha_1}q^{\alpha_2}$ for large $n$. A direct
comparison with Eq.~(\ref{smeq10}) immediately gives
$\alpha_1=1$ and $\alpha_2=-1/2$, in excellent agreement with previous
phenomenological predictions. Remarkably, the same result can be obtained
directly from the mapping equations by imposing
$\overline{I^2}_{n+1}=\overline{I^2}_n=\overline{I^2}_{sat}$, which leads to
$I_{sat}={{1}\over{\sqrt{2(2-q)}}}\epsilon\,q^{-1/2}$.

The limit of small $n$ constitutes the third regime of interest. Assuming an
initial action $I_0\cong0$, which is therefore negligible compared with
$\epsilon$, and performing a Taylor expansion of the exponential term in the
numerator of Eq.~(\ref{smeq9}), we obtain
$I_{rms}(n)\cong\sqrt{{\epsilon^2}n/2}$. This result demonstrates that, for
short times, an ensemble of particles diffuses along the chaotic attractor in a
manner analogous to a random walk, exhibiting normal diffusion with exponent
$\beta=1/2$. Accordingly, a scaling hypothesis in the short-time regime can be
written as $I_{rms}(n)\propto(n\epsilon^2)^{\beta}$, with $\beta=1/2$, in full
agreement with the theoretical prediction discussed above.

A fourth relevant regime arises for intermediate values of $n$ and non-negligible
initial action $I_0$, such that $0<I_0<I_{sat}$. In this parameter window, an
additional crossover is observed at
$n_x^{\prime}\cong2I_0^2/\epsilon^2$, marking the transition between distinct
growth regimes.

The fifth regime corresponds to the case $I_0\cong0$, where $I_{rms}$ initially
grows with $n$, followed by a crossover and subsequent bending toward the
saturation regime. This characteristic crossover occurs at
$n_x\cong[(2-q)q]^{-1}$. From the scaling approach, one assumes
$n_x\propto\epsilon^{z_1}q^{z_2}$, yielding $z_1=0$ and $z_2=-1$, in
agreement with the results obtained above.

The final regime of interest corresponds to the limit
$I_0\gg{\epsilon^2}/[2q(2-q)]$. In this case,
Eq.~(\ref{smeq8}) can be rewritten as
\begin{equation}
I_{rms}(n)=\sqrt{I_0^2e^{-(n+1)q(2-q)}+
\epsilon^2
{{1-e^{-(n+1)q(2-q)}}\over{2q(2-q)}}}.
\label{smeq11}
\end{equation}
For short times, the leading behaviour is
$I_{rms}(n)=I_0e^{-(n+1)q(2-q)/2}$, while the stationary state is
recovered in the limit $n\rightarrow\infty$, yielding
$I_{rms}={\epsilon}/[\sqrt{2(2-q)}\,q^{1/2}]$, in full agreement with
the previous results.

\begin{figure}[t]
\centerline{\includegraphics[width=1.0\linewidth]{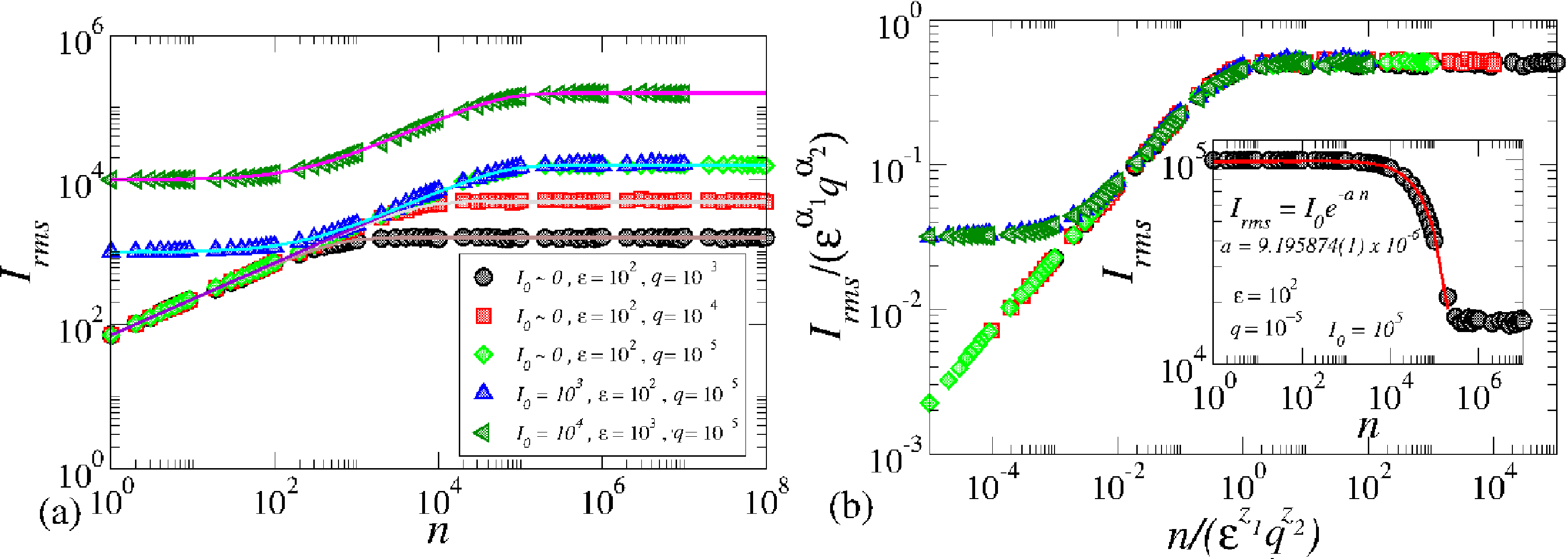}}
\caption{(a) plot of $I_{rms}$ versus $n$ for different control
parameters and initial conditions, as indicated in the figure. Filled symbols
correspond to numerical simulations obtained directly from iterating the
dynamical equations of the mapping for an ensemble of $M=10^3$ particles, all
starting with the same initial action $I_0$ and different initial phases
$\phi_0\in[0,2\pi]$. The analytical result given by Eq.~(\ref{smeq8}) is plotted
as a continuous line.}
\label{smFig2}
\end{figure}

Figure~\ref{smFig2}(a) shows a plot of $I_{rms}$ versus $n$ for different control
parameters and initial conditions, as indicated in the figure. Filled symbols
correspond to numerical simulations obtained directly from iterating the
dynamical equations of the mapping for an ensemble of $M=10^3$ particles, all
starting with the same initial action $I_0$ and different initial phases
$\phi_0\in[0,2\pi]$. The analytical result given by Eq.~(\ref{smeq8}) is plotted
as a continuous line. The agreement between analytical and numerical results is
remarkably good.

Figure~\ref{smFig2}(b) shows the collapse of the curves in
Fig.~\ref{smFig2}(a) onto a single universal curve after applying the scaling
transformations: (i) $I_{rms}\rightarrow
I_{rms}/(\epsilon^{\alpha_1}q^{\alpha_2})$ and
(ii) $n\rightarrow n/(\epsilon^{z_1}q^{z_2})$. The inset of
Fig.~\ref{smFig2}(b) illustrates the exponential decay predicted by
Eq.~(\ref{smeq11}). The parameters used in the inset were $\epsilon=10^2$,
$q=10^{-5}$, and $I_0=10^5$. The numerically measured decay rate,
$a=9.195874(1)\times10^{-6}$, is in close agreement with the theoretical value
$q(2-q)/2\cong9.99995\times10^{-6}$.

Taken together, the analytical treatment and the numerical simulations provide
a comprehensive and internally consistent description of the transition from
bounded to unbounded diffusion in the dissipative standard mapping. The
diffusion equation framework captures the full temporal evolution of
$I_{rms}(n)$ across all relevant regimes and naturally reveals the
underlying scaling invariance governing the approach to the stationary state.
The existence of well-defined crossover times, saturation values, and scaling
exponents demonstrates that the suppression of diffusion induced by
dissipation is not a mere quantitative effect, but rather a qualitative change
in the dynamical behavior of the system. In this sense, the conservative limit
$q\to0$ plays the role of a critical point separating two distinct
diffusive phases, characterized respectively by bounded and unbounded action
growth. This interpretation places the transition on firm theoretical grounds
and sets the stage for a unified treatment of diffusion suppression as a
continuous dynamical phase transition in low-dimensional chaotic systems.

\subsection{Application to billiard dynamics with time-dependent boundary}

\begin{figure}[b]
\centerline{\includegraphics[width=0.8\linewidth]{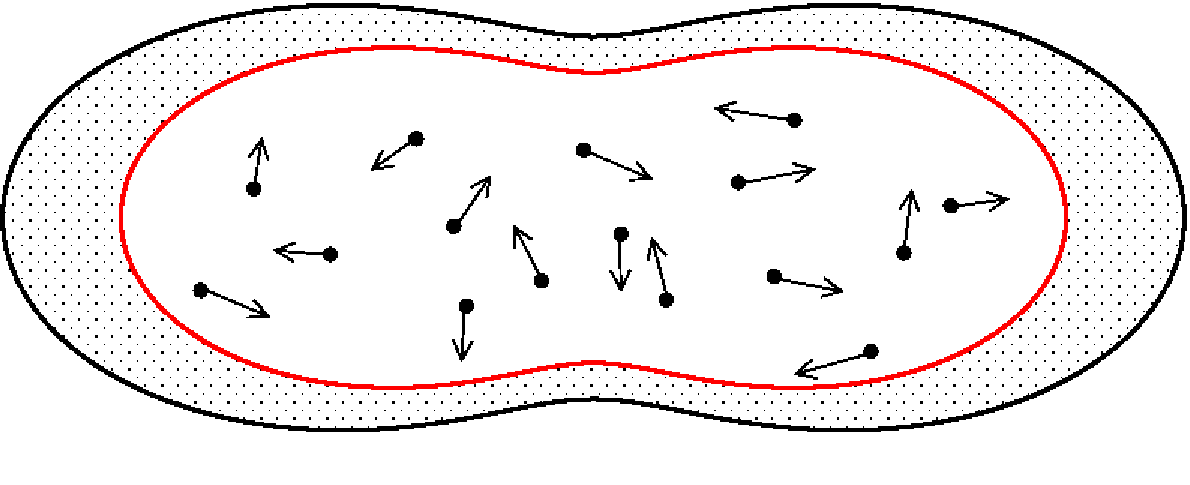}}
\caption{Sketch of an ensemble of particles moving in the time dependend billiard. The dashed area corresponds to the collision zone where the particles may change energy.}
\label{fig_bill}
\end{figure}

We now demonstrate the applicability of the formalism developed above to a more
complex system -- a billiard with a time-dependent boundary \cite{ref30}. The boundary confining an ensemble of non-interacting particles is described by $R(\theta,\eta,t)=1+\eta f(t)\cos(p\theta)$, where $p$ is an
integer. The case $\eta=0$ corresponds to the circular billiard, which is
integrable and exhibits a foliated phase space \cite{ref19}. For $\eta\neq0$
and $f(t)={\rm const.}$, the phase space becomes mixed, displaying chaotic
regions, invariant spanning curves, and periodic islands. The unlimited energy growth for an ensemble of particle is called as Fermi acceleration \cite{Fermi} and occurs when $f(t)=1+\epsilon\cos(\omega t)$. Figure \ref{fig_bill} shows a ilustrative sketch of the system.

Before we delve into the problem, let us first motivate the discussion and bring it to a physical application clear. Consider a rigid container in thermal contact with the environment at a temperature $T_b$. In this container, all particles are removed, creating a supposedly perfect vacuum. Next, we inject a set of low-density particles at a temperature $T \ll T_b$. From a thermodynamic perspective, thermal equilibrium will be reached when the temperature of the particle gas matches the temperature of the boundary. Once thermal equilibrium is achieved, changes -- by changes, we refer to the equilibrium average values, although fluctuations around the average value are always observed due to the statistical nature of the particles -- in temperature will only be observed if there is a change in the boundary temperature. It is also expected that if the particle gas is injected at a temperature $T \gg T_b$, its temperature will gradually decrease until thermodynamic equilibrium is reached. This represents the empirical knowledge of thermodynamics that has been accumulated over many years.

Now, we turn to the concept of billiards. The container where the vacuum was created consists of a crystalline lattice of particles. They are bound together by electronic forces, defining the mechanical properties of the system, such as rigidity, thermal conductivity, expansion properties, heat capacity, etc. When the container is at a temperature $T_b > 0$, the particles in the lattice oscillate around an equilibrium position with a characteristic frequency, forming a system comparable to the Einstein solid. Depending on the geometric shape of the boundary, the dynamics of the particles inside the billiard are expected to be chaotic. This is the necessary condition for the Loskutov-Ryabov-Akinshin (LRA) conjecture \cite{lra}. It clayms that the presence of chaotic dynamics in a billiard with fixed boundaries is a sufficient condition for the occurrence of Fermi acceleration in the system once a time perturbation to the boundary is introduced. Then it guarantees the unbounded growth of the energy of the particle ensemble, thus leading to the phenomenon of Fermi acceleration, hence unlimited energy diffusion.

On the other hand, the phenomenon of Fermi acceleration implies that the average velocity of the particle ensemble increases, which leads to an increase in the root mean square velocity. The equipartition theorem uses the root mean square velocity, which defines the kinetic energy of the particles, to establish the connection with temperature. Thus, the root mean square velocity is directly proportional to the temperature. If the root mean square velocity increases over time, the temperature also increases. This conclusion is immediate and arises from the formalism of time-dependent billiards.

However, this result contradicts the empirical knowledge of thermodynamics, which states that, in the stationary state, equilibrium occurs between the temperature of the particle ensemble and the boundary temperature. We will show that this contradiction between the expected result from time-dependent billiard theory and thermodynamics can be resolved by assuming that the collisions between the particles and the boundary are inelastic. This implies that a fractional energy loss occurs at each collision, ensuring the existence of attractors in phase space and suppressing the unlimited growth of energy. This suppression allows thermodynamic equilibrium to be ultimately achieved.

We therefore consider $f(t)=1+\epsilon\cos(\omega t+Z)$, where $Z\in[0,2\pi]$ is a random variable generated at each collision with the moving boundary. The dynamics is governed by a four-dimensional nonlinear mapping \cite{ref30} involving the particle velocity $V_n$, the collision time $t_n$, the polar angle $\theta_n$, and the trajectory angle $\alpha_n$. The boundary velocity at impact is
$\vec{V}_b(t)=d\vec{R}_b/dt\,(t+Z)$. The reflection laws are
$\vec{V}^{\prime}_{n+1}\cdot\vec{T}_{n+1}=
\vec{V}^{\prime}_{n}\cdot\vec{T}_{n+1}$ and
$\vec{V}^{\prime}_{n+1}\cdot\vec{N}_{n+1}=
-q\vec{V}^{\prime}_{n}\cdot\vec{N}_{n+1}$, where
$q\in[0,1]$ is the restitution coefficient. The prime stands for the moving referential frame, where the momentum conservation law applies. The case $q=1$ corresponds to the conservative dynamics, while $0<q<1$ introduces dissipation.

For $q=1$, the particle velocity undergoes unbounded diffusion, leading to
Fermi acceleration, hence an unbounded state. In contrast, for $q<1$, diffusion is suppressed and the phase space becomes bounded. The diffusion coefficient \cite{ref18} for this model is
\begin{equation}
D(\eta,\epsilon,q,n)=
{{\overline{V^2}_n}\over{4}}+
{{(1+q)^2\eta^2\epsilon^2}\over{16}},
\label{eq12}
\end{equation}
with
\begin{equation}
\overline{V^2}(n)=
V_0^2e^{-n(q^2-1)/2}+
{{(1+q)\eta^2\epsilon^2}\over{4(1-q)}}
\left(1-e^{-n(q^2-1)/2}\right).
\label{eq13}
\end{equation}

The behaviour of $V_{rms}(n)$ can be summarized as follows: (i) for short times, $V_{rms}(n)\propto n^{\beta}$; (ii) for long times, the stationary state scales as
$V_{sat}\propto(1-q)^{\alpha_1}(\eta\epsilon)^{\alpha_2}$; and (iii) the
crossover iteration number behaves as $n_x\propto(1-q)^{z_1}(\eta\epsilon)^{z_2}$. Applying the same analytical procedure developed throughout this work yields the critical exponents $\alpha_1=-0.5$, $z_1=-1$, $\alpha_2=1$, $z_2=0$, and $\beta=0.5$.

\begin{figure}[t]
\centerline{\includegraphics[width=0.475\linewidth]{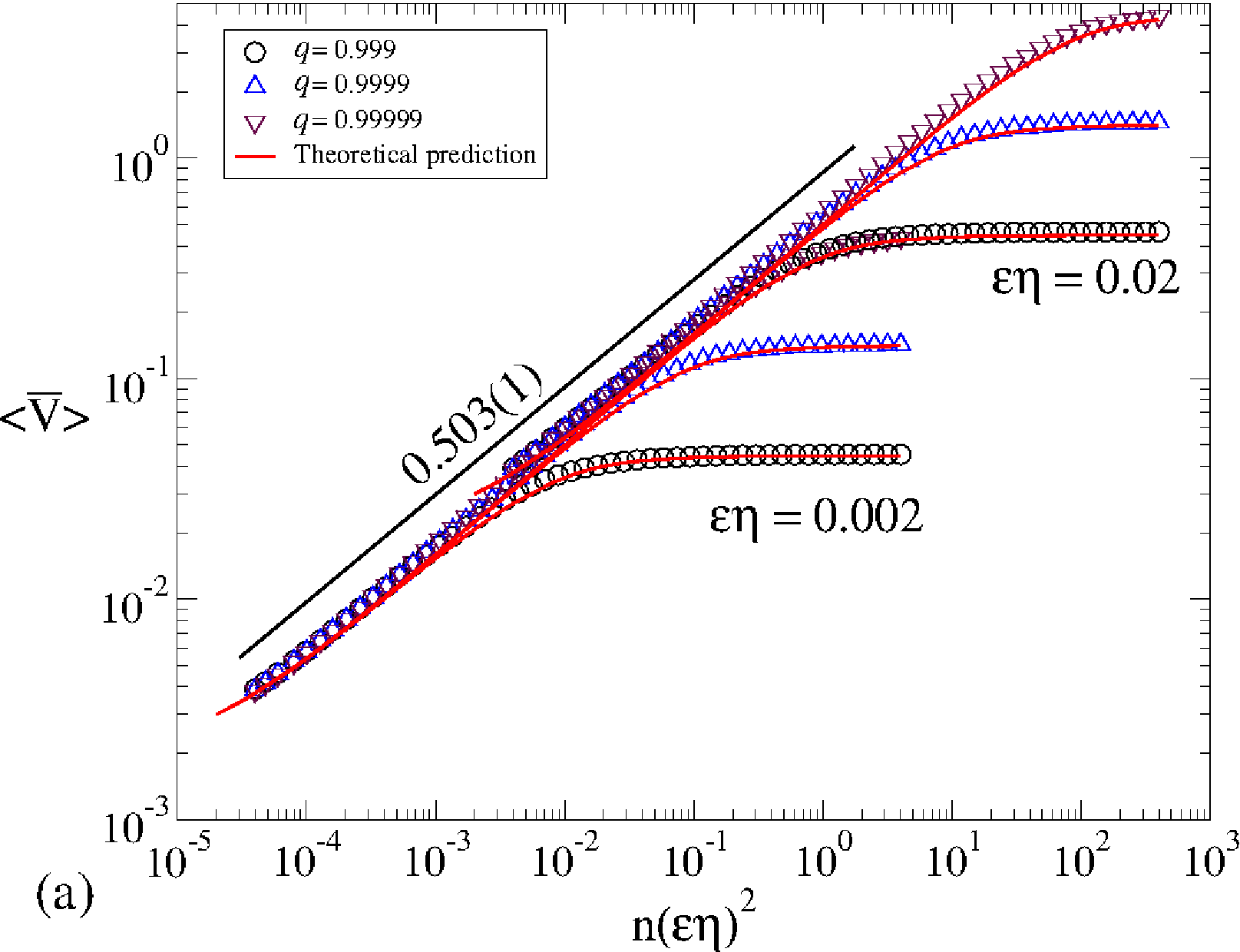}
\includegraphics[width=0.475\linewidth]{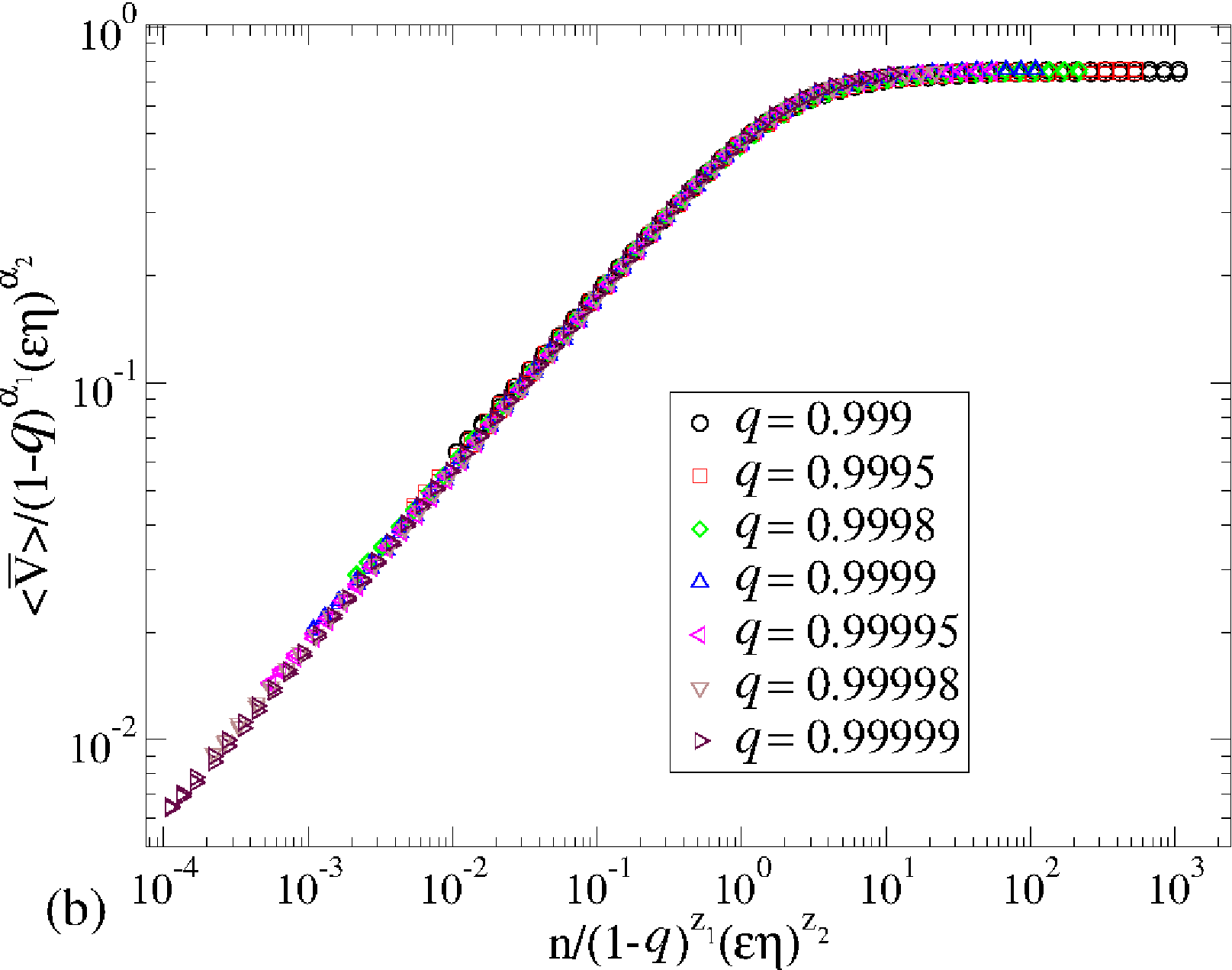}}
\caption{(a) Plot of $<\overline{V}>~vs.~n$ for different values of 
$q$ and two combinations of $\eta\varepsilon$. (b) Overlap of the curves 
shown in (a) onto a single and universal plot after the following scaling 
transformations: $n\rightarrow n/[(1-q)^{z_1}(\eta\varepsilon)^{z_2}]$ and 
$<\overline{V}>\rightarrow<\overline{V}>/[(1-q)^{\alpha_1}(\eta\varepsilon)^{
\alpha_2}]$. The straight line gives the theoretical prediction.}
\label{bill_Fig3}
\end{figure}

A clear confirmation of the scaling invariance of the diffusion is shown for the curves of $<\overline{V}>~vs.~n$, as shown in Fig. \ref{bill_Fig3}(a), that when plotted in scaling variables, overlap onto each other in a single and universal plot, as shown in Fig. \ref{bill_Fig3}(b).

The results discussed above provide a clear physical resolution of the apparent
contradiction between the predictions of time-dependent billiard theory and
the fundamental principles of thermodynamics. In conservative models
($q=1$), the unbounded growth of the particle energy leads to an ever-increasing root mean square velocity and, consequently, to a continuously increasing effective temperature. Such behaviour is incompatible with the thermodynamic expectation that a particle ensemble confined by a boundary at temperature $T_b$ must ultimately reach
thermal equilibrium.

The introduction of inelastic collisions ($q<1$) resolves this paradox
in a natural and physically consistent manner. Dissipation suppresses the
unbounded diffusion of energy, generates attractors in phase space, and leads
to a stationary state in which the average kinetic energy of the particle
ensemble saturates. In this regime, the system becomes compatible with the
thermodynamic picture of energy exchange between particles and a vibrating
boundary, allowing equilibrium to be achieved.

Remarkably, this entire scenario is quantitatively captured by the scaling
formalism. The short-time growth, long-time saturation, and crossover regimes
of the root mean square velocity are governed by well-defined critical
exponents, and the corresponding curves collapse onto universal scaling
curve when expressed in appropriate variables. This demonstrates that the
suppression of Fermi acceleration by dissipation is not a specific
feature, but rather a robust manifestation of scaling invariance in
nonequilibrium dynamical systems.

Taken together, these results highlight the power of the scaling approach when
combined with physically motivated modeling. The formalism reconciles
dynamical billiard theory with thermodynamics and provides a unified
framework to describe energy transport, diffusion suppression, and stationary
states in complex systems with time-dependent boundaries. This reinforces the
role of scaling invariance as a fundamental organizing principle in nonlinear
dynamics far from equilibrium.

As a short summary, the scaling approach is used with a great success to describe a 
transition from limited to unlimited diffusion in a time-dependent billiard. A set of critical exponents for a far more complex system in the billiard case is the same as those observed in a two-dimensional mapping like the Chirikov-Taylor map setting the two transitions in different models to the same universality class.

\section{Conclusions}
\label{sec5}

Scaling invariance emerges as one of the most powerful and unifying concepts in
contemporary physics, providing a common language to describe systems that are
very different in their microscopic details, dimensionality, and physical
origin. In this work, we have shown that the scaling formalism offers a coherent and
physically transparent framework for understanding a wide class of phenomena in
nonlinear dynamics, ranging from simple geometrical constructions to bifurcations,
chaotic diffusion, and genuine dynamical phase transitions.

By systematically analysing systems governed by one, two and three control parameters,
we demonstrated how power-law behaviour, crossover phenomena, and critical
exponents naturally arise whenever characteristic scales are absent. In the context
of local bifurcations, scaling provides a quantitative description of relaxation
processes, critical slowing down, and universality classes, revealing deep connections
between systems of different dimensionality and nonlinear structure.

More importantly, we have shown that transitions observed in chaotic dynamical
systems -- such as the loss of integrability, the suppression of diffusion, and the
onset of unbounded transport -- can be consistently interpreted as continuous
(second-order) phase transitions. Within this perspective, order parameters vanish
continuously at criticality, susceptibilities diverge, and scaling functions collapse
data obtained for different parameters onto universal curves. Concepts traditionally
associated with statistical mechanics, including symmetry breaking, elementary
excitations, and topological defects, acquire a clear dynamical interpretation in phase space, bridging deterministic chaos and nonequilibrium statistical physics.

The generality of the scaling approach is particularly striking. Area-preserving and
dissipative mappings, static and time-dependent billiards, and stochastic models of
the Fermi--Ulam type were shown to have sets of critical exponents, despite
their apparent differences. This universality highlights that scaling properties are
governed not by microscopic details, but by the underlying structure of the dynamics
near criticality.

Beyond its explanatory power, the scaling formalism also provides a practical and
predictive tool. It allows the identification of critical regimes, the classification of
transitions, and the anticipation of dynamical behaviour across different parameter
ranges. As such, it represents a natural bridge between nonlinear dynamics, chaos
theory, and the broader framework of statistical mechanics.

We hope that the unified perspective presented here will encourage the application
of scaling ideas to a wider class of problems, extending beyond dynamical systems to
fields such as soft matter, geophysics, biological systems, and complex networks,
where scale invariance and critical phenomena continue to play a central role.

\appendix
\section{Paper boat folding scheme}
\label{app1}

This appendix is dedicated to illustrate the procedure used in the folding paper boat.
\begin{figure}[t]
\centerline{\includegraphics[width=0.7\linewidth]{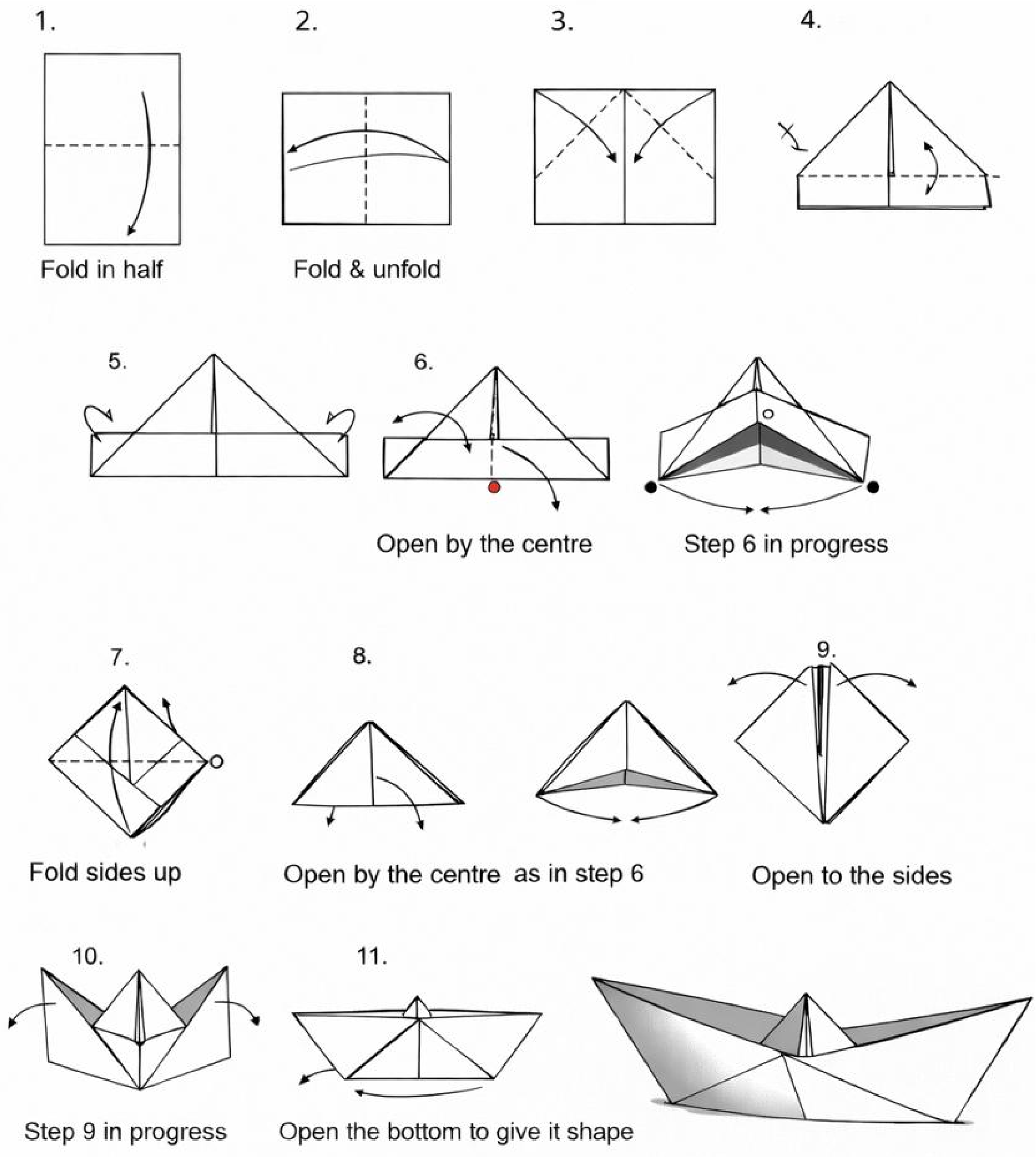}}
\caption{Sequence of steps used in the folding paper boat.}
\label{fig_app}
\end{figure}

Figure \ref{fig_app} is in charge of this and is self-contained.

\end{document}